\documentclass[a4paper,10pt, twoside, twocolumn]{article}
\usepackage{amsmath}
\usepackage{multicol}
\usepackage{graphicx}
\usepackage[english]{babel}
\usepackage{natbib}
\usepackage{fancyhdr}
\usepackage{xspace}

\makeatletter
\def\fnum@figure{\footnotesize{\bfseries \figurename\nobreakspace\thefigure}}
\def\fnum@table{\footnotesize{\bfseries \tablename\nobreakspace\thetable}}
\makeatother

\makeatletter
\renewcommand\section{\@startsection {section}{1}{\z@}%
	{-3.5ex \@plus -1ex \@minus -.2ex}%
	{2.3ex \@plus.2ex}%
	{\setcounter{equation}{0}\noindent\normalsize\bfseries\uppercase}}
\makeatother

\makeatletter
\renewcommand\subsection{\@startsection{subsection}{2}{\z@}%
	{-3.25ex\@plus -1ex \@minus -.2ex}%
	{1.5ex \@plus .2ex}%
	{\noindent\normalsize\bfseries}}
\makeatother

\makeatletter
\renewcommand\subsubsection{\@startsection{subsubsection}{2}{\z@}%
	{-3.25ex\@plus -1ex \@minus -.2ex}%
	{1.5ex \@plus .2ex}%
	{\noindent\normalsize\itshape}}
\makeatother

\renewenvironment{figure}{
	\begin{oldfigure}
	\begin{center}
	}{
	\end{center}
	\end{oldfigure}}
\newenvironment{figure2}{
	\begin{figure*}
	\begin{center}
	}{
	\end{center}
	\end{figure*}}

\renewenvironment{table}{
	\begin{oldtable}
	\begin{center}
	\begin{footnotesize}
	}{
	\end{footnotesize}
	\end{center}
	\end{oldtable}}
\newenvironment{table2}{
	\begin{table*}
	\begin{center}
	\begin{footnotesize}
	}{
	\end{footnotesize}
	\end{center}
	\end{table*}}

\renewenvironment{thebibliography}[1]{
	
	\begin{oldthebibliography}{#1}
	\setlength{\itemsep}{0pt}
	\begin{small}
	}{
	\end{small}
	\end{oldthebibliography}}
	
\newcommand{\dd}{\ensuremath{\mathrm{d}}}
\newcommand{\U}[1]{\ensuremath{\mathrm{~#1}}}

\newcommand{\yr}{\U{yr}}
\newcommand{\Myr}{\U{Myr}}
\newcommand{\Gyr}{\U{Gyr}}

\newcommand{\msun}{\U{M}_{\odot}}

\newcommand{\reft}[1]{Table~\ref{tab:#1}}
\newcommand{\reff}[1]{Figure~\ref{fig:#1}}
\newcommand{\tn}[1]{[\emph{T.N.}: #1]}

\newcommand{\fb}[1]{{\bf{#1}}}
\newcommand{\mc}[1]{\mathcal{#1}}
\newcommand{\e}[1]{\mathrm{e}^{#1}}

\begin{document}
\thispagestyle{empty}
\onecolumn

\noindent{\LARGE\bf On the dynamical evolution of globular clusters}

\vspace{0.7cm}
\noindent{\Large Michel~H\'enon$^{\star}$, translated by Florent~Renaud$^{1,2}$}

\vspace{0.2cm}
{\footnotesize \noindent \it
$^{\star}$ Institut d'Astrophysique, Paris (now at the Observatoire de Nice)\\
$^1$ Observatoire Astronomique and CNRS UMR 7550, Universit\'e de Strasbourg, 11 rue de l'Universit\'e, F-67000 Strasbourg, France\\
$^2$ Institute of Astronomy, University of Cambridge, Madingley Road, Cambridge, CB3 0HA, UK\\
\phantom{$^2$} \emph{florent.renaud@astro.unistra.fr}}\\

\noindent{\footnotesize Originaly published in October 1961; translated in October 2010}

\vspace{0.5cm}

\begin{flushright}
\begin{minipage}{12cm}
\setlength{\parindent}{15pt}

This paper is an English translation of Michel H\'enon's thesis, \emph{Sur l'\'evolution dynamique des amas globulaires} originally published in French in the Annales d'Astrophysique, Vol. 24, p.369 (1961).


Conventions and notations are as in the original version, for consistency. The English version is written so that it is as faithful to the French text as possible. The translator added some notes [\emph{T.N.}] for the sake of clarity, when required. Original French, English and Russian abstracts written by M.~H\'enon are available in the original version of the paper. The English part is reproduced below.

FR thanks Michel H\'enon for the enthusiasm and kindness he expressed when he was asked for permission to translate his work, as well as Douglas Heggie and Mark Gieles for their careful proofreading.

As the author explains himself on the cover page of the original version: \\
This work is the main thesis that M.M.~H\'enon presented to obtain the degree of \emph{Docteur \`es Sciences Physiques}, at the \emph{Falcult\'e  des Sciences de l'Universit\'e de Paris}. The thesis has been defended in Paris, on 1961 December 11, before a jury composed of Messieurs Danjon (president of the jury), Schatzman and Delcroix. This work is not related to a series of papers previously published and entitled \emph{L'amas isochrone}.

\end{minipage}
\end{flushright}

\vspace{0.5cm}
\begin{multicols}{2}

\section*{Original abstract}

Chapter~I: The structure and evolution of globular clusters are closely linked; we propose here to study them simultaneously, with a purely theoretical approach. The essential hypotheses are: (1) spherical symmetry; (2) quasi-permanent regime; (3) isotropy of the velocities at all points; (4) mass equality.

Chapter~II: We establish the system of fundamentals equations (2.25). The cluster model is reduced to a canonic form by a homology transformation with time-dependent parameters. We will look for a model of invariable canonic form, that is a model which remains similar to itself while evolving.

Chapter~III: We show that the galactic field imposes the relation (3.8): the cluster's radius is proportional to the cube root of its mass.

Chapter~IV: Preliminary calculations show that the density must be infinite at the center of the cluster. We obtain the asymptotic expression (4.14) for the distribution function near the center. The conservation of mass imposes condition (4.32). The existence of an energy flux toward the cluster center is predicted.

Chapter~V: The ensemble of equations and conditions obtained in the preceding chapters forms a system which is resolved numerically by means of an electronic calculator. We find a unique solution: the ``homology model'' \tn{homologous model} (\reft{1}, Figures~\ref{fig:4}, \ref{fig:5}, \ref{fig:6}, \ref{fig:15}, \ref{fig:16}). Its mass and radius are finite. The outer radius is approximately 10 times the mean radius. The mass decreases linearly as a function of time. About one-third of the negative energy of the cluster is carried off by the stars which escape; the other two-thirds accumulate in the center, in multiple stars.

Chapter~VI: We introduce a small number of stars of different mass in the homology model and we calculate their distribution (Figures~\ref{fig:8} and \ref{fig:9}) and their escape rate (\reff{10}). We find that as their mass increases, they are more concentrated, and escape less rapidly. In particular, if their mass is greater than 3/2 the mass of the normal stars, they are nearly all collected near the center of the cluster, and their escape rate is zero.

Chapter~VII: We show that should the initial central density of a cluster be finite, it will grow and become infinite within a finite time (\reff{12}). We study next the evolution of a cluster differing slightly from the homology model: we find that the differences decreases, whatever their form. Thus the homology model is very probably the final state toward which the clusters tend.

\end{multicols}\twocolumn\noindent 

Chapter~VIII: The theoretical results are compared with the observational data on globular clusters. First we construct an ``artificial cluster'' (Figures~\ref{fig:15} and \ref{fig:16}) which permits a general comparison. Then we compare the projected densities in detail, Sandage's star counts for the cluster M3, and the brightness measurements for 47~Tuc by Gascoigne \& Burr. The agreement is very satisfying (Figures~\ref{fig:18}, \ref{fig:19}, \ref{fig:21}). For the stars of M3, we find a definite mass-luminosity relation, which however is in disagreement with the relation indicated by stellar evolution theory (\reff{20}). The observed masses and radii are in good agreement with the theoretical relation (\reff{22}). The outer radii of the clusters are about twice as large as the observed radii. The absolute escape rate is constant and equals: $2.3\times 10^{-6} \msun/\yr$. This constancy is confirmed by the observed distribution of globular cluster masses (\reff{23}). We obtain the initial mass distribution. Observation shows that $\omega$~Cen is the only globular cluster whose central density is not yet infinite. From this we deduce that the age of the globular clusters must lie between $24 \times 10^{9} \yr$ and $44 \times 10^{9} \yr$, in good agreemen [\emph{sic}] with the age deduced from stellar evolution.

Chapter~IX: We list a number of directions in which development of research is to be hoped for.

\section{Introduction}

The problem of the dynamical evolution of a globular cluster can be stated very simply. The only important force is the mutual attraction between the stars of the cluster; the other forces (like radiative pressure, electromagnetic forces, relativistic effects, etc.) are negligible. Therefore, the topic is the classical $n$-body problem: finding the motion of $n$ points of given masses, mutually attracting themselves as the inverse square of their distance.

This exposition, whereas simple, relates to an extremely arduous mathematical problem. Despite a large number of studies, it has not been possible to find an explicit solution, which very likely does not exist. Hence, one can think of the numerical integration. This way, \citet{vonHoerner1960} computed the evolution of artificial clusters comprising up to 16 stars, thanks to an electronic device. But high values of $n$ are out of reach with such a method, as the computational time becomes rapidly extreme, even for a machine; the case of $n=16$ already corresponds to a system of 96 simultaneous differential equations.

In the globular clusters, $n$ is of the order of magnitude of $10^{5-6}$. Such a high value naturally suggests to give up following the individual motions, and to use a statistical method. The structure of the cluster is then defined at all times thanks to a distribution function in a 7-dimensional space: 3 position coordinates, 3 velocity coordinates, and the mass. One can write a system of equations that allows, in principle, to calculate the evolution of this function, given its initial form. Unfortunately, the numerical resolution seems absolutely out of reach with this general form: indeed, one faces an integro-differential system with 8 independent variables!

Thus, it is necessary to make simplifications, or additional hypotheses. Two of them are classical and well valid:\\
(H1) we assume that the cluster is spherically symmetric;\\
(H2) we assume that the cluster has reached the steady state;\\
The number of independent variables is then reduced from 8 to 4 (see Equation 2.2 below); but the equations are still too involved. Therefore, one must make new simplifications, much more arbitrary, that are often only justified by their practical utility. The main one consists in considering the structure and the evolution problems separately. This is how a series of studies (e.g. \citealt{Plummer1911, Plummer1915, Eddington1916, Jeans1916, Chandrasekhar1942} Section 5.8, \citealt{Camm1952, Woolley1956, Henon1959}) has focussed on the structure of the clusters, without considering the evolution. In this case, the distribution function can take any form; it is set by an additional, arbitrary hypothesis, changing from one author to the other. The models obtained this way represent well the observed clusters; however this agreement does not mean much, because one of the fundamental equation of the problem as been suppressed, and replaced by an \emph{ad hoc} hypothesis.

Other works have, on the contrary, studied the evolution by assuming a known structure for the cluster (among others, see \citealt{Spitzer1940, Chandrasekhar1943b, Chandrasekhar1943c, King1958a, King1958b, King1958c, Spitzer1958b, vonHoerner1958, Agekian1959, Henon1960, King1960, Michie1961}). In most of the cases, it is assumed that the cluster is homogeneous and that the gravitational potential is uniform; the advantage of this is to get rid completely of the space variables. However, the results obtained (the major one being the escape rate of the stars from the cluster) often disagree from one author to the next (see \reff{10}), as a consequence of the arbitrary hypothesis, that has been introduced here again.

In fact, the two aspects, structure and evolution, cannot be told apart; they are intimately related. Equations show this clearly (see Equation 2.22 below). First solving the problem of the structure of the cluster, and then finding out its evolution, or the other way round, is not possible. All the equations have to be treated and solved as a whole.

We hope that the present work constitutes a first step toward this. The homologous model presented in Chapter~V is obtained by solving simultaneously the equations of structure and evolution. However, this model is far from being a satisfactory solution to the problem. Indeed, it has been necessary to keep two arbitrary hypotheses, to avoid a too high complexity in the calculations:\\
(H3) we assume that the distribution function only depends on the total energy (this is equivalent to assuming that the velocity distribution is everywhere isotropic);\\
(H4) we assume that all the stars have the same mass.

It is difficult to known how these simplifications affect the correctness of the results. Therefore, the conclusions should be taken carefully, and considered as illustrations instead of definitive statements. In fact, the main goal of this work is not to immediately obtain useful informations, but rather to set a more rigorous theory by getting rid of one of the arbitrary postulates it used to have, and by pointing at those that remain to be narrowed (see Chapter~IX).

One can ask whether it would be possible to make progress thanks to direct observations, and thus to circumvent or at least diminish, the difficulties of the theoretical work. Unfortunately, it seems that such a method is not very successful. Observing globular clusters is difficult, because of a number of reasons: the stellar density varies a lot from the center to the outskirts, so that it is impossible to get a satisfactory image of all the parts of the cluster on a single plate; one only observes the projected density and not the spatial density; measurements of individual proper motions are impossible to make; measurements of radial velocities are very imprecise, because of large distances. Worst, one observes only the brightest stars of the cluster; these stars only contribute to a small fraction of the total mass. In other words, the essential structure of the cluster remains invisible.

Observations of globular clusters bring too little information if one is to derive the theoretical model, characterized by a multi-variable function. This point can be illustrated thanks to the following quotes about the comparison between theoretical models and observed clusters: \tn{in english in the original version} ``the law ... represents the structure of a globular cluster with close approach to accuracy everywhere, with the exception of the region immediately surrounding the center'' \citep{Plummer1911}; ``this model appears to fit the variations of brightness of the observed disk'' \citep{Camm1952}; ``calculations with this model give projected densities which agree much better with observations than do those of the simple isothermal case'' \citep{Woolley1956}; ``with the first models we can obtain a rather fair fit'' \citep{vonHoerner1957}; ``l'amas isochrone ... ressemble d'une mani\`ere surprenante aux amas r\'eels'' \tn{the isochrone cluster ... looks surprisingly similar to the real clusters} \citep{Henon1959}. But each time, the model is different...

Thus, we can say that the agreement with observations is a necessary, but not sufficient condition for the validity of a theory on the dynamics of the clusters. Observations can be a final test for the theory, but not a starting point.

However, the theory can precisely progress in a purely deductive way, only owing to its own strengths. As we have seen, the physical problem can be translated into a system of equations describing it, if not perfectly rigorously, at least with a very good precision. This system encloses all the information required to solve the problem. That is, we feel that efforts should be made in this direction. Obviously difficulties exist; but they are only technical, mathematical difficulties, not conceptual ones. Thanks to the powerful computing machines nowadays, it does not seem impossible to solve the problem anymore.

\section{Equations}

\subsection{Definitions and hypotheses}

Let
\begin{equation}
\varphi(x,y,z,v_x,v_y,v_z,m,t)\,\dd x\,\dd y\,\dd z\,\dd v_x\,\dd v_y\,\dd v_z\,\dd m
\end{equation}
be, at time $t$, the number of stars of mass between $m$ and $m+\dd m$ and whose the six spacial and velocity coordinates lie between $x$ and $x+\dd x$ ... $v_z$ and $v_z+\dd v_z$. As we will see, this distribution function $\varphi$ fully describes the structure of the cluster.

We assume that the cluster is spherically symmetric and that it has reached a steady state or, strictly speaking, a quasi-steady state because evolution makes it change slowly \citep[see][]{Spitzer1940, Kuzmin1957}. One can show \citep{Jeans1915, Kurth1955} that the distribution function depends on the position and velocity coordinates only through two invariants (or strictly speaking, quasi-invariant):\\
$A$, the angular momentum of the star;\\
$E$, the total energy of the star, per unit mass (we will refer to this quantity as total energy, for conciseness).\\
Therefore, we have:
\begin{equation}
\varphi(x,y,z,v_x,v_y,v_z,m,t) = f(E, A, m, t).
\end{equation}
$E$ is given by the fundamental relation:
\begin{equation}
E= U + \frac{v^2}{2},
\end{equation}
where $U$ is the gravitational potential, and $v$ is the velocity of the star.

Here comes our first arbitrary hypothesis. We assume, for the rest of the paper, that $f$ does not depend on the angular momentum $A$. This is equivalent to assuming that the velocity distribution is isotropic (i.e. spherically symmetric) everywhere in the cluster, as shown in (2.3). Thus, the distribution function shrinks to:
\begin{equation}
f(E, m, t).
\end{equation}

Thanks to this hypothesis, one can have the following reasoning. The mutual perturbations of the stars tend to establish a Maxwellian, therefore isotropic, velocity distribution everywhere. The escape phenomenon goes, as we know, against such a distribution: every star rising above a certain critical velocity leaves the cluster (see Chapter~III). But this limitation only concern the modulus of the velocity and not its orientation; in other words, it is itself spherically symmetric. We can therefore assume that the Maxwellian equilibrium, which is impossible to establish in term of the distribution of the velocity moduli, does occur in term of the distribution of the orientations, i.e. the velocity distribution becomes isotropic everywhere, after a sufficient time.

But this reasoning is ambiguous: the stars travel across the entire cluster, and thus, it is forbidden to consider the evolution of a small region in isolation. Therefore, the validity of our hypothesis is not ensured\footnote{\citet{Michie1961} highlights an increase of the anisotropy in the external regions, for a particular model.}. We will examine the perspectives about this point in Chapter~IX.

\subsection{Equations of structure}

These equations are well-known. The density at a given point comes from the distribution function through:
\begin{equation}
\rho = \int\int\int\int m \varphi\, \dd v_x\, \dd v_y\, \dd v_z\, \dd m,
\end{equation}
or, using the isotropy of the velocities and (2.3):
\begin{equation}
\rho = 4\pi \int_0^{\infty} m\, \dd m \int_U^{\infty} \left(2E-2U\right)^{1/2} f\, \dd E.
\end{equation}
Furthermore, $\rho$, $U$ and the distance to the center $r$ are linked through Poisson's equation:
\begin{equation}
\frac{\partial^2 U}{\partial r^2} + \frac{2}{r} \frac{\partial U}{\partial r} = 4\pi G \rho.
\end{equation}
We note that $U$ depends on both $r$ and the time $t$ because of the evolution of the cluster: $U = U(r,t)$. In the same way: $\rho = \rho(r,t)$. The equations (2.6) and (2.7) give $\rho$ and $U$, i.e. the structure of the cluster, once the distribution function $f$ is known. It remains to write an essential equation: the one ruling the variation of $f$ with time.

\subsection{Local evolution equations}

If the potential of the cluster was perfectly regular, every star would keep its total energy forever and no evolution would occur. In reality however, the potential is created by a discrete distribution of masses and yields irregular features, that are random and change with time. As a consequence, the total energy of the star experiences perturbations. This question has often been addressed \citep[see e.g.][Section 2.1]{Chandrasekhar1942}; one showed that the star mostly experiences many small perturbations, whose cumulative effect leads to a slow, quasi-continuous, variation of the total energy of the star.

This total effect is proportional to $\ln{(l/l_1)}$, where $l$ is the maximum distance to the perturber stars, and $l_1$ is the ``impact parameter'', i.e. the distance by which two stars have to be separated to be deviated by an angle of 90$^\circ$. That is, even the very distant stars have a non-negligible effect. However, $l_1$ is very small: one can show that:
\begin{equation}
l_1 \simeq \frac{r_e}{n},
\end{equation}
where $r_e$ is the radius of the cluster. (Recall that $n$ is the total number of stars.) Consequently, the majority of the perturbations are caused by relatively close stars: e.g., if the cluster counts $10^5$ stars, $l_1$ equals $r_e \times 10^{-5}$, and 80\% of the perturbations would come from stars closer than $r_e \times 10^{-1}$. Therefore, in order to calculate the perturbations at a given point in the cluster, we will admit that the distribution of the velocities is the same in this point as everywhere else. This approximation probably leads to an error of about 10\%, which is acceptable given the current state of the theory.

The distribution function depends on $r$ and $v$, through $E$; to emphasize this, we set:
\begin{equation}
f(E, m, t) = f\left(U+\frac{v^2}{2}, m, t\right) = a(r, v, m, t).
\end{equation}

In this entire Section, we will study what happens at a given point in the cluster, at the distance $r$ from the center. The function $a$ defined in (2.9) describes the velocity distribution at this point.

\citet{Rosenbluth1957} gave the equation of evolution of such a distribution, with the effects of perturbations, in the very general case where there is no symmetry for the velocities, as well as in the case of axisymmetry. In our case of spherical symmetry for the velocities, their equation becomes (after some calculations)\footnote{This equation has recently been cited and used by \citet{King1960}.}:
\begin{eqnarray}
\left(\frac{\partial a}{\partial t}\right)_p & = & 16 \pi^2 G^2\ \ln{(n)} \int_0^{\infty} m_1\, \dd m_1\\
& & \frac{1}{v^2}\frac{\partial}{\partial v}\bigg[m a \int_0^v a_1 v_1^2\, \dd v_1 \nonumber\\
& & + \frac{m_1}{3} \frac{\partial a}{\partial v} \bigg( \frac{1}{v} \int_0^v a_1 v_1^4\, \dd v_1 \nonumber \\
& & + v^2 \int_v^{\infty} a_1 v_1\, \dd v_1 \bigg)\bigg]\nonumber
\end{eqnarray}
where, for simplicity, we set $a(r,v_1,m_1,t) = a_1$. The maximum distance $l$ to the perturber stars has been taken equal to the radius of the cluster. The subscript $p$ reminds us that the variation of $a$ is due to perturbations.

\subsection{Global evolution equations}

It is now time to focus on the entire cluster. Let's consider the subset of stars of given mass $m$ and energy $E$, at time $t$. These stars are situated at different locations in the cluster; but, according to our hypotheses, they all corresponds to the same value of $f$, thus of $a$. And yet, after a very short time, the functional form of $a$ has changed. This modification, given by (2.10), is different from one position in the cluster to another, in general. As a consequence, $a$ does not take the same value over the entire subset of stars anymore. In other words, the perturbations instantaneously destroy the steady-state regime.

However, the steady-state re-appears almost immediately, because of the rapid circulation of the stars within the cluster; the perturbations become equal for all the stars of the subset, so that $a$ takes the same value for all of them, once again.

Thus, the actual variation of $a$, after the equalization, is obtained when computing the derivative of (2.9), i.e.:
\begin{equation}
\frac{\partial a}{\partial t} = \frac{\partial f}{\partial E} \frac{\partial U}{\partial t} + \frac{\partial f}{\partial t},
\end{equation}
and we write that $\partial a / \partial t$, given in (2.11), is equal in average to $\left(\partial a / \partial t\right)_p$, given in (2.10).

Let $g(E,m,t)\, \dd E\, \dd m$ be the total number of stars in the subset. One has:
\begin{equation}
g(E, m, t)\, \dd E\, \dd m = \int_0^{r_m} 4\pi r^2\, \dd r \cdot 4\pi a v^2\, \dd v\, \dd m,
\end{equation}
where the integral is with respect $\dd r$. The maximum value $r_m$ taken by $r$ for a given energy $E$, is set by:
\begin{equation}
U(r_m,t) = E,
\end{equation}
and the velocity $v$ is given in (2.3), i.e.:
\begin{equation}
v = \left(2E-2U\right)^{1/2}.
\end{equation}

Equation (2.12) becomes:
\begin{equation}
g = 16\pi^2 \int_0^{r_m} r^2\, \dd r \cdot av,
\end{equation}
and the number of stars in the subset changes as:
\begin{equation}
\frac{\partial g}{\partial t} = 16\pi^2 \int_0^{r_m} r^2\, \dd r \left[v\frac{\partial a}{\partial t} + \frac{\partial (av)}{\partial v}\left(\frac{\partial v}{\partial t}\right)_E\right].
\end{equation}
We write that this integral remains the same whether using (2.10) or (2.11). The second term in the integral is the same in both cases; therefore we get:
\begin{equation}
\int_0^{r_m} r^2\, \dd r\, v \left[\left(\frac{\partial a}{\partial t}\right)_p - \frac{\partial f}{\partial E}\frac{\partial U}{\partial t}-\frac{\partial f}{\partial t}\right] = 0,
\end{equation}
which is the equation of evolution we were looking for\footnote{This equation has already been presented, in a slightly different form however, by \citet[Equation 2.5]{Kuzmin1957}.}.

By changing $a$ and $a_1$ into $f$ and $f_1$ and using (2.14), it becomes:
\begin{eqnarray}
0 = \int_0^{r_m} r^2\, \dd r \Bigg\{ 16\pi^2 G^2\ \ln{(n)} \int_0^{\infty} m_1\, \dd m_1 \\
 \frac{\partial}{\partial E} \bigg[ m f \int_U^E f_1 \left(2E_1-2U\right)^{1/2} \dd E_1 \nonumber\\
 + \frac{m_1}{3}\frac{\partial f}{\partial E} \int_U^E f_1 \left(2E_1-2U\right)^{3/2} \dd E_1 \nonumber\\
 + \frac{m_1}{3} \frac{\partial f}{\partial E} \left(2E-2U\right)^{3/2} \int_E^{\infty} f_1\, \dd E_1 \bigg] \nonumber\\
 - \left(2E-2U\right)^{1/2} \left(\frac{\partial f}{\partial E}\frac{\partial U}{\partial t}+\frac{\partial f}{\partial t}\right)\Bigg\}.\nonumber
\end{eqnarray}
We can switch the operations $\int_0^{r_m} \dd r$ and $\frac{\partial}{\partial E}$, because the quantity between square-brackets becomes zero for $r=r_m$. Then, we switch the order of the integrations over $r$, $m_1$ and $E_1$, by recalling that $f$ and $f_1$ do not depend on $r$. Furthermore, we set:
\begin{equation}
\frac{1}{3}\int_0^{r_m} \left(2E-2U\right)^{3/2} r^2\, \dd r = q(E,t),
\end{equation}
(\tn{The upper limit of the integral is missing in the original version}) and thus,
\begin{eqnarray}
\frac{\partial q}{\partial E} &=& \int_0^{r_m} \left(2E-2U\right)^{1/2} r^2\, \dd r,\\
\frac{\partial q}{\partial t} &=& -\int_0^{r_m} \left(2E-2U\right)^{1/2} \frac{\partial U}{\partial t} r^2\, \dd r.
\end{eqnarray}
After these transformations, (2.18) takes its final form:
\begin{eqnarray}
0 &=& \Bigg\{ 16\pi^2 G^2\ \ln{(n)} \int_0^{\infty} m_1\, \dd m_1 \, \frac{\partial}{\partial E}\\
&& \bigg[ m f \int_{-\infty}^E f_1 q_1'\, \dd E_1 + m_1 f' \Big( \int_{-\infty}^E f_1 q_1\, \dd E_1 \nonumber\\
&& + q \int_E^{\infty} f_1\, \dd E_1 \Big) \bigg] + f' \frac{\partial q}{\partial t} - q' \frac{\partial f}{\partial t} \Bigg\}\nonumber
\end{eqnarray}
(Hereafter, the symbol $'$ indicates the partial derivative with respect to $E$: $\frac{\partial f}{\partial E} = f'$, and so on.).

We note that the structure of the cluster is present in this equation only via the function $q$, defined in (2.19). This function will play a very important role in the following.

We now have the complete system of fundamental equations that rule the structure and the evolution of the cluster: these are the four equations~(2.6), (2.7), (2.19) and (2.22), linking together the four quantities $f$, $\rho$, $U$, $q$. But before actually solving this system, we first have to modify it.

\subsection{Equal masses}

In the rest of this paper (except in Chapter~VI), we limit ourselves to the study of the case where all the stars have the same mass. This is our second arbitrary hypothesis, less critical than the first one however. Indeed, whereas it is obvious that the masses of the stars are different in real clusters, one can imagine a fictitious cluster where all masses would be equal. In other words, the system we study corresponds to a physically plausible situation, while being simpler than the real systems.

Therefore, the distribution function reads:
\begin{equation}
f(E,m,t) = \delta(m-m_1)\cdot F(E,t),
\end{equation}
where $m_1$ is the mass of each star, and $\delta$ is the Dirac's distribution \tn{Dirac delta function}. Performing the integrations over $m$ in (2.6) and (2.22) is immediate.

\subsection{Normalized variables}

In order to get rid of the numerical constants, we define ``normalized variables'':
\begin{eqnarray}
\rho &=& 4\pi m_1\ D,\nonumber\\
r & = & (16\pi^2\ G\ m_1)^{-1/2}\ R,\nonumber\\
q & = & (16\pi^2\ G\ m_1)^{-3/2}\ Q,\nonumber\\
\dd t & = & [16\pi^2\ G^2\ m_1^2\ \ln{(n)}]^{-1}\ \dd T.
\end{eqnarray}

The four fundamental equations become:
\begin{eqnarray}
D &=& \int_U^{\infty} (2E - 2U)^{1/2} F \, \dd E,\\
&& \frac{\partial^2 U}{\partial R^2} + \frac{2}{R}\frac{\partial U}{\partial R} = D,\nonumber\\
Q &=& \frac{1}{3}\int_0^{R_m} (2E - 2U)^{3/2} R^2\, \dd R,\nonumber\\
0 &=& \frac{\partial}{\partial E} \bigg[ F \int^E_{-\infty} F_1 Q_1' \, \dd E_1 + F' \Big( \int^E_{-\infty} F_1 Q_1 \, \dd E_1 \nonumber\\
&& + Q \int_E^{+\infty}F_1 \, \dd E_1 \Big) \bigg] + F'\frac{\partial Q}{\partial T} - Q' \frac{\partial F}{\partial T}. \nonumber
\end{eqnarray}

In addition, we set
\begin{equation}
F^{(-1)} = -\int_E^{\infty} F_1 \, \dd E_1.
\end{equation}
\begin{equation}
S= F^{(-1)} \int_{-\infty}^E F_1 Q_1'\, \dd E_1 - F \int_{-\infty}^E F_1^{(-1)} Q_1'\, \dd E_1.
\end{equation}

We have
\begin{equation}
\frac{\partial S}{\partial E} = S' =  F \int_{-\infty}^E F_1 Q_1'\, \dd E_1 - F' \int_{-\infty}^E F_1^{(-1)} Q_1'\, \dd E_1,
\end{equation}
i.e., thanks to a partial integration and by taking into account that $Q(-\infty) =0$:
\begin{eqnarray}
S' &=& F \int_{-\infty}^E F_1 Q_1'\, \dd E_1 \\
&& + F' \left( \int_{-\infty}^E F_1 Q_1\, \dd E_1 + Q \int_E^{\infty} F_1\, \dd E_1 \right), \nonumber
\end{eqnarray}
so that (2.25d) can be re-written:
\begin{equation}
0 = S'' + F' \frac{\partial Q}{\partial T} - Q' \frac{\partial F}{\partial T}.
\end{equation}

\subsection{Canonization}

We know \citep[see e.g.][]{Kurth1955} that every theoretical cluster model includes two dimensional parameters; or, in other words: one can apply to the model an homology based on two arbitrary parameters. This can be easily checked with the equations: at a given time $T_0$, the cluster is completely defined by the distribution function $F(E, T_0)$. The most general homologous transformation would be to multiply $F$ on the one hand, and $E$ on the other, with two arbitrary constants. (We assume that the masses of the stars are set; thus, it is not possible to change them through a homology.)

It is interesting to use this possibility of homology to set any cluster model to a \emph{canonical form} defined by two chosen conditions. This allows one to highlight the very structure of the model, in a way that is independent of its size. The two conditions could be, for example: masses and radius equal to defined values. However, we will choose others, more convenient (see Chapter~IV, Equation 4.6).

In the case of a model in evolution, which we study here, this model will always be set to the canonical form thanks to a proper homology. Therefore, the two parameters of the homology will be functions of time. This has the great advantage to allow us to split two aspects of the evolution of the model: (1) the evolution of the size, represented by the variation of the parameters of the homology; (2) the evolution of the structure itself, dimensionless, represented by the variation of the canonical form.

Therefore, we set:
\begin{equation}
\begin{array}{l}
E = \beta\ \fb{E}\\
F = \gamma\ \fb{F},
\end{array}
\end{equation}
where $\beta$ and $\gamma$ are the two fundamental parameters of the homology, chosen so that the new model, defined by $\fb{F}(\fb{E})$, is canonical. By replacing this in the equations, we find out that the transformation formulae of the other quantities are:
\begin{eqnarray}
U &=& \beta\ \fb{U},\\
D &=& \beta^{3/2}\ \gamma\ \fb{D},\nonumber\\
R &=& \beta^{-1/4}\ \gamma^{-1/2}\ \fb{R},\nonumber\\
Q &=& \beta^{3/4}\ \gamma^{-3/2}\ \fb{Q},\nonumber\\
\dd T &=& \gamma^{-1}\ \dd\fb{T},\nonumber\\
S &=& \beta^{7/4}\ \gamma^{1/2}\ \fb{S}.\nonumber
\end{eqnarray}

One can easily check that the first three equations of the system (2.25) remain the same with the new variables. The fourth one changes however, because of derivatives with respect to time. From (2.31a) and (2.32e), we get:
\begin{equation}
\frac{\partial (\fb{E},\fb{T})}{\partial (E,T)} = \left| \begin{array}{cc}\displaystyle\frac{1}{\beta} & \displaystyle-\frac{1}{\beta^2}\frac{\dd\beta}{\dd T}E\\ \displaystyle 0 & \displaystyle\gamma \end{array}\right|,
\end{equation}
which allows us to compute:
\begin{equation}
\frac{\partial F}{\partial E}\frac{\partial Q}{\partial T}-\frac{\partial Q}{\partial E}\frac{\partial F}{\partial T} = \frac{\gamma}{\beta}\left(\frac{\partial F}{\partial \fb{E}}\frac{\partial Q}{\partial \fb{T}}-\frac{\partial Q}{\partial \fb{E}}\frac{\partial F}{\partial \fb{T}}\right).
\end{equation}
By use of (2.31b) and (2.32d), this becomes:
\begin{eqnarray}
\frac{\partial F}{\partial E}\frac{\partial Q}{\partial T}-\frac{\partial Q}{\partial E}\frac{\partial F}{\partial T} = \\
\beta^{-1/4}\gamma^{1/2} \Bigg[\frac{\partial \fb{F}}{\partial \fb{E}}\frac{\partial \fb{Q}}{\partial \fb{T}}-\frac{\partial \fb{Q}}{\partial \fb{E}}\frac{\partial \fb{F}}{\partial \fb{T}}+\fb{Q}\frac{\partial \fb{F}}{\partial \fb{E}}\big(\frac{3}{4}\frac{1}{\beta}\frac{\dd\beta}{\dd\fb{T}}\nonumber\\
- \frac{3}{2}\frac{1}{\gamma}\frac{\dd\gamma}{\dd\fb{T}} \big) - \fb{F}\frac{\partial \fb{Q}}{\partial \fb{E}}\frac{1}{\gamma}\frac{\dd\gamma}{\dd\fb{T}}\Bigg]. \nonumber
\end{eqnarray}
Let:
\begin{eqnarray}
\frac{1}{\beta}\frac{\dd\beta}{\dd\fb{T}} &=& b,\\
\frac{1}{\gamma}\frac{\dd\gamma}{\dd\fb{T}} &=& c.\nonumber
\end{eqnarray}

With the new variables, the system (2.25) becomes:
\begin{eqnarray}
\fb{D} &=& \int_{\fb{U}}^{\infty} (2\fb{E} - 2\fb{U})^{1/2} \fb{F} \, \dd\fb{E},\\
&&\frac{\partial^2 \fb{U}}{\partial \fb{R}^2} + \frac{2}{\fb{R}}\frac{\partial \fb{U}}{\partial \fb{R}} = \fb{D},\nonumber\\
\fb{Q} &=& \frac{1}{3}\int_0^{\fb{R}_m} (2\fb{E} - 2\fb{U})^{3/2} \fb{R}^2\, \dd\fb{R},\nonumber\\
\fb{S}' &=& \fb{F} \int_{-\infty}^{\fb{E}} \fb{F}_1\fb{Q}_1'\, \dd\fb{E}_1 + \fb{F}' \bigg( \int_{-\infty}^{\fb{E}} \fb{F}_1 \fb{Q}_1\, \dd\fb{E}_1\nonumber\\
&&+ \fb{Q} \int_{\fb{E}}^{\infty}\fb{F}_1 \, \dd\fb{E}_1\bigg)\nonumber\\
0 &=& \fb{S}'' + \fb{F}' \frac{\partial \fb{Q}}{\partial \fb{T}} - \fb{Q'} \frac{\partial \fb{F}}{\partial \fb{T}} + \left(\frac{3}{4}b-\frac{3}{2}c\right)\fb{Q}\fb{F}-c\fb{F}\fb{Q}'.\nonumber
\end{eqnarray}
(The symbol $'$ indicates here the derivative with respect to $\fb{E}$, and not $E$ any more.)

\subsection{Secondary equations}

We write here the equations that allow us to compute several interesting quantities.

\subsubsection{Partial mass and partial kinetic energy}
Let $\mc{P}_E$ be the subset of stars whose total energy is less than $E$. The total mass of this subset $\mc{P}_E$ is, when using (2.14):
\begin{equation}
\mc{M} = \int_0^{\infty}m \, \dd m \int_0^{r_m} 4\pi r^2\, \dd r \int_U^E 4\pi (2E-2U)^{1/2} f_1\, \dd E_1,
\end{equation}
and its total kinetic energy is
\begin{equation}
\mc{L} = \int_0^{\infty}m\, \dd m \int_0^{r_m} 4\pi r^2\, \dd r \int_U^E 2\pi (2E-2U)^{3/2} f_1\, \dd E_1.
\end{equation}

Switching the order of the integrations and using (2.19) and (2.20), one obtains:
\begin{eqnarray}
\mc{M} &=& 16\pi^2 \int_0^{\infty}m\, \dd m \int_{-\infty}^E f_1 q_1'\, \dd E_1,\\
\mc{L} &=& 24\pi^2 \int_0^{\infty}m\, \dd m \int_{-\infty}^E f_1 q_1\, \dd E_1.\nonumber
\end{eqnarray}
We assume all the masses to be equal and we introduce the normalized variables defined as:
\begin{eqnarray}
\mc{M} &=& \left(16\pi^2\ m_1\right)^{-1/2} G^{-3/2}\ M,\\
\mc{L} &=& \left(16\pi^2\ m_1\right)^{-1/2} G^{-3/2}\ L,\nonumber
\end{eqnarray}
which results in:
\begin{eqnarray}
M &=& \int_{-\infty}^E F_1 Q_1'\, \dd E_1,\\
L &=& \frac{3}{2}\int_{-\infty}^E F_1 Q_1\, \dd E_1.\nonumber
\end{eqnarray}

These relations allow us to give a physical meaning to the integrals found in the evolution equation (2.25d). We also note that if we set $E = +\infty$, the subset $\mc{P}_E$ matches the entire cluster and the equations (2.40) and (2.42) give the total mass and the total kinetic energy of the cluster.

We switch to the canonical variables by setting:
\begin{eqnarray}
M &=& \beta^{3/4}\ \gamma^{-1/2}\ \fb{M},\\
L &=& \beta^{7/4}\ \gamma^{-1/2}\ \fb{L}.\nonumber
\end{eqnarray}

\subsubsection{Another partial mass}

The mass within the sphere of radius $r$ is
\begin{equation}
\mc{M}_r = \int_0^r 4\pi r_1^2\ \rho_1 \, \dd r_1 = \frac{1}{G}r^2\ \frac{\partial U}{\partial r},
\end{equation}
which becomes, after transformations as in (2.41a) and (2.43a):
\begin{equation}
\fb{M}_\fb{R} = \int_0^\fb{R} \fb{D}_1 \fb{R}_1^2\, \dd\fb{R}_1 = \fb{R}^2 \frac{\partial \fb{U}}{\partial \fb{R}}.
\end{equation}
One must be careful not to mix up $\fb{M}_\fb{R}$, the mass within the radius $\fb{R}$, with $\fb{M}$, the mass of the stars whose energy is less than $\fb{E}$.

\subsubsection{Projected density}

The projected density of the cluster, i.e. the mass observed per unit surface, is
\begin{equation}
\rho_P(r) = \int_{-\infty}^{+\infty} \rho\left(\sqrt{r^2+z^2}\right) \, \dd z.
\end{equation}

We switch to the normalized, and then canonical, variables with
\begin{eqnarray}
\rho_P(r) &=& m_1^{1/2}\ G^{-1/2}\ D_P,\\
D_P(r) &=& \beta^{5/4}\ \gamma^{1/2}\ \fb{D}_P,
\end{eqnarray}
and we obtain
\begin{equation}
\fb{D}_P = \int_{-\infty}^{+\infty} \fb{D}\left(\sqrt{\fb{R}^2+\fb{Z}^2}\right) \, \dd\fb{Z}.
\end{equation}

\subsubsection{Projected mass}

The mass enclosed within a circle of radius $r$ in the projection of the cluster is:
\begin{equation}
\mc{M}_P = \int_0^r 2\pi r_1\ \rho_P\, \dd r_1,
\end{equation}
i.e., through canonical variables and transformation as in (2.41a) and (2.43a):
\begin{equation}
\fb{M}_P = \frac{1}{2} \int_0^\fb{R} \fb{D}_P \fb{R}\, \dd\fb{R}.
\end{equation}

\subsection{Homologous evolution}

The system of equations (2.37) allows us to compute the evolution of the cluster from a given initial state. However, a few tests revealed that even with a powerful machine (IBM~704), this calculation is extremely long. Furthermore, the circumstances of the birth of a globular cluster (and even more, the initial energy distribution) remain unknown at present time. The choice of the function describing the initial state of the cluster is, therefore, quite arbitrary.

Consequently, it seems more rational to face the problem from the other side, and to first look for what the final evolution of the cluster will be. An analogy with gases suggests indeed that the stellar systems should tend toward a defined final state, independent of their initial state; and the similarity of the globular clusters as we observe them today \citep{vonHoerner1957} seems to confirm this assumption. This final state should be deduced from the system of equations (2.37) alone; it is the equivalent, in a way, to the Maxwellian distribution of velocities for a gas in equilibrium.

However, it is well-known that the final state of a cluster cannot be equilibrium. The stars which gain enough energy through perturbations escape forever from the cluster, whose total mass therefore decreases. Hence, a stationary state \emph{stricto sensu}, does not exist.

How then, can we imagine the final evolution of the cluster? It is time to go back to the distinction made above between the evolution of the size and the evolution of the structure. The escape of stars only implies the evolution of the size; it does not forbid the existence of a model for which only the dimensions would vary, while its structure would remain the same; in other words, a model where the evolution would only consist in ``expansions'' or ``contractions'' of the physical quantities. We call such a process \emph{homologous evolution}.

Therefore, it seems natural to seek whether the final evolution of the cluster would be of this kind. This is what we will do in the next Chapters. The answer will be yes: we will see first (Chapters~III to V) that the system (2.37) allows one and only one solution in homologous evolution, which we will name \emph{homologous model}; second we will show (Chapter~VII) that this model represents indeed the final state toward which all the clusters tend.

Writing of the above in mathematical terms is very simple. A model in homologous evolution allows an invariant canonical form: $\fb{F}$, $\fb{Q}$, and so on, are independent of the time. Thus, in particular:
\begin{equation}
\frac{\partial \fb{F}}{\partial \fb{T}} = 0, \qquad \frac{\partial \fb{Q}}{\partial \fb{T}} = 0,
\end{equation}
and the last equation of the system (2.37) shrinks to
\begin{equation}
0 = \fb{S}'' + \left(\frac{3}{4}b-\frac{3}{2}c\right)\fb{Q}\fb{F}'-c\fb{F}\fb{Q}'.
\end{equation}
This way, the system becomes independent of time.

Before focussing on the numerical solution, we still have to examine the boundary conditions. This will be the topic of the next two Chapters.

\section{Boundary conditions}

\subsection{Effect of the galactic field}

Globular clusters are not isolated systems: they are subject to the gravitational field of the galaxy, an effect that cannot be neglected.

Let $U_G$ be the galactic potential. The dimensions of a cluster being much smaller than those of the galaxy, this potential can be described with sufficient precision by a Taylor series truncated to the second order; in a reference frame centered on the cluster, moving with it, and with the right orientation, the series reads
\begin{equation}
U_G = U_G(0) + \frac{1}{2}\left(\frac{\partial^2 U_G}{\partial x^2}x^2 + \frac{\partial^2 U_G}{\partial y^2}y^2 + \frac{\partial^2 U_G}{\partial z^2}z^2\right).
\end{equation}
Furthermore, we almost have:
\begin{equation}
\frac{\partial^2 U_G}{\partial x^2} + \frac{\partial^2 U_G}{\partial y^2} + \frac{\partial^2 U_G}{\partial z^2} =0,
\end{equation}
because the density of the galaxy is very small in the regions where the globular clusters are found. As a consequence, at least one of these three main curvatures is negative. We assume that the one along $x$ is the most negative.

The potential created by the cluster is, in the external regions:
\begin{equation}
U_A = -\frac{G\mc{M}_e}{r},
\end{equation}
\tn{the subscript $A$ stands for \emph{amas}: cluster.}, where $\mc{M}_e$ is the total mass of the cluster. Thus, the total potential along the $x$-axis is (\reff{1}):
\begin{equation}
U = U_G + U_A = U_G(0) + \frac{1}{2}\frac{\partial^2 U_G}{\partial x^2}x^2 -\frac{G\mc{M}_e}{x}.
\end{equation}
It reaches a maximum for
\begin{eqnarray}
x_e = (G\ \mc{M}_e)^{1/3} \left(-\frac{\partial^2 U_G}{\partial x^2}\right)^{-1/3},\\
U_e = U_G(0) - \frac{3}{2}(G\ \mc{M}_e)^{2/3} \left(-\frac{\partial^2 U_G}{\partial x^2}\right)^{1/3}.\nonumber
\end{eqnarray}

\begin{figure}
\includegraphics[width=\columnwidth]{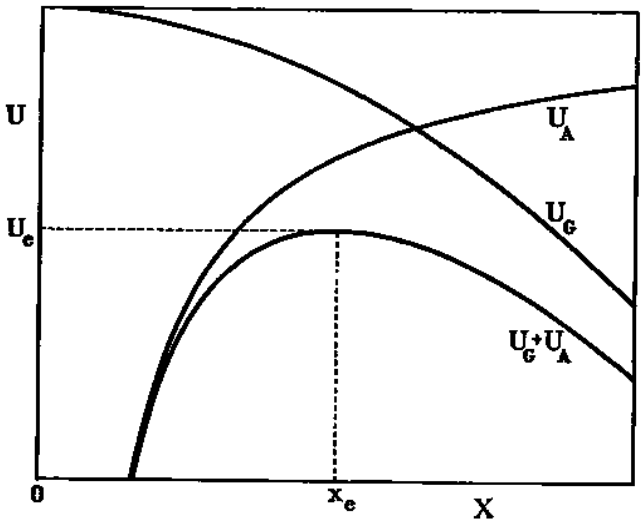}
\caption{Galactic potential $U_G$, cluster potential $U_A$ and total potential $U_G+U_A$.}
\label{fig:1}
\end{figure}

It follows that every star that goes beyond this point escapes from the cluster forever. This implies that the stars that have a radial, oscillatory, trajectory along the $x$-axis cannot have a total energy greater than $U_e$; therefore, for these stars:
\begin{equation}
f = 0 \qquad\textrm{for } E \ge U_e.
\end{equation}
But, as we supposed that the distribution function depends only on $E$, (3.6) is, in fact, a general property: no star can have an energy higher $U_e$. In $U_e$, the distribution function has what \citet{Chandrasekhar1943d} calls an ``absorbing barrier'', and that we call here, in a more illustrative way, a ``leak''. We know that, in this case, the function $f$ is zero for $U_e$, but its derivative is not; the value of the latter is proportional to the escape rate of the stars \citep[Equations 25 and 33]{Chandrasekhar1943d}.

The galactic field has also the effect of modifying the potential inside the cluster. However, the equations (3.1) and (3.2) show that $U_G-U_G(0)$ is zero in average on a sphere centered on the cluster; as a consequence, in our approximation of spherical symmetry, the galactic field has no effect inside the cluster. Its effect reduces to the creation of a ``leak'' at $U_e$.

The boundary of the cluster is defined by the equation $U=U_e$; it is a closed surface, with a shape of a lemon, elongated along the $x$-axis (on which it has two conic points for $x=\pm x_e$). Therefore, the ``radius'' of the cluster is a function of the angle; but in the case of the spherical symmetry approximation, we will limit ourselves to considering its mean value. It is obtained from (3.3) and (3.4), by replacing the galactic potential $U_G$ with its mean value $U_G(0)$, which gives out
\begin{eqnarray}
r_e &=& \frac{G\ \mc{M}_e}{U_G(0)-U_e} \\
&=& \frac{2}{3} (G\ \mc{M}_e)^{1/3}\left(-\frac{\partial^2 U_G}{\partial x^2}\right)^{-1/3} = \frac{2}{3}x_e.\nonumber
\end{eqnarray}
This is this mean value that we will refer to as the radius of the cluster, from now on.

The relation (3.7) links the radius to the mass of the cluster. The curvature of the galactic field can be considered as constant: we know that the dynamical evolution of the galaxy is much slower than that of the clusters. It is true that the cluster travels along an orbit in the galaxy; but this motion is much faster than the evolution of the cluster, and we can use the mean value of the curvature of the field along the orbit in (3.7). Thus, the relation becomes
\begin{equation}
r_e \propto \mc{M}_e^{1/3}.
\end{equation}

\emph{Isolated cluster}

For future applications, we will also consider the case of an isolated cluster, which would experience no external gravitational field. Indeed, it seems that several systems (galaxies, galaxy clusters) are almost in this situation. In this case, at a sufficiently large distance from the center, the total potential reduces to $U_A$, given in (3.3). It does not yield a maximum but rather increases with distance. Therefore, escapes can occur only if the distribution function ranges up to $E=0$, corresponding to an infinite radius for the cluster. Furthermore, we find out from (2.20) and (3.3):
\begin{equation}
q' = \frac{\pi\ G^3\ \mc{M}_e^3}{8\sqrt{2}\ (-E)^{5/2}}.
\end{equation}
In order to have a finite total mass (which is given by Equation 2.40), the distribution function must decrease steeper than $(-E)^{3/2}$ for $E\to 0$. It follows that its derivative is zero at the leak point $E=0$, leading to an escape rate of zero. This result has already be shown by means of another method \citep{Henon1960}; it is confirmed by the behavior of the artificial clusters of \citet{vonHoerner1960} from which no stars escape, even after a time much longer than the relaxation time. Therefore, for an isolated cluster, one has:
\begin{equation}
\mc{M}_e = \textrm{cst}.
\end{equation}

\subsection{Mass-radius relation}

The two relations (3.8) and (3.10) can be represented by the unique form:
\begin{equation}
r_e \propto \mc{M}_e^\lambda,
\end{equation}
with $\lambda = 1/3$ for globular clusters (and for any system in a non-uniform external field in general) and $\lambda = \infty$ for isolated clusters. We will use this general form (3.11), so that the formulae obtained could be eventually applied to the case of isolated clusters; but all the numerical calculations, in the present paper, will be done for $\lambda = 1/3$.

Using (2.32c), (2.43a) and (2.36), we switch to the canonical variables, and the relation (3.11) becomes:
\begin{equation}
-\frac{b}{4}-\frac{c}{2}+\frac{\dd\ln{(\fb{R}_e)}}{\dd\fb{T}} = \lambda \left(\frac{3b}{4}-\frac{c}{2}+\frac{\dd\ln{(\fb{M}_e)}}{\dd\fb{T}}\right).
\end{equation}

If the model is in homologous evolution, $\fb{R}_e$ and $\fb{M}_e$ are independent of time, and we get:
\begin{equation}
0 = (3\lambda +1) b + (2-2\lambda)c.
\end{equation}

We see that the mass-radius relation translates into a definite relation between the two parameters of the homology.

\subsection{Choice of a reference level for the potential}

The potential $U$ is defined up to an additive constant. We set this constant by writing:
\begin{equation}
U_e = 0.
\end{equation}

Therefore, the potential is zero at the boundary of the cluster, negative inside, positive outside. The total energy of a star is always negative, $E=0$ corresponding to the escape threshold.

\section{Central conditions}

\subsection{Infinite central density}

\emph{A priori}, one could expect the density profile of the model we seek to have the classical shape of \reff{2} with, in particular, a well-defined central density; and our first attempts have been set to find this. However, the calculations of the evolution from such an initial state revealed an unexpected phenomenon: no matter how big the initial value of the central density is, it always increases, without limits. As an example, \reff{2} shows the spacial density computed at two successive times. The horizontal axis is normalized to the external radius; this way, we see that the increase only concerns the central region of the cluster\footnote{\citet{Michie1961} has also observed a flux of stars toward the center of his model.}. This increase cannot be diminished by means of an homologous transformation; indeed, one can measure it by computing the ratio between the central density and the mean density of the cluster, and this ratio does not change in an homology.

\begin{figure}
\includegraphics[width=\columnwidth]{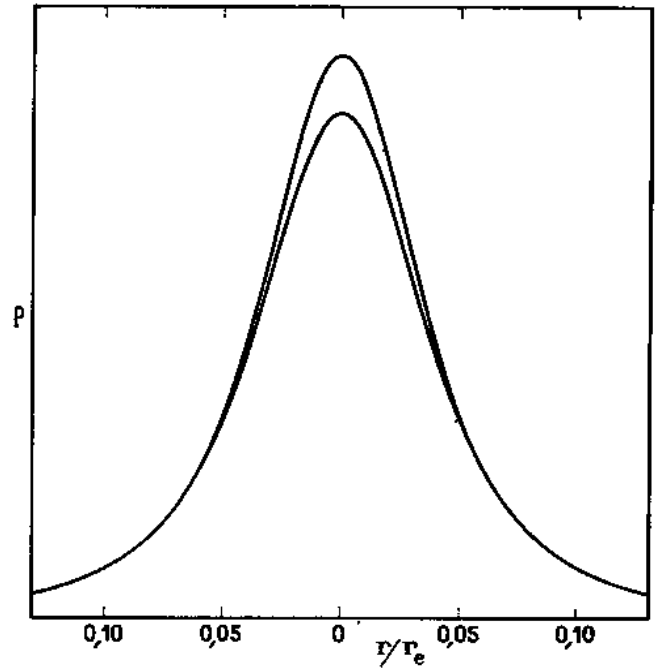}
\caption{Increase of the density in the central region of a cluster}
\label{fig:2}
\end{figure}

Therefore, we have been led to consider, for the final state of the cluster, a \emph{model with infinite central density}. This can seem physically absurd at first sight. However, one should not forget that the density considered here is a probability density. The cluster is not made of a continuous medium, but rather of separated particles. For the model to have a physical meaning, the integral of the probability density must be finite over any finite volume; and we will see it is indeed the case.

In order to better emphasize this important point, we have created, from a table of random numbers, an ``artificial cluster'' , that follows the projected density law
\begin{equation}
\rho_P \propto \frac{1}{r}.
\end{equation}

\begin{figure}
\includegraphics[width=\columnwidth]{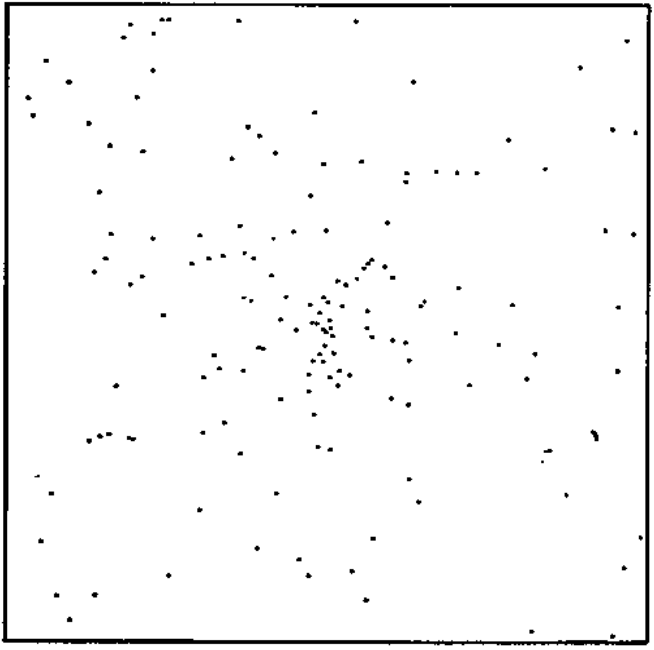}
\caption{Artificial cluster with infinite central density.}
\label{fig:3}
\end{figure}

This cluster, shown in \reff{3}, looks like a real galactic cluster; the fact that its theoretical density is infinite in its center is not visible, and if we would plot the density profile from \reff{3}, as one would do for real cluster, we would likely state that the curve shows a finite maximum in the center!

The functional form of (4.1) is indeed the one we will find near the center of the model (see Equation 5.9). Note that the number of stars inside a circle of radius $r$ is proportional to $r$, and thus tends toward zero when the circle becomes smaller and smaller.

In Chapter~VIII we will study in detail how this central singularity of the density arises. For the time being, we only focus on the final state, and thus we assume that the singularity exists.

\subsection{Asymptotic expressions near the center}

$\rho$ being infinite at the center, the function $f$ must become infinite too, as shown in (2.6). As a consequence, the first two terms of the equation (2.53), which are of the order of $\fb{F}\fb{Q}$, become negligible with respect to the first one which is of the order of $\fb{F}^2\fb{Q}$, and we have:
\begin{equation}
\fb{S}'' = 0,
\end{equation}
i.e., after integration,
\begin{equation}
\fb{S} = \alpha_1 + \alpha_2\ \fb{E},
\end{equation}
where $\alpha_1$ and $\alpha_2$ are two constants.

Let's assume first that $\alpha_1$ and $\alpha_2$ are zero. In this case, using (2.27) and (2.28), (4.3) reduces to
\begin{equation}
\frac{\fb{F}}{\fb{F}^{(-1)}} = \frac{\fb{F}'}{\fb{F}},
\end{equation}
which immediately integrates as
\begin{equation}
\fb{F} = C_1\ \e{-C_2\fb{E}},
\end{equation}
where $C_1$ and $C_2$ are two constants, necessarily positive, because $\fb{F}$ must be positive and increasing for $\fb{E} \to -\infty$.

This way, we find a Maxwellian distribution of the energies near the center, independently of the particular functional form of $\fb{Q}$, i.e. of the structure of the cluster. This result is in agreement with what one could expect: in the central regions, which are very dense, the perturbations are very active and almost establish the Maxwellian equilibrium\footnote{often called ``isothermal equilibrium'', which is a misuse of language, because the concept of temperature is meaningless in stellar dynamics.}. We are now to set the two conditions that define the canonical form of the cluster (see Equation 2.31 and the following): we assume that $\fb{E}$ and $\fb{F}$ have been normalized so that the constants $C_1$ and $C_2$ equal unity. The distribution function near the center is then:
\begin{equation}
\fb{F} = \e{-\fb{E}}.
\end{equation}
This choice of the canonical conditions has the advantage of simplifying the formulae and the calculations a lot.

From (4.6), using successively (2.37a), (2.37b), (2.37c), we easily find the expressions of the various quantities near the center:
\begin{eqnarray}
\fb{D} &=& \left(\frac{\pi}{2}\right)^{1/2} \e{-\fb{U}},\\
\fb{R} &=& \left(\frac{8}{\pi}\right)^{1/4} \e{\fb{U}/2},\nonumber\\
\fb{Q} &=& K_D\ \e{3\fb{E}/2},\nonumber\\
\bigg(K_D &=& \frac{16}{9\sqrt{3}\ (8\pi)^{1/4}} = 0.45841\bigg).\nonumber
\end{eqnarray}
Note that (4.7b) is obtained as the unique solution of the differential equation (2.37b) for which $\fb{D}$ is infinite at the center of the cluster. We have, near the center:
\begin{equation}
\fb{D} = \frac{2}{\fb{R}^2}.
\end{equation}
Also note that the potential is, near the center:
\begin{equation}
\fb{U} = \frac{1}{2}\ln{\left(\frac{\pi}{8}\right)} + 2 \ln{(\fb{R})}.
\end{equation}
Therefore, it tends logarithmically toward $-\infty$ in the center: the cluster yields an infinite ``potential well''. As a consequence, the energy of the stars can take any value from $\fb{E}=-\infty$ to $\fb{E}=0$.

Finally, we have
\begin{eqnarray}
\int_{-\infty}^{\fb{E}} \fb{F}_1\fb{Q}_1\, \dd\fb{E_1} &=& 2 K_D\ \e{\fb{E}/2},\\
\int_{-\infty}^{\fb{E}} \fb{F}_1\fb{Q}_1'\, \dd\fb{E_1} &=& 3 K_D\ \e{\fb{E}/2}.\nonumber
\end{eqnarray}

Let's go back to (4.3) in the general case where $\alpha_1$ and $\alpha_2$ are not zero. For small values of $\alpha_1$ and $\alpha_2$, the solution of (4.3) can be expanded as:
\begin{equation}
\fb{F} = \e{-\fb{E}} + \alpha_1 Y_1 + \alpha_2 Y_2,
\end{equation}
where $Y_1$ and $Y_2$ are function yet to be determined. By setting this in (4.3) and using (4.7c), we obtain two differential equations for $Y_1$ and $Y_2$:
\begin{eqnarray}
Y_1^{(-1)} + 2 Y_1 + Y_1' &=& \frac{\e{-\fb{E}/2}}{3K_D},\\
Y_2^{(-1)} + 2 Y_2 + Y_2' &=& \frac{(1+ \fb{E})\ \e{-\fb{E}/2}}{3K_D},\nonumber
\end{eqnarray}
which can be solved as
\begin{eqnarray}
Y_1 &=& C_{11} \e{-\fb{E}} + C_{12} \fb{E}\ \e{-\fb{E}} - \frac{2}{3K_D} \e{-\fb{E}/2},\\
Y_2 &=& C_{21} \e{-\fb{E}} + C_{22} \fb{E}\ \e{-\fb{E}} + \frac{2}{3K_D} (5-\bf{E})\  \e{-\fb{E}/2}.\nonumber
\end{eqnarray}

The $C_{ij}$ are arbitrary constants. The terms $C_{ij}$ are, in fact, parasite solutions coming from a bad normalization of $\fb{F}(\fb{E})$; we can get rid of them by modifying this normalization. Then, $\fb{F}$ reads
\begin{equation}
\fb{F} = \e{-\fb{E}} + \frac{2}{3K_D} \left(5\alpha_2 - \alpha_1 - \alpha_2\ \fb{E}\right) \e{-\fb{E}/2}.
\end{equation}

The second term of this expansion becomes negligible with respect to the first one, for all $\alpha_1$ and $\alpha_2$, when $\fb{E}$ is small enough. As a consequence, the expansion (4.14) is valid near the center, not only for small $\alpha_1$ and $\alpha_2$, but for any values they can take. Thus, we have found the general solution of (4.3), i.e. the functional form of the distribution function, for $\fb{E}\to -\infty$.\footnote{The expression (4.14) can be considered as an expansion of $\fb{F}$ up to the second order in $\e{\fb{E}/2}$; however, it has been obtained using the expansion of $\fb{Q}$ up to the first order, (4.7c). This method is valid because $\fb{S}$ cancels when $\fb{F}$ takes the form (4.6), for all $\fb{Q}$'s. We easily find out that the expansion of $\fb{S}$ up to the $n$-th order depends on the expansion of $\fb{F}$ up to the $n$-th order, and on the expansion of $\fb{Q}$ up to the $(n-1)$-th order only. Therefore, it is possible to consider one term less in the expansion of $\fb{Q}$ than in those of $\fb{F}$.}

\subsection{Flux of matter and flux of energy toward the center}

We are going to calculate the flux of matter and the flux of energy through the surface $E=$ cst, in the six-dimensional space of positions-velocities. This calculation will allow us to give a physical meaning to the constants $\alpha_1$ and $\alpha_2$, and to specify the conditions the model must fulfill near the center of the cluster. Let's go back a second to the study of the local properties and let's consider, in one point in the cluster, the subset of stars whose velocity is smaller than a given $v$. The density of this population is
\begin{eqnarray}
\rho_v &=& \int_0^\infty m\, \dd m \int_0^v 4\pi a_1\ v_1^2\, \dd v_1\\
&=& m_1 \int_0^v 4\pi A_1\ v_1^2\, \dd v_1,\nonumber
\end{eqnarray}
where $a$ is the local distribution function, defined in (2.9). In analogy with (2.23), we set
\begin{equation}
a = \delta(m-m_1) A.
\end{equation}

From this, and using (2.10), we find
\begin{eqnarray}
\left(\frac{\partial \rho_v}{\partial t}\right)_p &=& 64 \pi^3\ G^2\ m_1^2\ \ln{(n)} \\
&& \bigg[ A \int_0^v A_1\ v_1^2\, \dd v_1 + \frac{1}{3} \frac{\partial A}{\partial v} \bigg(\frac{1}{v} \int_0^v A_1\ v_1^4\, \dd v_1\nonumber\\
&& + v^2 \int_0^v A_1\ v_1\, \dd v_1 \bigg) \bigg]\nonumber.
\end{eqnarray}

This quantity can be physically interpreted as the flux of stars, in velocity-space, through the surface $v=$ cst.

The total mass of the stars in the subset $\mc{P}_E$ (stars whose total energy is less than a given $E$) is (see Equation 2.38):
\begin{equation}
\mc{M} = \int_0^{r_m} 4\pi \rho_v\ r^2\, \dd r,
\end{equation}
which becomes, after differentiation while keeping $E$ constant:
\begin{equation}
\left(\frac{\partial \mc{M}}{\partial t}\right)_p = 4\pi \int_0^{r_m} \left[ \left(\frac{\partial \rho_v}{\partial t}\right)_p + \frac{\partial \rho_v}{\partial v} \left(\frac{\partial v}{\partial t}\right)_E\right] r^2\, \dd r.
\end{equation}

This is the flux of the stars through the surface $E= $ cst, in the six-dimensional space. The subscript $p$ of the first term can be removed, because the equalization of the perturbation occurs between stars of same energy, and does not affect the value of $\mc{M}$. Furthermore, from (2.14):
\begin{equation}
\left(\frac{\partial v}{\partial t}\right)_E = -(2E - 2U)^{-1/2}\ \frac{\partial U}{\partial t}.
\end{equation}

Substituing into (4.19), then changing the variables, switching the operations the same way as in Chapter~II, and using (2.19), (2.20) and (2.21), we obtain:
\begin{eqnarray}
\frac{\partial \mc{M}}{\partial t} &=& 16 \pi^2\ F \frac{\partial q}{\partial t} + 256 \pi^4\ G^2\ m_1^2\ \ln{(n)}\\
&& \bigg[ F \int_{-\infty}^E F_1\ q_1'\, \dd E_1 + F' \bigg(\int_{-\infty}^E F_1\ q_1\, \dd E_1\nonumber\\
&& + q \int_E^{\infty} F_1\, \dd E_1 \bigg) \bigg].\nonumber
\end{eqnarray}

We notice that when using (2.40a), we can easily go from this relation to the fundamental equation of the evolution (2.22), which is thus proved again. On the other hand, it is not possible to go from (2.22) to (4.21), because an integration constant remains undefined; that is why we had to go back to the basic equations to establish (4.21).

We switch to normalized variables thanks to (2.24) and (2.41a), which gives, using (2.29):
\begin{equation}
\frac{\partial M}{\partial T} = S' + F \frac{\partial Q}{\partial T}.
\end{equation}

Let's compute, in the same way, the flux of energy through a surface $E=$ cst. At a point in the cluster, the subset of stars which velocity is less than $v$ has a total energy density
\begin{equation}
h_v = m_1 \int_0^v 4\pi A\ Ev^2\, \dd v,
\end{equation}
whose variation is
\begin{equation}
\left(\frac{\partial h_v}{\partial t}\right)_p = \int_0^v 4\pi \left[ \left(\frac{\partial A}{\partial t}\right)_p E + A \frac{\partial U}{\partial t} \right] v^2 \, \dd v.
\end{equation}
The total energy of the subset $\mc{P}_E$ is
\begin{equation}
\mc{H} = \int_0^{r_m} 4\pi h_v\ r^2 \, \dd r.
\end{equation}

We will not give the details of the calculation, which is similar to the previous one, but slightly longer; one must do several integrations by parts. We set, in analogy with (2.41b):
\begin{equation}
\mc{H} = (16\pi^2\ m_1)^{-1/2}\ G^{-3/2}\ H,
\end{equation}
and we obtain the expression of the flux of total energy through the surface $E=$ cst:
\begin{equation}
\frac{\partial H}{\partial T} = ES' - S + EF\ \frac{\partial Q}{\partial T} - \int_{-\infty}^E F_1 \frac{\partial Q_1}{\partial T} \, \dd E_1.
\end{equation}

It is interesting to note that by computing the derivative (4.22) and (4.27) with respect to $E$, we find:
\begin{equation}
\frac{\partial H'}{\partial T} = E\ \frac{\partial M'}{\partial T}.
\end{equation}

This relation can be proved more directly. Indeed, let $\dd\mc{P}_E$ be the subset of stars whose total energy is between $E$ and $E+\dd E$; the mass of this subset is: $\dd M = M' \dd E$, and its total energy is: $\dd H = H' \dd E$. Thus comes
\begin{equation}
H' = E M',
\end{equation}
and the relation (4.28). However, here again, it is not possible to use (4.22) and (4.28) to demonstrate (4.27), because an integration constant would remain undefined.

(The quantity $\mc{H}$, that we called ``total energy of the subset $\mc{P}_E$'', has been computed by simply adding the total energies $E$ of its members, which is not correct, because by doing this, we count twice the mutual potential energy of the stars of $\mc{P}_E$. However, this effect is negligible if $\mc{P}_E$ only counts a small fraction of the stars of the cluster, which is the case for $E\to -\infty$.)

Near the center, one can neglect the last term of (4.22), and the two last terms of (4.27), which are of the order of $\e{E/2}$; we also replace $S$ with its value, given in (4.3) and (2.32f), and we get
\begin{eqnarray}
\frac{\partial M}{\partial T} &=& \beta^{3/4}\ \gamma^{1/2}\ \alpha_2,\\
\frac{\partial H}{\partial T} &=& -\beta^{7/4}\ \gamma^{1/2}\ \alpha_1.\nonumber
\end{eqnarray}

Thus, the two constants $\alpha_1$ and $\alpha_2$ represent, up to a factor, the fluxes of mass and of energy toward the center of the cluster. This result is unexpectedly simple.

Note that we can switch to the canonical variables thanks to (2.32e) and (2.43); $H$ has still the same dimension as $L$. By neglecting once more the terms in $\e{E/2}$, we obtain:
\begin{eqnarray}
\frac{\partial \fb{M}}{\partial \fb{T}} &=& \alpha_2,\\
\frac{\partial \fb{H}}{\partial \fb{T}} &=& -\alpha_1.\nonumber
\end{eqnarray}

We are now going to examine the physical meaning of these fluxes. Let's focus first on the flux of mass. (4.30a) shows that, near the center, it takes a constant value, independent of $E$; this means that a flow of matter enters or exits the cluster (depending on the sign of $\alpha_2$) through the central singularity. The fundamental equations do not rule out such a flow; one can even, as we will see, create an infinite number of models obeying the equations and yielding a non-zero flow of matter in the center. However, from a physical point of view, such a flow obviously does not make sense. This leads us to write the additional condition
\begin{equation}
\alpha_2 = 0.
\end{equation}

The need of writing a separate condition for the mass conservation in the center can be more easily explained: at the center of the cluster, the quantities become infinite and the fundamental equations are meaningless; thus an additional condition is required for this particular point. For that matter, we have already noted, just above, that it is impossible to go directly from the equation of evolution (2.22) to the Equation (4.21) which states the conversation of the mass; one constant is missing, which is precisely the value of the central flow.

However, the models with a non-zero central flow are useful: we will see them again in Chapter~VII, while studying the initial stages of the evolution, during which the central density slowly increases until it becomes infinite. Then, the flow of matter toward the center exists and simply corresponds to the slow ``filling'' of the central part of the density profile (see \reff{2}).

Let's consider now the flux of energy, given by (4.30b). This flux is also constant near the center: therefore the center of the cluster creates or absorbs energy. One could think that, in analogy with what is true for the mass, the flow of energy must be zero in the center. But the numerical computation shows (see the next Chapter) that the previously written condition (4.32) completed the definition of the solution of the system of equations, which is now unique; and this solution corresponds to a value of $\alpha_1$ that is positive and non zero. Thus, we have to admit that a flow of energy toward the center exists; more precisely: \emph{the center of the cluster absorbs some negative energy}.

We will come back with more details to this strange and very interesting phenomenon in the Chapter~V, and we will see how it can be physically interpreted.

\subsection{Follow-up on the expansions near the center}

To prepare the numerical computation, it is useful to push further the expansions of the various quantities near the center. Taking the relation (4.32) into account and setting
\begin{equation}
-\frac{2\alpha_1}{3K_D} = K,
\end{equation}
the expansion (4.14) of $\fb{F}$ reduces to
\begin{equation}
\fb{F} = \e{-\fb{E}} + K\ \e{-\fb{E}/2}.
\end{equation}

By using successively the fundamental equations (2.37a), (2.37b), (2.37c), we find out the two-terms expansions:
\begin{eqnarray}
\fb{D} &=& \left(\frac{\pi}{2}\right)^{1/2} \left(\e{-\fb{U}} + 2\sqrt{2}\ K\ \e{-\fb{U}/2} \right),\\
\fb{R} &=& \left(\frac{8}{\pi}\right)^{1/4} \left(\e{\fb{U}/2} - \frac{K}{\sqrt{2}} \ \e{\fb{U}} \right),\nonumber\\
\fb{Q} &=&  K_D \left( \e{3\fb{E}/2} - \frac{9\sqrt{3}}{8\sqrt{2}}\ K\ \e{2\fb{E}}\right),\nonumber
\end{eqnarray}
that make (4.7) more precise. From a remark made above, we can compute the expansion of $\fb{F}$ at the third order by putting in (2.53) the expansion of $\fb{Q}$ at the second order only, given by (4.35c). The last two terms of (2.53) must now be taken into account. We find:
\begin{eqnarray}
\fb{F}^{(-1)} &=& -\e{-\fb{E}} - 2K\ \e{-\fb{E}/2} + \left(\frac{3}{2}b - \frac{3\sqrt{3}}{8\sqrt{2}}K^2 \right),\nonumber\\
\fb{F} &=& \e{-\fb{E}} + K\ \e{-\fb{E}/2} + 0.
\end{eqnarray}
The third term in the expansion of $\fb{F}$ is zero.

We stop the expansions here because the calculation of the terms of higher order is much more involved: the third term of $\fb{D}$ is not constant but proportional to $(-\fb{U})^{1/2}$ and its coefficient depends on the entire function $\fb{F}$. Therefore, we will use the expansions (4.35) for $\fb{D}$, $\fb{R}$, $\fb{Q}$. Finally, we note the expansions of the two integrals:

\begin{eqnarray}
\int_{-\infty}^\fb{E} \fb{F}_1 \fb{Q}_1\, \dd\fb{E}_1 = \\
K_D \left[2\e{\fb{E}/2} + \left(1-\frac{9\sqrt{3}}{8\sqrt{2}}\right)K\ \e{\fb{E}} - \frac{3\sqrt{3}}{4\sqrt{2}}K^2\ \e{3\fb{E}/2} \right],\nonumber\\
\int_{-\infty}^\fb{E} \fb{F}_1 \fb{Q}_1'\, \dd\fb{E}_1 =  \nonumber\\
 \frac{3}{2}K_D \left[2\e{\fb{E}/2} + \left(1-\frac{3\sqrt{3}}{2\sqrt{2}}\right)K\ \e{\fb{E}} - \frac{\sqrt{3}}{\sqrt{2}}K^2\ \e{3\fb{E}/2} \right].\nonumber
\end{eqnarray}

\section{The homologous model}

\subsection{Summary of the equations}

The model that we propose to compute is defined by the set of equations and conditions (2.37a, b, c, d), (2.53), (3.6), (3.13), (4.34), (4.35b) obtained in the previous Chapters. Some transformations are yet necessary to set the equations in a form that most favors the numerical computation:

\begin{enumerate}
\item In the differential equation (2.37b), we will consider $\fb{U}$ as the independent variable. Furthermore, $\fb{R}$ varies rapidly near the boundary, and it is useful to switch to a new variable $\fb{Z}$ defined as:
\begin{equation}
\fb{R} = \frac{1}{\fb{Z}}.
\end{equation}
\item The integral (2.37c) is transformed by means of an integration by parts.
\item The equation (2.53) is replaced with its integrated form with respect to $\fb{E}$; we have seen in Chapter~IV that the integration constant is zero.
\end{enumerate}

All the equations, modified this way, and the boundary conditions are gathered below:

\emph{Equations:}
\begin{eqnarray}
\fb{D} &=& \int_\fb{U}^\infty (2\fb{E}-2\fb{U})^{1/2} \fb{F}\, \dd\fb{E}, \\
\frac{\dd^2 \fb{Z}}{\dd\fb{U}^2} &=& -\fb{D} \left(\frac{\dd\fb{Z}}{\dd\fb{U}}\right)^3 \fb{Z}^{-4},\nonumber\\
\fb{R} &=& \fb{Z}^{-1},\nonumber\\
\fb{Q} &=& \frac{1}{3}\int_{-\infty}^{\fb{E}} (2\fb{E}-2\fb{U})^{1/2}\ \fb{R}^3\, \dd\fb{U},\nonumber\\
0 &=& \bigg\{ \fb{F} \int_{-\infty}^{\fb{E}} \fb{F}_1 \fb{Q}_1'\, \dd\fb{E}_1\nonumber\\
&& + \fb{F}' \bigg( \int_{-\infty}^{\fb{E}} \fb{F}_1 \fb{Q}_1\, \dd\fb{E}_1 - \fb{Q}\fb{F}^{(-1)} \bigg) \nonumber\\
&& + \left( \frac{3}{4}b-\frac{3}{2}c \right) \fb{F}\fb{Q}\nonumber\\
&& + \left( \frac{1}{2}c-\frac{3}{4}b \right) \int_{-\infty}^{\fb{E}} \fb{F}_1 \fb{Q}_1'\, \dd\fb{E}_1 \bigg\},\nonumber\\
0 &=& (3\lambda + 1)b + (2-2\lambda) c.\nonumber
\end{eqnarray}

\emph{Boundary conditions:}
\begin{eqnarray}
\fb{F}(0) &=& 0,\\
\fb{F}^{(-1)}(0) &=& 0,\nonumber\\
\fb{F} &\simeq& \e{-\fb{E}} + K \e{-\fb{E}/2} \quad \textrm{for } \fb{E} \to -\infty,\nonumber\\
\fb{R} &\simeq& \left(\frac{8}{\pi}\right)^{1/4} \left(\e{\fb{U}/2} - \frac{K}{\sqrt{2}} \e{\fb{U}}\right) \quad \textrm{for } \fb{U} \to -\infty.\nonumber
\end{eqnarray}

The unknown functions are: $\fb{F}(\fb{E})$, $\fb{Q}(\fb{E})$, $\fb{D}(\fb{U})$, $\fb{Z}(\fb{U})$, $\fb{R}(\fb{U})$, defined from $-\infty$ to 0; the unknown constants are: $b$, $c$, $K$. $\lambda$ equals 1/3.

\subsection{Numerical solving method}

The form of the system does not allow the solution to be computed directly; one has to proceed by trial and error. Experience leads us to adopt the following iterative method:
\begin{enumerate}
\item Choose a temporary form for the function $\fb{F}$, that fulfills (5.3a) and (5.3c), with a temporary value of $K$;
\item compute $\fb{D}$ using (5.2a);
\item compute $\fb{Z}$ by integrating (5.2b) from the center to the boundary; the initial conditions are given in (5.3d);
\item compute $\fb{Q}$ by means of (5.2c) and (5.2d);
\item choose temporary values for $c$ and $K$; compute $b$ using (5.2f);
\item integrate (5.2e), which is equivalent to a differential system of the fourth order, from the center to the boundary; the initial conditions are given in (5.3c), (4.36a) and (4.37);
\item in general, this integration gives the final values of $\fb{F}(0)$ and $\fb{F}^{(-1)}(0)$ that are different from zero; go back to point [5.], modify $c$ or $K$ and begin the integration again; grope around this way for the values of $c$ and $K$ until $\fb{F}(0)$ and $\fb{F}^{(-1)}(0)$ vanish;
\item go back to point [2.] with the new function $\fb{F}$.
\end{enumerate}

This way, we obtain a series of approximations for $\fb{F}$; we stop when two successive approximations are equal, up to the desired precision. In practice, the convergence is quite fast: the errors are divided by about 5 at each iteration.

The computation has been done thanks to the IBM 650 device of the Observatoire de Meudon; 8 hours of computation are required to get the solution with a precision of 1/1000.

\subsection{Results: structure}

The four fundamental functions $\fb{F}$, $\fb{D}$, $\fb{R}$, $\fb{Q}$ are given in \reft{1} and plotted in \reff{4}. The table covers the range $(-5,0)$; below $\fb{E} = -5$ (or $\fb{U} = -5$), the functions are represented with a sufficient precision by the expressions (4.34) and (4.35). The values found for the constants are
\begin{eqnarray}
K &=& -0.9499,\\
c &=& + 0.4078,\nonumber\\
b &=& -\frac{2}{3}c = - 0.2719.\nonumber
\end{eqnarray}

\begin{figure}
\includegraphics[width=\columnwidth]{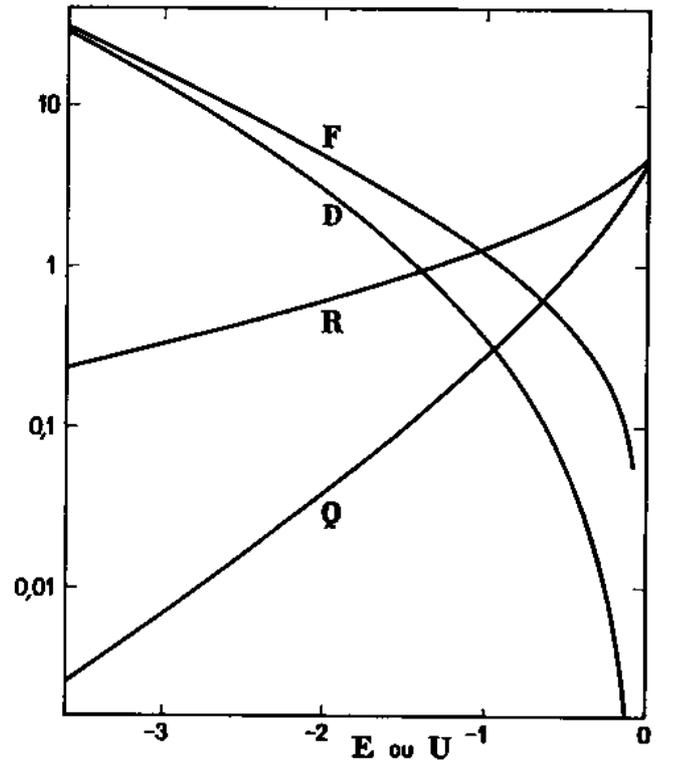}
\caption{Homologous model: the four fundamental functions $\fb{F}(\fb{E})$, the distribution function; $\fb{D}(\fb{U})$, the spacial density; $\fb{R}(\fb{U})$, the distance to the center; $\fb{Q}(\fb{E})$, see (2.19).}
\label{fig:4}
\end{figure}

\begin{table2}
\caption{} 
\label{tab:1} 
\begin{tabular}{r@{}lr@{}lr@{}lr@{}lr@{}lr@{}lr@{}l}
\multicolumn{2}{c}{$\fb{E}$ or $\fb{U}$} & \multicolumn{2}{c}{$\fb{F}$} & \multicolumn{2}{c}{$\fb{D}$} & \multicolumn{2}{c}{$\fb{R}$} & \multicolumn{2}{c}{$\fb{Q}$} & \multicolumn{2}{c}{$\fb{D}_P$} & \multicolumn{2}{c}{$\fb{M}_P$} \\
\hline
\hline
-5& & 136&.96 & 148&.86 & 0&.1097 & 0&.0002805 & 44&.18 & 0&.2988 \\
-4&.9 & 123&.38 & 133&.12 & 0&.1156 & &\phantom{000.}3280 &  &  \\
-4&.8 & 111&.11 & 118&.97 & 0&.1220 & &\phantom{000.}3834 & 38&.80 & 0&.3282 \\
-4&.7 & 100&.05 & 106&.26 & 0&.1286 & &\phantom{000.}4487 &  &  \\
-4&.6 & 90&.06 & 94&.83 & 0&.1357 & &\phantom{000.}5249 & 34&.02 & 0&.3603 \\
-4&.5 & 81&.05 & 84&.57 & 0&.1432 & &\phantom{000.}6148 &  &  \\
-4&.4 & 72&.91 & 75&.36 & 0&.1511 & &\phantom{000.}7198 & 29&.73 & 0&.3954 \\
-4&.3 & 65&.57 & 67&.10 & 0&.1595 & &\phantom{000.}8436 &  &  \\
-4&.2 & 58&.95 & 59&.69 & 0&.1684 & &\phantom{000.}9885 & 25&.90 & 0&.4337 \\
-4&.1 & 52&.98 & 53&.04 & 0&.1778 & 0&.001159 &  &  \\
-4& & 47&.60 & 47&.09 & 0&.1878 & &\phantom{00.}1360 & 22&.48 & 0&.4754 \\
-3&.9 & 42&.75 & 41&.76 & 0&.1984 & &\phantom{00.}1596 &  &  \\
-3&.8 & 38&.37 & 36&.99 & 0&.2097 & &\phantom{00.}1874 & 19&.42 & 0&.5208 \\
-3&.7 & 34&.43 & 32&.73 & 0&.2217 & &\phantom{00.}2202 &  &  \\
-3&.6 & 30&.87 & 28&.91 & 0&.2345 & &\phantom{00.}2588 & 16&.70 & 0&.5702 \\
-3&.5 & 27&.67 & 25&.51 & 0&.2480 & &\phantom{00.}3045 &  &  \\
-3&.4 & 24&.79 & 22&.47 & 0&.2625 & &\phantom{00.}3584 & 14&.28 & 0&.6239 \\
-3&.3 & 22&.19 & 19&.76 & 0&.2779 & &\phantom{00.}4222 &  &  \\
-3&.2 & 19&.86 & 17&.35 & 0&.2943 & &\phantom{00.}4977 & 12&.13 & 0&.6821 \\
-3&.1 & 17&.75 & 15&.20 & 0&.3119 & &\phantom{00.}5872 &  &  \\
-3& & 15&.86 & 13&.30 & 0&.3306 & &\phantom{00.}6933 & 10&.24 & 0&.7452 \\
-2&.9 & 14&.16 & 11&.60 & 0&.3507 & &\phantom{00.}8195 &  &  \\
-2&.8 & 12&.63 & 10&.10 & 0&.3721 & &\phantom{00.}9695 & 8&.562 & 0&.8133 \\
-2&.7 & 11&.25 & 8&.769 & 0&.3951 & 0&.01148 &  &  \\
-2&.6 & 10&.02 & 7&.593 & 0&.4198 & &\phantom{0.}1362 & 7&.091 & 0&.8868 \\
-2&.5 & 8&.905 & 6&.554 & 0&.4464 & &\phantom{0.}1617 &  &  \\
-2&.4 & 7&.908 & 5&.639 & 0&.4750 & &\phantom{0.}1922 & 5&.806 & 0&.9658 \\
-2&.3 & 7&.013 & 4&.834 & 0&.5058 & &\phantom{0.}2288 &  &  \\
-2&.2 & 6&.211 & 4&.128 & 0&.5391 & &\phantom{0.}2729 & 4&.688 & 1&.0504 \\
-2&.1 & 5&.491 & 3&.509 & 0&.5751 & &\phantom{0.}3260 &  &  \\
-2& & 4&.847 & 2&.969 & 0&.6142 & &\phantom{0.}3901 & 3&.722 & 1&.1407 \\
-1&.9 & 4&.269 & 2&.498 & 0&.6566 & &\phantom{0.}4679 &  &  \\
-1&.8 & 3&.753 & 2&.090 & 0&.7029 & &\phantom{0.}5623 & 2&.894 & 1&.2884 \\
-1&.7 & 3&.290 & 1&.736 & 0&.7535 & &\phantom{0.}6774 &  &  \\
-1&.6 & 2&.877 & 1&.431 & 0&.8089 & &\phantom{0.}8183 & 2&.192 & 1&.3371 \\
-1&.5 & 2&.508 & 1&.170 & 0&.8699 & &\phantom{0.}9915 &  &  \\
-1&.4 & 2&.179 & 0&.9469 & 0&.9372 & 0&.1205 & 1&.605 & 1&.4420 \\
-1&.3 & 1&.885 & 0&.7576 & 1&.012 & &\phantom{.}1470 &  &  \\
-1&.2 & 1&.623 & 0&.5981 & 1&.095 & &\phantom{.}1801 & 1&.123 & 1&.5497 \\
-1&.1 & 1&.390 & 0&.4648 & 1&.189 & &\phantom{.}2216 &  &  \\
-1& & 1&.182 & 0&.3545 & 1&.294 & &\phantom{.}2741 & 0&.7369 & 1&.6578 \\
-0&.9 & 0&.9978 & 0&.2642 & 1&.414 & &\phantom{.}3410 &  &  \\
-0&.8 & 0&.8337 & 0&.1913 & 1&.552 & &\phantom{.}4273 & 0&.4392 & 1&.7622 \\
-0&.7 & 0&.6879 & 0&.1336 & 1&.712 & &\phantom{.}5396 &  &  \\
-0&.6 & 0&.5583 & 0&.08888 & 1&.900 & &\phantom{.}6880 & 0&.2232 & 1&.8571 \\
-0&.5 & 0&.4429 & 0&.05534 & 2&.125 & &\phantom{.}8873 & 0&.1442 &  \\
-0&.4 & 0&.3397 & 0&.03128 & 2&.400 & 1&.161 & 0&.08554 & 1&.9319 \\
-0&.3 & 0&.2468 & 0&.01514 & 2&.744 & 1&.546 & 0&.04268 & 1&.9646 \\
-0&.2 & 0&.1619 & 0&.00550 & 3&.193 & 2&.107 & 0&.01526 & 1&.9861 \\
-0&.1 & 0&.0817 & 0&.00098 & 3&.805 & 2&.968 & 0&.00235 & 1&.9956 \\
0& & 0& & 0& & 4&.703 & 4&.384 & 0& & 1&.9968 \\
\hline
\end{tabular}
\end{table2}

Furthermore, the interesting physical quantities take the following values:
\begin{eqnarray}
\textrm{external radius:} & \fb{R}_e & = 4.703,\\
\textrm{total mass:} & \fb{M}_e & = 1.996,\nonumber\\
\textrm{total kinetic energy:} & \fb{L}_e & = 1.423.\nonumber
\end{eqnarray}

\begin{figure}
\includegraphics[width=\columnwidth]{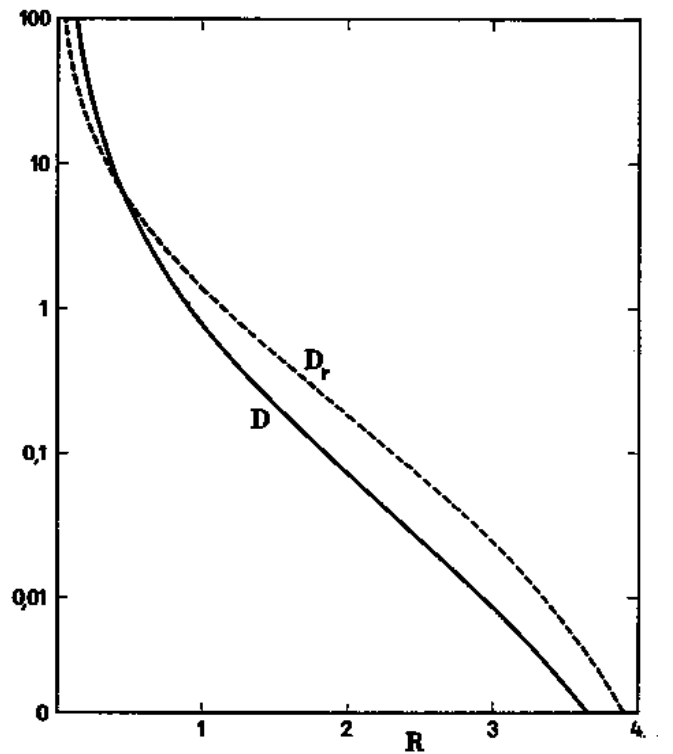}
\caption{Homologous model: spacial density and projected density as a function of the radius.}
\label{fig:5}
\end{figure}

The spacial density $\fb{D}$ is plotted in \reff{5} as a function of the radius. It increases very rapidly toward the center, as expected from (4.8). The structure of the model is perhaps better rendered in \reff{6}, which shows the mass fraction $\fb{M}_\fb{R}$ enclosed in a sphere of radius $\fb{R}$ (see Equation 2.45). Near the center, $\fb{M}_\fb{R}$ is proportional to $\fb{R}$. We note that half of the total mass is enclosed inside the radius $\fb{R} = 0.6800$, i.e. only about 1/7 of the external radius.

\begin{figure}
\includegraphics[width=\columnwidth]{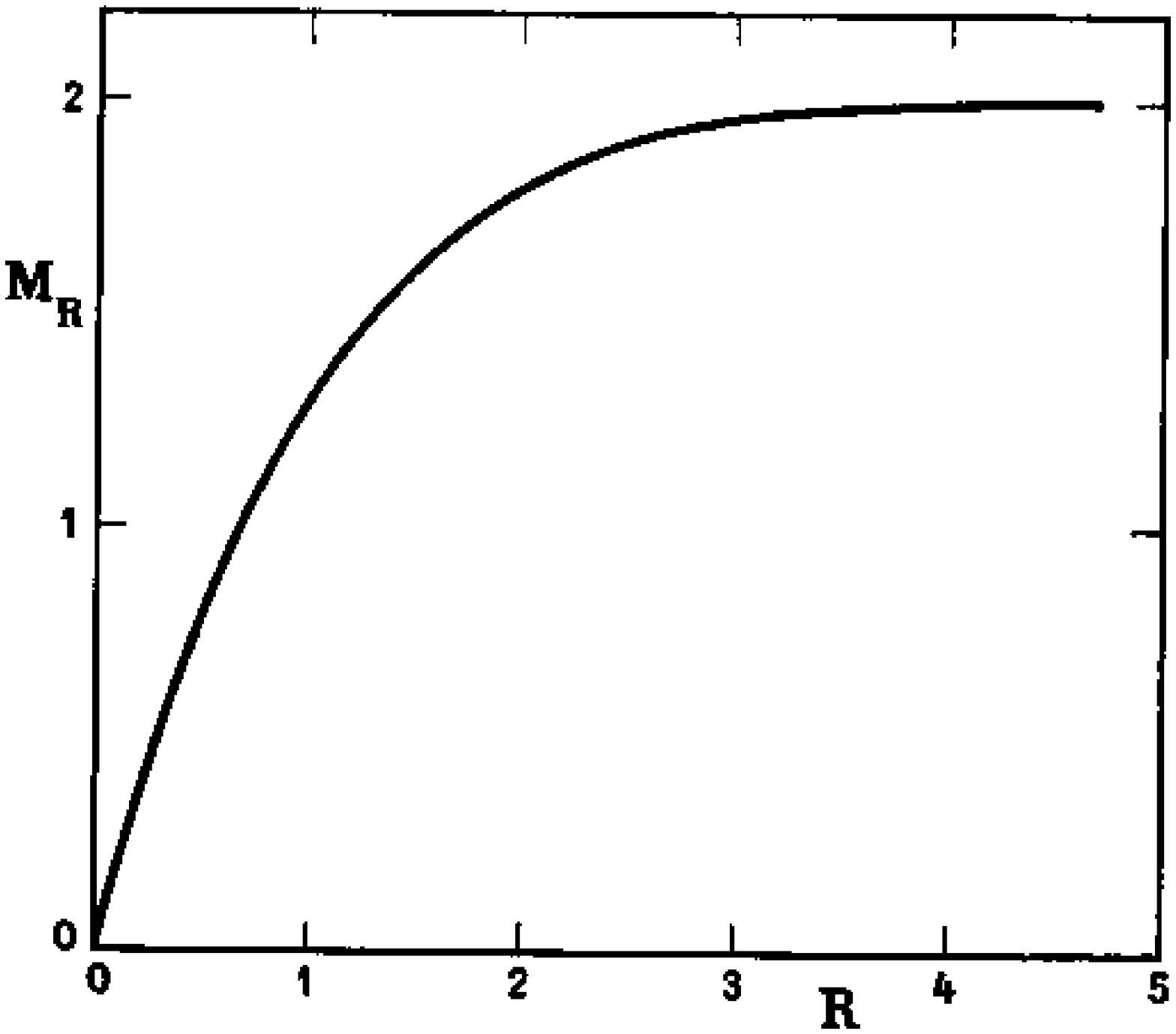}
\caption{Homologous model: mass enclosed in a sphere of radius $\fb{R}$.}
\label{fig:6}
\end{figure}

The potential $\fb{U}$ vanishes at the boundary of the cluster, according to our conventions; outside the cluster, its form is obtained by integrating (2.45):
\begin{equation}
\fb{U} = \fb{M}_e \left(\frac{1}{\fb{R}_e} - \frac{1}{\fb{R}}\right) \qquad \textrm{for } \fb{R} > \fb{R}_e.
\end{equation}

In particular, for $\fb{R} \to \infty$, the potential tends toward:
\begin{equation}
\fb{U}_\infty =  \frac{\fb{M}_e}{\fb{R}_e} = 0.4243.
\end{equation}
(This is the potential created by the cluster only.)

The projected density, computed from (2.49) is given in \reft{1}, column 6, and plotted in \reff{5} as a function of the distance to the center (see also \reff{15} and \reff{16}). Near the center, from (4.35a) et (4.35b), the spatial density is expressed as a function of the radius as:
\begin{equation}
\fb{D} =  \frac{2}{\fb{R}^2} + \frac{(8\pi)^{1/4}\ K}{\fb{R}}.
\end{equation}
We find
\begin{equation}
\fb{D}_P =  \frac{2\pi}{\fb{R}} - 2(8\pi)^{1/4}\ K\ \ln{(\fb{R})} + K_P  \qquad \textrm{for } \fb{R} \to 0,
\end{equation}
where $K_P$ is a constant which depends on the entire function \fb{D}. By linking the formula (5.9) with the values of \reft{1}, one finds $K_P \simeq -3.82$.

The projected mass (mass enclosed, in projection, inside a circle of radius $\fb{R}$), computed from the formula (2.51), is given in \reft{1}, column 7. In particular, we find out the value of the \emph{median radius} $\fb{R}_0$ of the cluster, defined as the radius of the circle which contains, in projection, half of the total mass. This quantity has the advantage of being easily measured for real clusters, while the external radius is, on the contrary, almost impossible to observe. We find
\begin{equation}
\fb{R}_0 = 0.4997.
\end{equation}

We note that this median radius $\fb{R}_0$ is about 10 times smaller than the external radius $\fb{R}_e$.

\subsection{Evolution}

When integrating (2.36), we get
\begin{eqnarray}
\beta &=& \beta_0\ \e{b\bf{T}},\\
\gamma &=& \gamma_0\ \e{c\bf{T}},\nonumber
\end{eqnarray}

The ``time'' $\fb{T}$ is defined by the differential equation (2.32e); it is not proportional to the physical time $T$. It is, somehow, the ``proper time'' of the cluster; its variation is measured with a scale which is proper to the cluster, and which always varies according to the evolution. To avoid any confusion, it is preferable to consider $\fb{T}$ as a simple parameter which measures the level of evolution of the cluster, as it is in (5.11).

The relation between $\fb{T}$ and the physical time $T$, found from (2.32e) and (5.11b), is
\begin{equation}
\fb{T} = -\frac{1}{c} \ln{(1-\gamma_0\ c\ T)},
\end{equation}
(by setting the origin of time at $T=0$ for $\fb{T}=0$). This relation is plotted in \reff{7}. We see that the evolution does not last forever, but rather ends abruptly at a time $T_1$, given by
\begin{equation}
T_1 = \frac{1}{\gamma_0 c}.
\end{equation}

\begin{figure}
\includegraphics[width=\columnwidth]{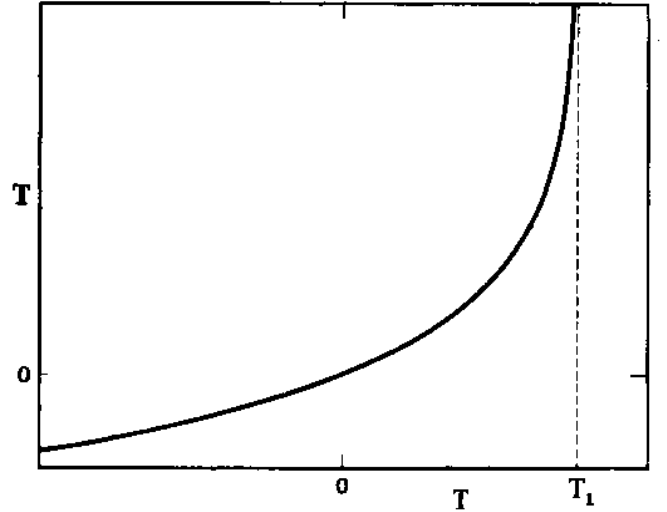}
\caption{Homologous model: evolution of the parameter $\fb{T}$, as a function of time.}
\label{fig:7}
\end{figure}

For $T=T_1$, the mass of the cluster becomes zero, as we will see below; it is therefore the time when the cluster disappears, after the escape of the last stars.

However, the curve extends forever in the past: the age of the cluster can be anything, it is not possible to give it an upper limit.

When introducing (5.12) in (5.11), and taking (5.4c) into account, we obtain the variation laws of the two parameters of the homology:
\begin{eqnarray}
\beta &=& \beta_0 \left(1-\frac{T}{T_1}\right)^{2/3},\\
\gamma &=& \gamma_0 \left(1-\frac{T}{T_1}\right)^{-1}.\nonumber
\end{eqnarray}

We derive, from (2.32), (2.43), (2.48), the variations of the various physical quantities as functions of time. In particular, for the total mass, we get
\begin{equation}
M_e = M_{e_0} \left(1-\frac{T}{T_1}\right),
\end{equation}
which shows that \emph{the mass decreases linearly with time}. The absolute escape rate is thus constant.

The radius is proportional to $(1-T/T_1)^{1/3}$, thus decreases quite slowly. The density is constant, and as a consequence, the orbital period of the stars within the cluster is also constant. The velocities decrease as $(1-T/T_1)^{1/3}$. The total energy of the cluster decreases (in absolute value) as $(1-T/T_1)^{5/3}$, thus faster than the mass.

\subsection{Accumulation of negative energy in the center}

As we have seen in the previous Chapter, a non-zero value of $K$ implies the existence of a flow of energy toward the center. This energy cannot vanish; we must admit that it constitutes an energy reservoir.

Thus, the total energy of the cluster is always made of two parts:
\begin{itemize}
\item a ``point'' energy $H_1$, accumulated in the center of the cluster;
\item a ``diffuse'' energy $H_2$, distributed within the entire cluster.
\end{itemize}
Let's look at how these two fractions of the energy vary with time. The variation of $H_1$ equals the flux of energy toward the center, i.e., from (4.30b) and (4.33):
\begin{equation}
\frac{dH_1}{dT} = \beta^{7/4}\ \gamma^{1/2}\ \frac{3}{2} K_D\ K.
\end{equation}
The diffuse energy $H_2$ is, from the virial theorem:
\begin{equation}
H_2 = -L_e,
\end{equation}
where $L_e$ is the total kinetic energy of the cluster; this relation remains valid as soon as the potential is set to zero at infinity (instead of the convention $U_e = 0$ adopted up to now). Therefore, from (2.43b) and (5.2f), we have:
\begin{eqnarray}
\frac{\dd H_2}{\dd T} &=& -\gamma \frac{\dd L_e}{\dd\fb{T}} \\
&=& - \gamma\ L_e \left(\frac{7}{4}b-\frac{1}{2}c\right) = \beta^{7/4}\ \gamma^{1/2}\ \frac{4-2\lambda}{3\lambda+1}\ \fb{L}_e\ c.\nonumber
\end{eqnarray}

We can finally compute the total variation of energy $H_1+H_2$ of the cluster. This variation is only due to the fact that the stars escape taking some energy away with them. The difference of gravitational potential between the boundary of the cluster and infinity is $U_\infty$ given by (5.7). Therefore, each star that escapes takes away a mass $m_1$ and a negative energy, equal to $-m_1 U_\infty$. The escape rate is, from (2.43a) and (5.2f):
\begin{eqnarray}
\frac{\dd M_e}{\dd T} = \gamma \frac{\dd M_e}{\dd\fb{T}} &=& \gamma\ M_e \left(\frac{3}{4}b-\frac{1}{2}c\right)\\
&=& - \beta^{3/4}\ \gamma^{1/2}\ \frac{2}{3\lambda+1}\ \fb{M}_e\ c,\nonumber
\end{eqnarray}
and thus,
\begin{equation}
\frac{\dd(H_1 + H_2)}{\dd T} = \beta^{7/4}\ \gamma^{1/2}\ \frac{2}{3\lambda+1}\frac{\fb{M}_e^2}{\fb{R}_e}\ c.
\end{equation}

The comparison of (5.16), (5.18), (5.20) shows that we must have:
\begin{equation}
\frac{3}{2}K_D\ K + \frac{4-2\lambda}{3\lambda+1}\ \fb{L}_e\ c = \frac{2}{3\lambda+1}\ \frac{\fb{M}_e^2}{\fb{R}_e}\ c.
\end{equation}
which expresses the conservation of energy. This relation between the parameters of the model provides a useful verification of the calculations. When using the numerical values (5.4) and (5.5), and when omitting the factor $\beta^{7/4}\ \gamma^{1/2}$, we get:
\begin{eqnarray}
\frac{\dd H_1}{\dd T} &=& -0.6464,\\
\frac{\dd H_2}{\dd T} &=& +0.9672,\nonumber\\
\frac{\dd(H_1+H_2)}{\dd T} &=& +0.3455.\nonumber
\end{eqnarray}

The conservation of energy is quite well verified; the discrepancy that remains comes from the errors in the computation (mainly the error made in $\fb{R}_e$).

Thus, in the course of the evolution, the diffuse energy $H_2$, always negative according to (5.17), decreases in absolute value; the numerical values (5.22) show that about one third of the negative energy is taken away by the stars that escape, while the other two thirds go in the center.

It remains to explain the mechanism of this accumulation of energy in the center. It is not associated with an accumulation of matter; we have supposed that the flux of matter toward the center of the cluster is zero (and this is indeed necessary, because a central condensation of mass would create an additional potential that would modify the structure of the cluster; for example the potential would vary as $1/r$ and not $\ln{(r)}$ anymore, near the center). Thus, the negative energy $H_1$ must be stored without any increase of the number of stars. Apparently, there is only one process that allows this: \emph{the formation of tight binary or multiple stars in the center of the cluster}.

The direct observation cannot confirm the existence of this phenomenon: the images of the stars are sorely separated \tn{resolved} in the central region of the globular clusters, and it would be impossible to discover there the presence of a particularly compact group of stars. On the other hand, \citet{vonHoerner1960} has computed numerically the evolution of artificial clusters; by this means, it is possible to observe the mechanism of the evolution, in as much detail as desired. von Hoerner has indeed noted the frequent formation of binaries in the center of the cluster; in one case, a compact group of 4 stars appeared. This seems an excellent confirmation of the process that we have been led to admit.

Note again that this central accumulation of energy does not affect the structure of the cluster, except for the few stars that support it; thus, it can develop independently of the global evolution of the cluster. In fact, we have seen that the absolute value of the central energy increases, while that of the diffuse energy decreases. At the end of the evolution, about two thirds of the initial negative energy is in the central condensation, and one third only has left the cluster, carried by the stars. It would be very interesting, although not possible here, to study in more detail the process of accumulation and to answer in particular the following questions: how do multiple stars form? How many are there? What happens to them after the cluster has disappeared?

\section{Stars with different masses}

The case of a cluster with an arbitraty mass distribution seems much more involved than the simple case of equal masses that we have considered so far; thus we shall not seek to treat it in the general case. But we are going to see that some simplified cases allow easy calculations, and provide, not a complete and rigorous solution, but at least several indications about the effect of the dispersion of the masses. We will first assume that we add a small number of stars of different masses to a cluster that contains equal mass stars, and will study the behavior of this secondary population. Later, this study will help us to obtain an approximate solution to the general case.

\subsection{Simplified model made of two populations}

We suppose that the cluster is made of the mix of a main population 1 of stars of mass $m_1$, and of a secondary population 2, numerically negligible with respect to the first one, of stars of mass $m_2$. The distribution function is not (2.23) anymore but rather:
\begin{equation}
f(E,m,t) = \delta(m-m_1) F_1(E,t) + \delta(m-m_2) F_2(E,t),
\end{equation}
with:
\begin{equation}
F_2 \ll F_1.
\end{equation}

The population 1 is almost not affected by the presence of the population 2; it alone determines the structure and the evolution of the cluster; as a consequence, all the results obtained in the previous Chapters still work for it. We assume that the population 1 has reached the final state represented by the homologous model of the Chapter~V.

Furthermore, by setting $m=m_2$ in (2.22), one gets a new equation, that describes the evolution of $F_2$. One can neglect the perturbations between the stars of the population 2, and the equation becomes, after applying the transformation of (2.24):
\begin{eqnarray}
0 &=& \frac{\partial}{\partial E} \bigg[ \mu F_2 \int_{-\infty}^E F_1 Q_1'\, \dd E_1 \\
&& + F_2' \left(\int_{-\infty}^E F_1 Q_1\, \dd E_1 + Q \int_E^{\infty} F_1\, \dd E_1 \right) \bigg] \nonumber\\
&& +F_2' \frac{\partial Q}{\partial T} - Q' \frac{\partial F_2}{\partial T},\nonumber
\end{eqnarray}
where we set:
\begin{equation}
\frac{m_2}{m_1} = \mu.
\end{equation}

This equation has the dimension of $F_2$; therefore, we can apply an homologous transformation to $F_2$, independently of those of $F_1$. Thus, we set
\begin{equation}
F_2 = \gamma_2\ \fb{F}_2.
\end{equation}
Therefore the homology now depends on the three parameters $\beta$, $\gamma$, $\gamma_2$. We set
\begin{equation}
\frac{1}{\gamma_2}\frac{\dd\gamma_2}{\dd\fb{T}} = c_2,
\end{equation}
and when continuing the calculation, as in the Chapter~II, we obtain the equation
\begin{eqnarray}
0 &=& \Bigg\{ \mu \fb{F}_2 \int_{-\infty}^{\fb{E}} \fb{F}_1 \fb{Q}_1'\, \dd\fb{E}_1 \\
&& + \fb{F}_2' \bigg( \int_{-\infty}^{\fb{E}} \fb{F}_1 \fb{Q}_1\, \dd\fb{E}_1 + \fb{Q} \int_\fb{E}^{\infty} F_1\, \dd E_1 \bigg) \nonumber\\
&& + \left( \frac{3}{4}b-\frac{3}{2}c \right) \fb{F}_2\fb{Q}\nonumber\\
&& + \left( \frac{3}{2}c-\frac{3}{4}b - c_2 \right) \int_{-\infty}^{\fb{E}} \fb{F}_2 \fb{Q}_1'\, \dd\fb{E}_1\Bigg\}.\nonumber
\end{eqnarray}

In this equation, $\fb{F}_1$, $\fb{Q}$ or $\fb{Q}_1$, $b$, $c$ are the functions and constants of the homologous model, given in the Chapter~V; the function $\fb{F}_2$ and the constant $c_2$ are unknown.

The boundary conditions are only:
\begin{equation}
\fb{F}_2(0) = 0,
\end{equation}
which tells us that the stars of the population 2 escape when they reach the boundary of the cluster.

Near the center, on the other hand, we can neglect the last terms of (6.7) in a first order approximation, and use the asymptotical expressions (4.6) and (4.7c) for $\fb{F}_1$ and $\fb{Q}$; the equation becomes:
\begin{equation}
\mu\ \fb{F}_2 + \fb{F}_2' = 0.
\end{equation}
and thus
\begin{equation}
\fb{F}_2 = C\ \e{-\mu \fb{E}},
\end{equation}
where $C$ is an arbitrary constant. The equation (6.10) is as expected: indeed, it shows that the Maxwellian equilibrium is realised in the central region of the cluster between the two populations.

\subsubsection{The $\mu \ge 3/2$ case}

This functional form for $\fb{F}_2$ has a strange consequence. The mass of the subset of stars of the population 2 whose energy is below $E$ is, from (2.40a):
\begin{equation}
\mc{M}_2 = 16 \pi^2\ m_2 \int_{-\infty}^E F_2 q_1'\, \dd E_1,
\end{equation}
i.e., when switching to normalized variables, and then to the canonical variables,
\begin{equation}
\fb{M}_2 = \mu \int_{-\infty}^{\fb{E}} \fb{F}_2 \fb{Q}_1'\, \dd\fb{E}_1.
\end{equation}
$\fb{Q}'$ is given near the center by (4.7c), i.e.,
\begin{equation}
\fb{Q}' = \frac{3}{2} K_D\ \e{3\fb{E}/2}.
\end{equation}

When using this form and (6.10), we immediately see that the integral (6.12) diverges if
\begin{equation}
\mu \ge \frac{3}{2}.
\end{equation}

In other words, in this case, the mass of the population 2 is infinite, because of a too rapid increase of the distribution function (and of the density) toward the center.

This anomaly is easy to explain. We have assumed in (6.2) that $F_2$ is always negligible with respect to $F_1$; however, when comparing the asymptotical forms (4.6) and (6.10) of these two functions, we can see that when $\mu$ is greater than unity, $F_2$ increases faster than $F_1$ for $E\to -\infty$, and that, a value of $E$ below which $F_2$ becomes larger than $F_1$ always exists, however small the constant $C$. This critical value is (from Equations 2.31b and 6.5)
\begin{equation}
\fb{E}_c = \frac{\ln{(C\ \gamma_2/\gamma)}}{\mu-1}.
\end{equation}

Using the formula (6.13) for $\fb{Q}'$ is only meaningful when $\fb{E} \gg \fb{E}_c$. On the contrary, for $\fb{E} \ll \fb{E}_c$, $\fb{F}_1$ becomes negligible with respect to $\fb{F}_2$. Then, we can obtain the new forms of $\fb{D}$, $\fb{R}$, $\fb{Q}$ by using (6.10) and the fundamental equations. In particular, we get
\begin{equation}
\fb{Q}' = \frac{C\ \gamma_2}{\gamma} \mu^{5/4}\ \frac{3}{2}K_D\ \e{3\mu\ \fb{E}/2} \qquad \textrm{for } \fb{E} \ll \fb{E}_c
\end{equation}
\tn{The $c$ subscript is missing in the original version.} and with this correct form, the integral (6.12) does not diverge anymore. The quantity $\fb{F}_2\fb{Q}'$ yields a maximum in the vicinity of $\fb{E} = \fb{E}_c$ and exponentially decreases in both sides. We can compute the mass $\fb{M}_2$ in an approximate way by assuming that $\fb{Q}'$ is given by (6.16) when $\fb{E}<\fb{E}_c$ and by (6.13) when $\fb{E}>\fb{E}_c$; we find
\begin{eqnarray}
\fb{M}_2 &=& 3 K_D \bigg( \mu^{-5/4}\\
&& + \frac{\mu}{2\mu - 3}\bigg) \frac{\gamma}{\gamma_2} \left(\frac{C\ \gamma_2}{\gamma}\right)^{1/(2\mu-2)} \qquad \textrm{for } \fb{E} \gg \fb{E}_c,\nonumber
\end{eqnarray}

$\fb{M}_2$ does not depend on $\fb{E}$ anymore; indeed, because of the form of $\fb{F}_2\fb{Q}'$, almost all the stars of the population 2 have an energy of the order of magnitude of $\fb{E}_c$, therefore are part of $\fb{M}_2$ for $\fb{E} \gg \fb{E}_c$.

We focus here on the extreme case where the population 2 is negligible with respect to the population 1; therefore $C$ must be very small compared to unity. For $C\to 0$, (6.15) shows that $\fb{E}_c \to -\infty$. This way, we obtain the following result: \emph{the stars whose mass is greater than $3/2$ times the mean mass are almost all gathered near the center of the cluster}. Obviously, this conclusion is related to the simplified mass distribution that we have adopted, and should not be extended without further study to the case of an arbitrary mass distribution.

We also notice that $\fb{M}_2$ is not proportional to $C$: the power of $C$ in (6.17) is less than unity. As a consequence, for $C\to 0$, the last term of the equation (6.7) decreases slower than the others, which are proportional to $C$. In the limit, the equation becomes
\begin{equation}
\frac{3}{2}c-\frac{3}{4}b - c_2 = 0.
\end{equation}

Let's divide all the terms of (6.7) by $C$; the last term is then, for $C \to 0$, an indeterminate form $0\times \infty$. Let $p$ be its value; $p$ is a new constant yet to be determined, which replaces $c_2$.

\subsubsection{Expansion near the center}

Substituting in the equation for the expansions (4.35c), (4.36a), (4.37), we get, after some calculations, the expansion of $\fb{F}_2$ near the center. For $\mu < 3/2$:
\begin{eqnarray}
\fb{F}_2 &=& C\ \e{-\mu\fb{E}} \bigg\{ 1+ \mu K\ \e{\fb{E}/2}\\
&& + \bigg[ \frac{\mu(\mu-1)}{2}K^2 + \frac{c_2 - \mu c + \mu (\mu-1) b}{3-2\mu} \bigg] \e{\fb{E}} \bigg\}.\nonumber
\end{eqnarray}
For $\mu > 3/2$:
\begin{eqnarray}
\fb{F}_2 &=& C\ \e{-\mu\fb{E}} \bigg\{ 1+ \mu K\ \e{\fb{E}/2}\\
&& + \bigg[ \frac{\mu(\mu-1)}{2}K^2 + \frac{1}{2}c - \frac{1}{2} \bigg(\mu+\frac{1}{2}\bigg)b \bigg] \e{\fb{E}} \bigg\}.\nonumber
\end{eqnarray}
Finally, for the special case $\mu = 3/2$:
\begin{eqnarray}
\fb{F}_2 &=& C\ \e{-3\fb{E}/2} \bigg\{ 1+ \frac{3}{2} K\ \e{\fb{E}/2}\\
&& + \bigg[ \frac{3}{8}K^2 + \frac{1}{2}c - b - \frac{p}{3K_D\ C} \bigg] \e{\fb{E}} \bigg\}.\nonumber
\end{eqnarray}

\subsubsection{Method of solution}

Equation (6.7) is integrated from the center toward the boundary. The initial conditions are given by one of the expansions (6.19) to (6.21). The value of the factor $C$ does not matter here; in practice, one sets $C=1$. Proceeding by trial and error, we find the value of the parameter $c_2$ or $p$ for which the condition (6.8) is fulfilled at the boundary.

\subsection{Results: structure}

The calculation has been done for several values of the relative mass $\mu$. The value of the parameter is given in \reft{2}, column 2 or 3. \reff{8} plots the ratio of the distribution function $\fb{F}_2$ to its asymptotical form $C\ \e{-\mu\fb{E}}$. This ratio tends toward unity when $\fb{E}$ goes to $-\infty$; toward the boundary, it becomes smaller and smaller, which expresses the discrepancy between the real distribution and a Maxwellian distribution\footnote{\citet{Spitzer1958b} have computed these functions in a simpler case: the structure of the cluster was supposed to remain constant and was represented by a constant potential inside the cluster, zero outside; furthermore, the distribution function $F_1$ of the main population was supposed to be Maxwellian. The curves obtained by these authors have some similarities with ours but are arranged in reverse order! This peculiarity is probably linked to the too rough approximation made to the potential.}.

\begin{table} 
\caption{} 
\label{tab:2}
\begin{tabular}{r@{}lccr@{}l}
\multicolumn{2}{c}{$\mu$} & $c_2$ & $p$ & \multicolumn{2}{c}{$\theta$} \\
\hline
\hline
0& &  -0.6240&  &  1&.4396\\ 
0&.2&  -0.3998&  &  1&.2154\\ 
0&.4&  -0.1838&  &  0&.9994\\ 
0&.6&  +0.0233&  &  0&.7923\\ 
0&.8&  +0.2208&  &  0&.5948\\ 
1& &  +0.4078&  &  0&.4078\\ 
1&.2&  +0.5829&  &  0&.2327\\ 
1&.4&  +0.7430&  &  0&.0726\\ 
1&.5&  &  0.4796&  0 & \\ 
1&.6&  &  0.4670&  0 & \\ 
1&.8&  &  0.4307&  0 & \\ 
2& &  &  0.3859&  0 & \\ 
2&.5&  &  0.2685&  0 & \\ 
3& &  &  0.1726&  0 & \\ 
4& &  &  0.0685&  0 & \\ 
\hline
\end{tabular}
\end{table}

\begin{figure}
\includegraphics[width=\columnwidth]{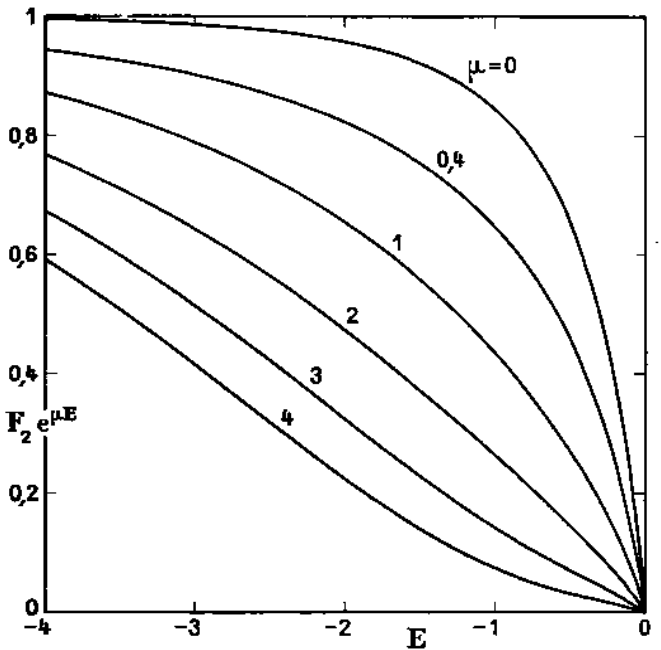}
\caption{Stars with different masses: distribution functions.}
\label{fig:8}
\end{figure}

\reff{9} and \reft{3} give, for several values of $\mu$, the product $\fb{D}_P\fb{R}$ of the projected density and the radius, as a function of the radius; we will see in the Chapter~VIII that this function is the one that best allows for a comparison with the observations. The factors have been adjusted so that all the curves cross at the same point: $\fb{D}_P\fb{R} = 1$ for $\fb{R} = 1$. The values of $C$ are given in the last row of \reft{3}.

\begin{figure}
\includegraphics[width=\columnwidth]{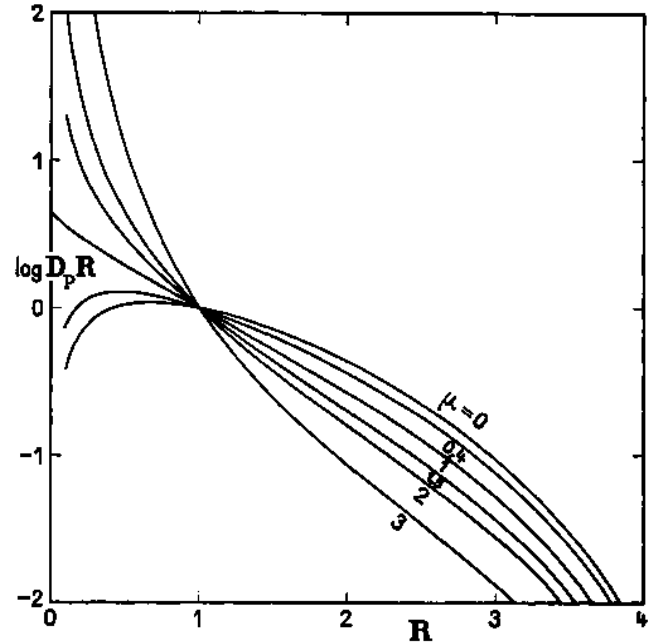}
\caption{Stars with different masses: projected density times radius, as a function of the radius.}
\label{fig:9}
\end{figure}

\begin{table2}
\caption{$\log{(\fb{D}_P \fb{R})}$} 
\label{tab:3}
\begin{tabular}{r@{}lr@{}lr@{}lr@{}lr@{}lr@{}lr@{}lr@{}lr@{}lr@{}lr@{}lr@{}lr@{}lr@{}lr@{}l}
\multicolumn{2}{c}{$\fb{R}$} & \multicolumn{2}{c}{$\mu = 0$} & \multicolumn{2}{c}{0.2} & \multicolumn{2}{c}{0.4} & \multicolumn{2}{c}{0.6} & \multicolumn{2}{c}{0.8} & \multicolumn{2}{c}{1} & \multicolumn{2}{c}{1.2} & \multicolumn{2}{c}{1.4} & \multicolumn{2}{c}{1.6} & \multicolumn{2}{c}{1.8} & \multicolumn{2}{c}{2} & \multicolumn{2}{c}{2.5} & \multicolumn{2}{c}{3} & \multicolumn{2}{c}{4} \\
\hline
\hline
0& & \multicolumn{2}{c}{$-\infty$} & \multicolumn{2}{c}{$-\infty$} & \multicolumn{2}{c}{$-\infty$} & \multicolumn{2}{c}{$-\infty$} & \multicolumn{2}{c}{$-\infty$} & 0.&655 & \multicolumn{2}{c}{$+\infty$}  & \multicolumn{2}{c}{$+\infty$} & \multicolumn{2}{c}{$+\infty$} & \multicolumn{2}{c}{$+\infty$} & \multicolumn{2}{c}{$+\infty$} & \multicolumn{2}{c}{$+\infty$} & \multicolumn{2}{c}{$+\infty$} & \multicolumn{2}{c}{$+\infty$}\\
0&.110 & 1.&645 & 1.&763 & 1.&910 & 0.&090 & 0.&301 & & 542 & 0.&809 & 1.&094 & 1.&376 & 1.&659 & 1.&963 & 2.&783 & 3.&652 & 5.&433 \\
0&.168 & & 784 & & 883 & 0.&002 & & 145 & & 310 & & 497 & & 703 & 0.&923 & & 139 & & 355 & & 589 & & 228 & 2.&915 & 4.&332 \\
0&.262 & & 906 & & 983 & & 073 & & 178 & & 297 & & 431 & & 578 & & 735 & 0.&887 & & 036 & & 200 & 1.&657 & & 158 & 3.&203\\
0&.331 & & 958 & 0.&022 & & 097 & & 182 & & 279 & & 386 & & 505 & & 631 & & 751 & 0.&869 & 0.&999 & & 364 & 1.&769 & 2.&623\\
0&.420 & 0.&000 & & 051 & & 109 & & 175 & & 249 & & 331 & & 421 & & 516 & & 606 & & 693 & & 789 & & 063 & & 372 & & 030\\
0&.539 & & 029 & & 065 & & 106 & & 152 & & 204 & & 260 & & 321 & & 386 & & 447 & & 505 & & 569 & 0.&753 & 0.&965 & 1.&422\\
0&.703 & & 037 & & 058 & & 081 & & 107 & & 135 & & 165 & & 199 & & 234 & & 267 & & 297 & & 331 & & 429 & & 543 & 0.&794\\
0&.937 & & 012 & & 016 & & 020 & & 025 & & 029 & & 034 & & 041 & & 047 & & 052 & & 057 & & 063 & & 080 & & 100 & & 143\\
1&.095 & 1.&979 & 1.&974 & 1.&968 & 1.&962 & 1.&954 & 1.&947 & 1.&940 & 1.&931 & 1.&924 & 1.&917 & 1.&910 & 1.&889 & 1.&864 & 1.&806\\
1&.294 & & 924 & & 909 & & 893 & & 876 & & 857 & & 836 & & 816 & & 793 & & 774 & & 755 & & 736 & & 679 & & 613 & & 457\\
1&.552 & & 835 & & 811 & & 784 & & 755 & & 724 & & 691 & & 657 & & 620 & & 589 & & 560 & & 529 & & 439 & & 334 & & 087\\
1&.900 & & 688 & & 652 & & 614 & & 574 & & 530 & & 484 & & 438 & & 388 & & 344 & & 305 & & 263 & & 144 & & 003 & 2.&676\\
2&.125 & & 568 & & 528 & & 487 & & 443 & & 393 & & 343 & & 293 & & 239 & & 188 & & 146 & & 101 & 2.&968 & 2.&812 & & 455\\
2&.400 & & 413 & & 369 & & 325 & & 278 & & 224 & & 169 & & 113 & & 054 & & 001 & 2.&958 & 2.&908 & & 764 & & 594 & & 208\\
2&.744 & & 177 & & 133 & & 086 & & 036 & 2.&982 & 2.&925 & 2.&867 & 2.&805 & 2.&751 & & 704 & & 652 & & 505 & & 333 & 3.&930\\
3&.193 & 2.&797 & 2.&753 & 2.&706 & 2.&656 & & 602 & & 545 & & 487 & & 425 & & 371 & & 324 & & 272 & & 125 & 3.&953 & & 550\\
3&.805 & & 060 & & 016 & 3.&969 & 3.&919 & 3.&865 & 3.&808 & 3.&750 & 3.&688 & 3.&634 & 3.&587 & 3.&535 & 3.&388 & & 216 & 4.&813\\
4&.703 & \multicolumn{2}{c}{$-\infty$} & \multicolumn{2}{c}{$-\infty$} & \multicolumn{2}{c}{$-\infty$} & \multicolumn{2}{c}{$-\infty$} & \multicolumn{2}{c}{$-\infty$} & \multicolumn{2}{c}{$-\infty$} & \multicolumn{2}{c}{$-\infty$} & \multicolumn{2}{c}{$-\infty$} & \multicolumn{2}{c}{$-\infty$} & \multicolumn{2}{c}{$-\infty$} & \multicolumn{2}{c}{$-\infty$} & \multicolumn{2}{c}{$-\infty$} & \multicolumn{2}{c}{$-\infty$} & \multicolumn{2}{c}{$-\infty$}\\
\multicolumn{2}{c}{$\log{(C)}$} & 1.&737 & 1.&769 & 1.&797 & 1.&821 & 1.&841 & 1.&857 & 1.&870 & 1.&877 & 1.&858 & 1.&827 & 1.&805 & 1.&769 & 1.&741 & 1.&651\\
\hline
\end{tabular}
\end{table2}

For $\mu=1$, we naturally retrieve the distribution of the homologous model; near the center, the quantity $\fb{D}_p\fb{R}$ tends toward a constant (see Equation 5.9). For $\mu <1$, it tends toward zero; for $\mu > 1$, it tends to infinity. Toward the boundary, the curves tend to become parallel.

\subsection{Escape rate}

From the formula (6.11) we find out that the total mass $\mc{M}_{2e}$ of the population 2 is proportional to
\[
\beta^{3/4}\ \gamma^{-3/2}\ \gamma_2.
\]

As a consequence, the relative escape rate of the stars of the population 2 is given by:
\begin{equation}
\theta= -\frac{1}{\mc{M}_{2e}}\frac{\dd\mc{M}_{2e}}{\dd\fb{T}} = -\frac{3}{4}b + \frac{3}{2}c - c_2.
\end{equation}
$\theta$ is given in \reft{2} column 4, and plotted in \reff{10} as a function of $\mu$. From (6.18), we have
\begin{equation}
\theta= 0 \qquad \textrm{for } \mu \ge \frac{3}{2}.
\end{equation}
Therefore, in our simplified model, \emph{the stars more massive than $3/2$ time the mean mass do not escape}.

\begin{figure}
\includegraphics[width=\columnwidth]{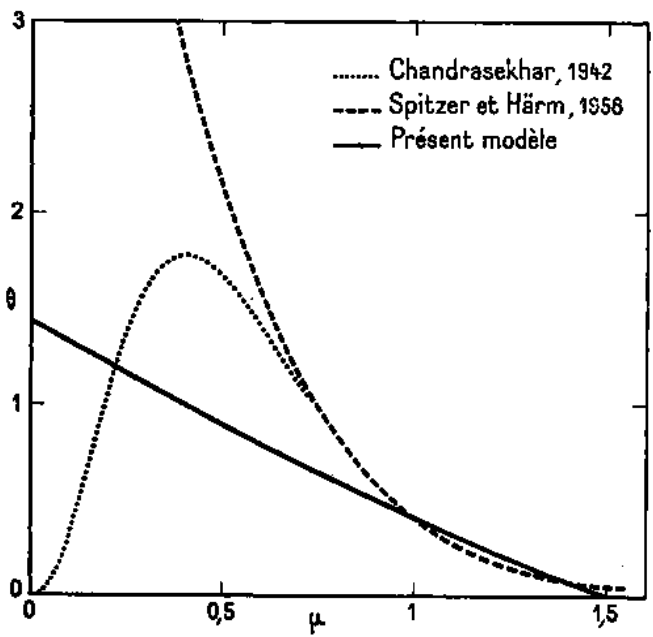}
\caption{Escape rate as a function of mass. \tn{The legend reads \emph{Pr\'esent mod\`ele}: this model.}}
\label{fig:10}
\end{figure}

Between $\mu=0$ and $\mu=3/2$, the escape rate decreases regularly, as one would expect. The very low-mass stars ($\mu \simeq 0$) escape about 3.5 times faster than the mean-mass stars.

The escape rates computed by \citet{Chandrasekhar1942} and by \citet{Spitzer1958b}, normalized so that they all take the same value for $\mu=1$, are also plotted in \reff{10}. We can see that the three curves are very different, especially for the low masses. This shows strikingly the lack of rigor of the cluster theory, in its present state.

It is quite obvious, however, that the less massive stars must escape faster. The fact that the curve from Chandrasekhar yields a maximum and then decreases for the low-mass stars is due to excessive simplifications. Indeed, in his computation he assumes that the stars must first adopt a Maxwellian distribution (and thus, high velocities for the low-mass stars), before being able to escape. In reality, as we have seen, the stars escape almost as soon as they have reach their escape velocity.

Previous works \citep{vandenBergh1957, Takase1960} used the escape rate provided by Chandrasekhar to calculate the initial mass function of open clusters from their present-day mass function; it would be advisable to do these calculations again with a more exact escape rate. In particular, the abnormal lack of low-mass stars, found by Takase in the initial distribution of the Pleiades, is likely to disappear.

\subsection{Approximation in the general case}

We are going to show that the results obtained above in the case of the very special mass distribution (6.1) allow us to solve in an approximate way the general case of any mass distribution.

In general, the distribution function takes the form (2.4): $f(E,m,t)$. Let $n_m\ \dd m$ be the number of stars of the cluster whose mass is between $m$ and $m+\dd m$; we easily derive (see Equations 2.38 and 2.40)
\begin{equation}
n_m = 16 \pi^2 \int_{-\infty}^0 f q'\, \dd E.
\end{equation}
$n_m$ depends on $m$ and also on $t$, because of the escape.

The total number of stars and the total mass are given by
\begin{eqnarray}
n &=& \int_0^\infty n_m\, \dd m,\\
\mc{M}_e &=& \int_0^\infty n_m\ m\, \dd m.\nonumber
\end{eqnarray}

The mean mass and the quadratic mean mass are defined by means of the classical formulae:
\begin{eqnarray}
\overline{m} &=& \frac{1}{n} \int_0^\infty n_m\, \dd m = \frac{\mc{M}_e}{n},\\
\overline{m^2} &=& \frac{1}{n}\int_0^\infty n_m\ m^2\, \dd m.\nonumber
\end{eqnarray}
This done, let's consider the fundamental equations (2.6) and (2.22). We can switch the order of integration with respect to $m$ and $E$. Then, we notice that $m$ can be removed if we introduce the two functions:
\begin{eqnarray}
\int_0^\infty f m\, \dd m &=& J_1(E,t),\\
\int_0^\infty f m^2\, \dd m &=& J_2(E,t).\nonumber
\end{eqnarray}
$J_1$ and $J_2$ are ``mean distribution functions'', obtained by weighting in two ways the distribution functions corresponding to the different values of the mass. To come back to equations already solved, we have to make an approximation: we suppose that theses two mean distribution functions are similar up to a factor. In this case, the equations (6.24) and (6.26) show that $J_1$ and $J_2$ must be proportional to $\overline{m}$ and $\overline{m^2}$ respectively, and thus we set
\begin{eqnarray}
J_1(E,t) &=& \overline{m}\ F(E,t),\\
J_2(E,t) &=& \overline{m^2}\ F(E,t).\nonumber
\end{eqnarray}

When using these expressions in the fundamental equations, they become
\begin{eqnarray}
\rho &=& 4\pi \overline{m} \int_U^{\infty} (2E-2U)^{1/2} F\,  \dd E\\
0 &=& \Bigg\{ 16 \pi^2\ G^2\ \ln{(n)} \frac{\partial}{\partial E} \bigg[ \\
&& \overline{m} mf \int_{-\infty}^{E} F_1 q_1'\, \dd E_1\nonumber\\
&& + \overline{m^2} f' \bigg( \int_{-\infty}^{E} F_1 q_1\, \dd E_1 \nonumber \\
&& + q \int_E^{\infty} F_1\, \dd E_1 \bigg) \bigg] + f' \frac{\partial q}{\partial t}-q'\frac{\partial f}{\partial t}\Bigg\}.\nonumber
\end{eqnarray}

The other two fundamental equations, (2.7) and (2.19), do not involve the mass, and thus are not modified. We now define the normalized variables by means of transformations, slightly different from (2.24):
\begin{eqnarray}
\rho &=& 4\pi\ \overline{m}\ D\\
r &=& (16 \pi^2\ G\ \overline{m})^{-1/2}\ R\nonumber\\
q &=& (16 \pi^2\ G\ \overline{m})^{-3/2}\ Q\nonumber\\
\dd t &=& [16 \pi^2 G^2\ \overline{m^2}\ \ln{(n)}]^{-1}\ \dd T.\nonumber
\end{eqnarray}

We easily check that, this way, we retrieve the equations (2.25) which have been obtained for the case of equal mass stars. (The equation 6.30 must be multiplied by $m\, \dd m$ and integrated.) As a consequence, the structure of the cluster, given by the functions $D$, $R$, $Q$, is that of the homologous model, and the mean distribution function $F$, introduced in (6.28), also matches that of the homologous model.

Furthermore, when setting
\begin{equation}
\frac{\overline{m}}{\overline{m^2}}\ m = \mu
\end{equation}
and replacing $f$ with $F_2$, we find that the equation (6.30) transforms into (6.3), i.e. the equation obtained at the beginning of this Chapter for the simplified model. It follows that the detailed distribution function $f$ is represented, for the different masses, by the solutions $F_2$ of the simplified model taking (6.32) into account.

From the previous relation, we note that $\mu$ is the relative mass computed by considering the mass unit to be, not the mean mass $\overline{m}$, but rather
\begin{equation}
m_0 = \frac{\overline{m^2}}{\overline{m}},
\end{equation}
which is different, in practice, by a factor greater than 2 (see Equation 8.9, below).

Finally, we note that the transformation equations (2.41) and (2.47) must be replaced with:
\begin{eqnarray}
\mc{M} &=& (16 \pi^2\ \overline{m})^{-1/2}\ G^{-3/2}\ M\\
\mc{L} &=& (16 \pi^2\ \overline{m})^{-1/2}\ G^{-3/2}\ L\nonumber\\
\rho_P &=& \overline{m}^{1/2}\ G^{-1/2}\ D_P.\nonumber
\end{eqnarray}

\section{Approach of the homologous model}

In this Chapter, we are back to the hypothesis of equal masses, and we are going to try to extend the results of the homologous model in another direction: we will study the evolution of a cluster whose shape is close to those of the homologous model, but not identical. We will first consider the case of a cluster that differs from the homologous model because its central density is finite; then the case of a cluster with infinite central density but with differences in the global structure. In both cases, the calculation will only be approximate. Finally, the results will be combined to draw a general picture of the evolution of the cluster.

\subsection{Formation of the central singularity}

Let's consider a cluster that matches the homologous model everywhere, except in a small central region, where it differs so that its central density is finite. The central potential is then also finite; let's call it $\fb{U}_0$. Near the center, the radius $\fb{R}$ is proportional to $(\fb{U} - \fb{U}_0)^{1/2}$; From (5.2d), we easily derive that $\fb{Q}$ is proportional to $(\fb{E} - \fb{U}_0)^3$. If we assume that the distribution function $\fb{F}$ also remains finite in the center, we find that $\fb{F}\fb{Q}'$ is proportional to $(\fb{E} - \fb{U}_0)^2$ near the center. It will be useful, for the next calculations, to have a formula for $\fb{F}\fb{Q}'$ as simple as possible; therefore, we define a simplified model by means of the following conditions: below a given value $\fb{E}=\fb{E}_1$, $\fb{F}\fb{Q}'$ is proportional to $(\fb{E} - \fb{U}_0)^2$; above this value, it is as the homologous model. In addition, the function and its derivative must be continuous for $\fb{E}=\fb{E}_1$ (\reff{11}). These conditions translate into:
\begin{eqnarray}
\fb{E}_1 &=& \fb{U}_0 + 4\\
\fb{F}\fb{Q}' &=& \left\{\begin{array}{ll}
\displaystyle
\frac{3}{32}K_D\ \e{\fb{E}_1/2}\ (\fb{E}-\fb{U}_0)^2 & \textrm{for } \fb{U}_0<\fb{E}<\fb{E}_1\\
\\
\displaystyle
\frac{3}{2}K_D\ \e{\fb{E}/2} & \textrm{for } \fb{E}>\fb{E}_1
\end{array}\right.\nonumber
\end{eqnarray}

\begin{figure}
\includegraphics[width=\columnwidth]{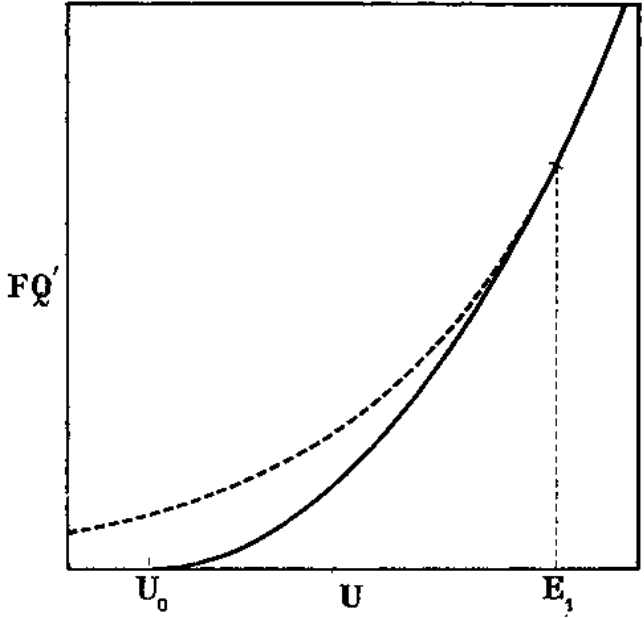}
\caption{Homologous model (dashed line) and simplified model, with finite central density (solid line).}
\label{fig:11}
\end{figure}

We now suppose that the formula (7.1) is valid not only initially, but always, although the central potential $\fb{U}_0$ is a function of the time. Thus the evolution of the cluster consists in two phenomena that overlap: the normal homologous evolution, and the variation of the structure of a small central region. We will see that the latter can quite easily be computed when considering the fluxes of mass and energy toward the center of the cluster.

The partial mass of the stars whose energy is less than $\fb{E}_1$ is, for the model considered here (see Equation 2.42)
\begin{equation}
\fb{M} = \int_{\fb{U}_0}^{\fb{E}_1} \fb{F}\fb{Q}'\, \dd\fb{E} = 2 K_D\ \e{\fb{E}_1/2}.
\end{equation}
For the homologous model, this mass would be, from (4.10b)
\begin{equation}
\fb{M}_0 = 3K_D\ \e{\fb{E}_1/2}.
\end{equation}

Furthermore, the mass of the stars whose energy is greater than $\fb{E}_1$ is the same for both models. The difference of mass between the present model and the homologous one is therefore
\begin{equation}
\Delta\fb{M} = \fb{M} - \fb{M}_0 = -K_D\ \e{\fb{E}_1/2}.
\end{equation}
In the same way, we compute the total energy of the stars whose \tn{individual} energy is less than $\fb{E}_1$:
\begin{eqnarray}
\fb{H} &=& \int_{\fb{U}_0}^{\fb{E}_1} \fb{E}\fb{F}\fb{Q}' \, \dd\fb{E} = (6 + 2\fb{U}_0) K_D\ \e{\fb{E}_1/2}\\
\fb{H}_0 &=& (6 + 3\fb{U}_0) K_D\ \e{\fb{E}_1/2},\nonumber
\end{eqnarray}
thus,
\begin{equation}
\Delta\fb{H} = \fb{H} - \fb{H}_0 = -K_D\ \fb{U}_0\ \e{\fb{E}_1/2}.
\end{equation}

The definition (7.1) of the model is partly arbitrary. One can redo the calculations above for other definitions (for example, when supposing that $\fb{Q}'$ rather than $\fb{F}\fb{Q}'$ is represented by a parabola near the center); we note that the results are always
\begin{eqnarray}
\Delta\fb{M} &=& k_1\ \e{\fb{U}_0/2}\\
\Delta\fb{H} &=& (k_1 \fb{U}_0 + k_2) \e{\fb{U}_0/2},\nonumber
\end{eqnarray}
where $k_1$ and $k_2$ are two numerical constants, the values of which are slightly different from one model to the other. In the present case, we have
\begin{eqnarray}
k_1 &=& - K_D\ \e{2} = -3.387\\
k_2 &=& 0.\nonumber
\end{eqnarray}

By computing the derivative of (7.7) with respect to time, we obtain the fluxes of mass and energy toward the center:
\begin{eqnarray}
\frac{\partial \fb{M}}{\partial \fb{T}} &=& \frac{k_1}{2} \e{\fb{U}_0/2}\ \frac{\dd \fb{U}_0}{\dd\fb{T}}\\
\frac{\partial \fb{H}}{\partial \fb{T}} &=& \frac{1}{2} (k_1 \fb{U}_0 + 2k_1 + k_2) \e{\fb{U}_0/2}\ \frac{\dd\fb{U}_0}{\dd\fb{T}}.\nonumber
\end{eqnarray}

These fluxes are related, through the equations (4.31), to the factors $\alpha_1$ and $\alpha_2$ that appear in the expansion (4.14) of $\fb{F}$. It is more handy to write this expansion as
\begin{equation}
\fb{F} = \e{-\fb{E}} + (K+K_2 \fb{E})\ \e{-\fb{E}/2},
\end{equation}
by setting
\begin{eqnarray}
\frac{2}{3K_D}(5\alpha_2-\alpha_1) &=& K\\
-\frac{2}{3K_D} \alpha_2 &=& K_2.\nonumber
\end{eqnarray}
Then, we obtain the relations:
\begin{eqnarray}
K &=& \frac{k_1 \fb{U}_0 + 7 k_1 + k_2}{3K_D} \e{\fb{U}_0/2}\ \frac{\dd\fb{U}_0}{\dd\fb{T}}\\
K_2 &=& -\frac{k_1}{3K_D} \e{\fb{U}_0/2}\ \frac{\dd\fb{U}_0}{\dd\fb{T}}\nonumber
\end{eqnarray}

Furthermore, we have seen in Chapter~V that the boundary conditions require two relations for the parameters of the model. In the case of the homologous model, $K_2$ is zero, and the other two parameters $K$ and $c$ are determined uniquely. Here, there are three parameters: $K$, $K_2$ and $c$. Therefore, the boundary conditions require, after canceling of $c$, a relation between $K$ and $K_2$. As the models considered here are close to the homologous one, $K_2$ is slightly different from zero, and we can assume that the relation is linear, i.e.
\begin{equation}
K = K_0 + d\ K_2,
\end{equation}
where $K_0$ is the value taken by $K$ in the homologous model, and $d$ is a constant.

To find this constant, two models have been computed with a non-zero $K_2$. The procedure is as in Chapter~V, but the initial condition (5.3c) has to be replaced with (7.10); the other initial expansions have, of course, to be modified accordingly. While doing this, we assume that the first order expansions (4.7) are still valid; indeed these new models differ significantly from the homologous model only in the small central region, i.e. for very small values of $\fb{E}$ or $\fb{U}$. Near the boundary, the modification only affects the second order terms in the expansions, as visible for example in (7.10).

The results about the relation between $K$ and $K_2$ are:
\begin{center}
\begin{tabular}{r@{}lr@{}l}
\multicolumn{2}{c}{$K_2$} & \multicolumn{2}{c}{$K$} \\
\hline
\hline
0&& -0.&938464 \\
-0&.01& -0.&931757 \\
-0&.1& -0.&87028 \\
\hline
\end{tabular}
\end{center}

($K_2 = 0$ corresponds to the homologous model; the value obtained here is slightly different from that of Chapter~V, Equation 5.4a, because of a larger integration step). We verify that these figures match well a linear relation of the type (7.13) with
\begin{equation}
d = -0.6707.
\end{equation}

When combining the relations (7.12) and (7.13), we get
\begin{eqnarray}
\frac{\dd\fb{U}_0}{\dd\fb{T}} &=& \frac{3K_D\ K_0}{k_1}\ \frac{\e{-\fb{U}_0/2}}{\fb{U}_0 + 7 + d + k_2/k_1}\\
&=& \frac{0.3816\ \e{-\fb{U}_0/2}}{\fb{U}_0 + 6.329}.\nonumber
\end{eqnarray}

This differential equation allows to compute the variation of the central potential $\fb{U_0}$ as a function of time, and thus to solve our problem. Its explicit solution is:
\begin{equation}
\fb{T}-\fb{T}_2 = 5.241 (\fb{U}_0 + 4.329)\ \e{\fb{U}_0/2},
\end{equation}
where $\fb{T}_2$ is a constant. This relation is plotted in \reff{12}. We first note that the central potential $\fb{U}_0$ decreases with time, which corresponds to an increase of the central density, and to getting closer to the homologous model. Furthermore, this decrease gets faster and faster, so that the central potential becomes $-\infty$ after a finite timelapse: the cluster reaches the structure of the homologous model at a given time $\fb{T} = \fb{T}_2$, and keeps it afterwards.

\begin{figure}
\includegraphics[width=\columnwidth]{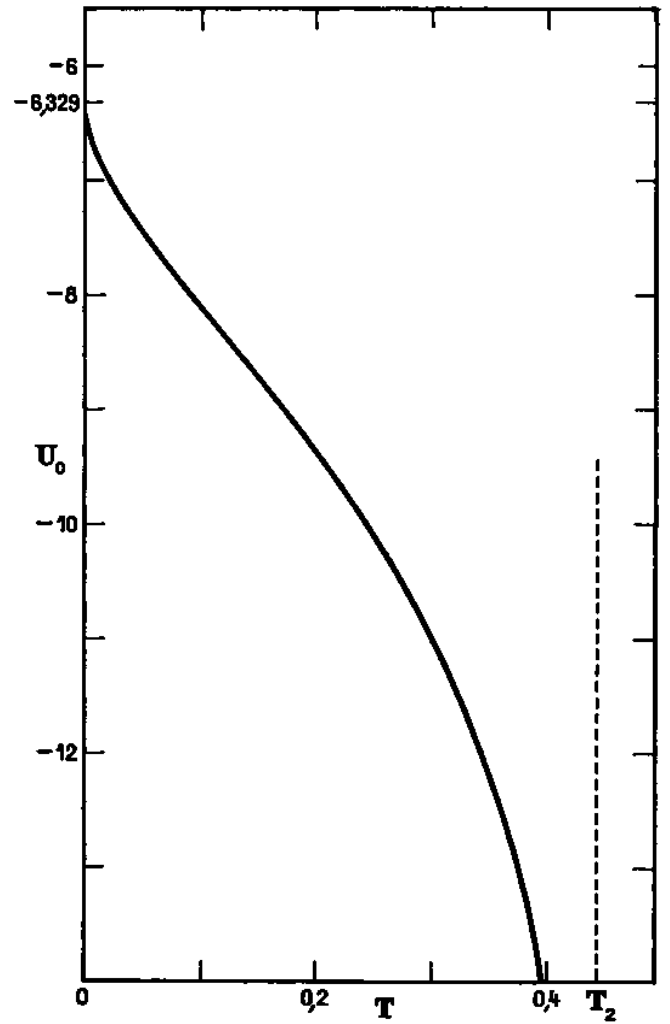}
\caption{Evolution of the central potential.}
\label{fig:12}
\end{figure}

Substituting (7.15) into (7.9), we obtain the explicit expressions of the fluxes near the center:
\begin{eqnarray}
\frac{\partial \fb{M}}{\partial \fb{T}} &=& -\frac{0.6464}{\fb{U}_0 + 6.329}\\
\frac{\partial \fb{H}}{\partial \fb{T}} &=& -\frac{0.6464\ (\fb{U}_0 + 2)}{\fb{U}_0 + 6.329}\nonumber
\end{eqnarray}

The flux of mass is first positive and then decreases; it becomes zero at the time $\fb{T}_2$ and remains null afterwards. The flux of energy is always negative; its absolute value is first decreasing; at $\fb{T}_2$, it reaches the value of -0.6464, that its keeps afterwards (see 5.22a). These results confirm the reasoning made in Chapter~IV.

At the beginning of its life, the cluster is likely not very concentrated; let's assume that $\fb{U}_0$ has initially the largest value allowed by (7.15), i.e. $\fb{U}_0 = -6.329$. By assuming that $\fb{T}=0$ at the initial time, we find
\begin{equation}
\fb{T}_2= 0.4427.
\end{equation}
$\fb{T}_2$ is, in canonical variables, the time needed to form the central singularity.

\subsection{Proper differences}

We now suppose that the central density has already become infinite, but that the cluster as a whole has not completely reached the final state represented by the homologous model.

Let $\fb{F}_0$ be the distribution function of the homologous model; the distribution function of the cluster considered here would be
\begin{equation}
\fb{F} = \fb{F}_0 + \Delta\fb{F},
\end{equation}
where $\Delta\fb{F}$ is small with respect to $\fb{F}_0$. $\Delta\fb{F}$ is time-dependent. In the same way, we set
\begin{equation}
\fb{Q} = \fb{Q}_0 + \Delta\fb{Q}, \textrm{ etc.}
\end{equation}

The function $\fb{F}$ is assumed to be in the usual canonical form, defined by (4.6). The parameters $K$, $b$, $c$, would all have slightly different values than those of the homologous model:
\begin{equation}
K = K_0 + \Delta K, \textrm{ etc.}
\end{equation}

The cluster obeys the relations (5.2) and (5.3), as soon as the terms including time derivatives, which were zero for the homologous model (see Equations 2.37e, 2.53, 3.12 and 3.13), are put back in (5.2e) and (5.2f). These equations become
\begin{eqnarray}
0 &=& S' + \left(\frac{3}{4}b-\frac{3}{2}c\right) \fb{F}\fb{Q} + \left(\frac{1}{2}c-\frac{3}{4}b\right)\\
&&\int_{-\infty}^{\fb{E}}\fb{F}_1\fb{Q}_1'\, \dd\fb{E}_1 + \int_{-\infty}^{\fb{E}} \left( \fb{F}_1'\frac{\partial \fb{Q}_1}{\partial \fb{T}} - \fb{Q}_1'\frac{\partial \fb{F}_1}{\partial \fb{T}}\right)\, \dd\fb{E}_1,\nonumber\\
0 &=& (3\lambda + 1) b + (2-2\lambda)c\nonumber\\
&&+4\lambda\frac{\dd \ln{(\fb{M}_e)}}{\dd\fb{T}}-4\frac{\dd \ln{(\fb{R}_e)}}{\dd\fb{T}}.\nonumber
\end{eqnarray}

By substituting the expressions (7.19) to (7.21) into the previous equations, and neglecting the second order terms, we obtain a set of linear equations that have the dimensions of $\Delta\fb{F}$, $\Delta\fb{Q}$, etc., and their derivatives with respect to time. We know that the general solution of such a system is (unless a degeneracy exists) a linear combination of particular solutions of the form
\begin{eqnarray}
\Delta\fb{F} &=& \Delta\fb{F}_0\ \e{-s \fb{T}},\\
\Delta\fb{Q} &=& \Delta\fb{Q}_0\ \e{-s \fb{T}}, \textrm{ etc},\nonumber
\end{eqnarray}
where $\Delta\fb{F}_0$, $\Delta\fb{Q}_0$, ... do not depend on the time, and $s$ is a constant, real or complex. As in the classical terminology, we shall call such a solution \emph{proper difference}, and $s$,  the associated \emph{proper value} \tn{eigenvalue}. By substituting (7.23) into the equations, and dividing by $\e{-s\fb{T}}$, we obtain a system that is independent of time, but which involves a new parameter: $s$.

\subsubsection{Method of solution}

In practice, instead of writing and solving the system of equations of $\Delta\fb{F}$ and $\Delta\fb{Q}$, etc ..., it is easier to keep the equations of $\fb{F}$, $\fb{Q}$, etc ..., to compute a slightly different model from the homologous one by means of these equations, and to obtain the differences by simply computing the differences between the values of the two models. The differences should not be too large (so that the second order terms are indeed negligible), nor too small (so that one can get them with a sufficient accuracy); experiment led us to fix the amplitude of these differences by setting
\begin{equation}
\Delta K = -0.001.
\end{equation}

Taking into account that the device we use computes with 8 significant digits, we can obtain the differences with a precision of the order of 1/1000.

Hence, the system to solve is as (5.2) and (5.3), as soon as (5.2e) and (5.2f) are replaced with
\begin{eqnarray}
0 &=& \Bigg\{ \fb{F} \int_{-\infty}^{\fb{E}}\fb{F}_1\fb{Q}_1'\, \dd\fb{E}_1\\
&& + \fb{F}'\left(\int_{-\infty}^{\fb{E}}\fb{F}_1\fb{Q}_1\, \dd\fb{E}_1 - \fb{Q}\fb{F}^{(-1)}\right) \nonumber\\
&& + \left(\frac{3}{4}b-\frac{3}{2}c\right)\fb{F}\fb{Q}\nonumber\\
&& +\left(\frac{1}{2}c-\frac{3}{4}b\right) \int_{-\infty}^{\fb{E}}\fb{F}_1\fb{Q}_1'\, \dd\fb{E}_1 \nonumber\\
&& + s\int_{-\infty}^{\fb{E}} [ \fb{Q}_1' (\fb{F}_1-\fb{F}_0) - \fb{F}_1' (\fb{Q}_1 - \fb{Q}_0)]\, \dd\fb{E}_1 \Bigg\}\nonumber\\
0 &=& (3\lambda + 1)b + (2-2\lambda)c \\
&& + 4 s\left( \frac{\fb{R}_e-\fb{R}_{e_0}}{\fb{R}_{e_0}} - \lambda \frac{\fb{M}_e-\fb{M}_{e_0}}{\fb{M}_{e_0}}\right)\nonumber
\end{eqnarray}
and when (7.24) is added.

The method of solution is an extension of the one used for the homologous model (Chapter~V):
\begin{enumerate}
\item choose a value for $s$;
\item choose a temporary form for the function $\fb{F}$ that satisfies (5.3a) and (5.3c), with $K$ set by (7.21) and (7.24);
\item compute $\fb{D}$, $\fb{Z}$, $\fb{R}$, $\fb{Q}$;
\item adopt temporary value for $b$ and $c$, and integrate (7.25);
\item re-do after changing $b$ and $c$, until the final conditions (5.3a) and (5.3b) are fulfilled;
\item go back to point 3 with the new function $\fb{F}$;
\item when two consecutive approximations of $\fb{F}$ are equal: compute the right-hand side term of (7.26). It is not zero in general. Modify the value of $s$, go back to point 2; grope around this way with $s$ until (7.26) is true.
\end{enumerate}

The computation is quite long, because of the trial and error required on three parameters: $b$, $c$, $s$; that is why we limited ourselves to the exact computation of the proper differences corresponding to the two smallest proper values $s$. (We will see later that the possible values of $s$ are real, positive and form a discrete series.) These differences are the most interesting in practice, because they are those which decay the slowest. The other differences will be treated in a more approximate way later in this Chapter.

\subsubsection{Results: first proper difference}

We find that the smallest proper value is:
\begin{equation}
s = 1.81.
\end{equation}

The differences $\Delta\fb{F}$, $\Delta\fb{D}$, $\Delta\fb{R}$, $\Delta\fb{Q}$, $\Delta\fb{D}_P$ of the four fundamental functions and the projected density are given in \reft{4}. The differences of the parameters and the characteristic quantities are:
\begin{eqnarray}
\frac{\Delta b}{\Delta K} = +1.30 \qquad  \frac{\Delta c}{\Delta K} = +5.22,\qquad\\
\frac{\Delta \fb{R}_e}{\Delta K} = -6.62 \quad \frac{\Delta \fb{M}_e}{\Delta K} = -0.537 \quad \frac{\Delta \fb{L}_e}{\Delta K} = +0.065.\nonumber
\end{eqnarray}

\begin{table2}
\caption{}
\label{tab:4}
\begin{tabular}{r@{}lr@{}lr@{}lr@{}lr@{}lr@{}l}
\multicolumn{2}{c}{$\fb{E}$ or $\fb{U}$} & \multicolumn{2}{c}{$\frac{\Delta \fb{F}}{\Delta K}$} & \multicolumn{2}{c}{$\frac{\Delta \fb{D}}{\Delta K}$} & \multicolumn{2}{c}{$-\frac{\Delta \fb{R}}{\Delta K}$} & \multicolumn{2}{c}{$-\frac{\Delta \fb{Q}}{\Delta K}$} & \multicolumn{2}{c}{$\frac{\Delta \fb{D_P}}{\Delta K}$} \vspace{0.1cm}\\
\hline
\hline
-5& & 12&.18 & 33&.4 & 0&.00673 & 0&.0000287 & 8&.84\\
-4&.9 & 11&.62 & 31&.2 & &\phantom{.00}748 & &\phantom{.0000}361 & &\\
-4&.8 & 11&.08 & 29&.4 & &\phantom{.00}831 & &\phantom{.0000}448 & 8&.42\\
-4&.7 & 10&.55 & 27&.6 & &\phantom{.00}923 & &\phantom{.0000}565 & &\\
-4&.6 & 10&.04 & 25&.8 & 0&.0102 & &\phantom{.0000}703 & 7&.98\\
-4&.5 & 9&.56 & 24&.2 & &\phantom{.0}114 & &\phantom{.0000}885 & &\\
-4&.4 & 9&.10 & 22&.6 & &\phantom{.0}126 & 0&.000110 & 7&.54\\
-4&.3 & 8&.66 & 21&.1 & &\phantom{.0}140 & &\phantom{.000}138 & &\\
-4&.2 & 8&.24 & 19&.7 & &\phantom{.0}155 & &\phantom{.000}171 & 7&.10\\
-4&.1 & 7&.84 & 18&.4 & &\phantom{.0}172 & &\phantom{.000}213 & &\\
-4& & 7&.45 & 17&.2 & &\phantom{.0}191 & &\phantom{.000}265 & 6&.66\\
-3&.9 & 7&.08 & 16&.0 & &\phantom{.0}212 & &\phantom{.000}329 & &\\
-3&.8 & 6&.73 & 14&.8 & &\phantom{.0}235 & &\phantom{.000}408 & 6&.21\\
-3&.7 & 6&.40 & 13&.8 & &\phantom{.0}261 & &\phantom{.000}508 & &\\
-3&.6 & 6&.07 & 12&.8 & &\phantom{.0}290 & &\phantom{.000}629 & 5&.77\\
-3&.5 & 5&.77 & 11&.9 & &\phantom{.0}323 & &\phantom{.000}782 & &\\
-3&.4 & 5&.47 & 11&.0 & &\phantom{.0}359 & &\phantom{.000}970 & 5&.33\\
-3&.3 & 5&.20 & 10&.1 & &\phantom{.0}399 & 0&.00121 & &\\
-3&.2 & 4&.93 & 9&.33 & &\phantom{.0}445 & &\phantom{.00}150 & 4&.89\\
-3&.1 & 4&.68 & 8&.59 & &\phantom{.0}495 & &\phantom{.00}186 & &\\
-3& & 4&.44 & 7&.89 & &\phantom{.0}552 & &\phantom{.00}232 & 4&.45\\
-2&.9 & 4&.21 & 7&.23 & &\phantom{.0}616 & &\phantom{.00}289 & &\\
-2&.8 & 3&.99 & 6&.61 & &\phantom{.0}688 & &\phantom{.00}360 & 4&.02\\
-2&.7 & 3&.78 & 6&.03 & &\phantom{.0}769 & &\phantom{.00}449 & &\\
-2&.6 & 3&.58 & 5&.49 & &\phantom{.0}861 & &\phantom{.00}562 & 3&.59\\
-2&.5 & 3&.39 & 4&.98 & &\phantom{.0}965 & &\phantom{.00}703 & &\\
-2&.4 & 3&.21 & 4&.50 & 0&.108 & &\phantom{.00}882 & 3&.18\\
-2&.3 & 3&.04 & 4&.06 & &122 & 0&.0111 & &\\
-2&.2 & 2&.87 & 3&.64 & &137 & &\phantom{.0}139 & 2&.77\\
-2&.1 & 2&.71 & 3&.25 & &154 & &\phantom{.0}176 & &\\
-2& & 2&.56 & 2&.90 & &174 & &\phantom{.0}223 & 2&.37\\
-1&.9 & 2&.42 & 2&.56 & &197 & &\phantom{.0}282 & &\\
-1&.8 & 2&.28 & 2&.25 & &223 & &\phantom{.0}359 & 1&.98\\
-1&.7 & 2&.15 & 1&.97 & &253 & &\phantom{.0}458 & &\\
-1&.6 & 2&.02 & 1&.71 & &288 & &\phantom{.0}587 & 1&.62\\
-1&.5 & 1&.90 & 1&.47 & &329 & &\phantom{.0}755 & &\\
-1&.4 & 1&.78 & 1&.25 & &377 & &\phantom{.0}975 & 1&.27\\
-1&.3 & 1&.66 & 1&.06 & &433 & 0&.127 & &\\
-1&.2 & 1&.55 & 0&.877 & &500 & &165 & 0&.95\\
-1&.1 & 1&.44 & 0&.718 & &580 & &217 & &\\
-1& & 1&.33 & 0&.576 & &675 & &287 & 0&.665\\
-0&.9 & 1&.22 & 0&.452 & &791 & &382 & &\\
-0&.8 & 1&.11 & 0&.345 & &933 & &514 & 0&.418\\
-0&.7 & 1&.00 & 0&.253 & 1&.11 & &700 & &\\
-0&.6 & 0&.886 & 0&.177 & 1&.34 & &965 & 0&.220\\
-0&.5 & 0&.769 & 0&.116 & 1&.63 & 1&.35 & &\\
-0&.4 & 0&.646 & 0&.0684 & 2&.01 & 1&.93 & 0&.0816\\
-0&.3 & 0&.512 & 0&.0346 & 2&.55 & 2&.83 & &\\
-0&.2 & 0&.364 & 0&.0130 & 3&.32 & 4&.28 & 0&.0111\\
-0&.1 & 0&.195 & 0&.0024 & 4&.53 & 6&.80 & &\\
0& & 0 && 0 && 6&.62 & 11&.61 & 0&\\
\hline
\end{tabular}
\end{table2}

\reff{13} compares the projected densities of the usual homologous model ($\Delta K=0$) with those of the homologous model modified by the first proper difference; here we have set $\Delta K=0.2$ so that the difference is well-visible. (As for its amplitude, the sign of $\Delta K$ is arbitrary; we could have chosen a negative $\Delta K$; in this case, the difference with the homologous model would have been in the opposite sense.) We see that the difference mostly affects the external regions of the cluster: \emph{the first proper difference mainly consists in a variation of the external radius of the cluster}. The structure of the internal region differs little; the total mass and kinetic energy vary much less than the radius, as shown by the values (7.28).

\begin{figure}
\includegraphics[width=\columnwidth]{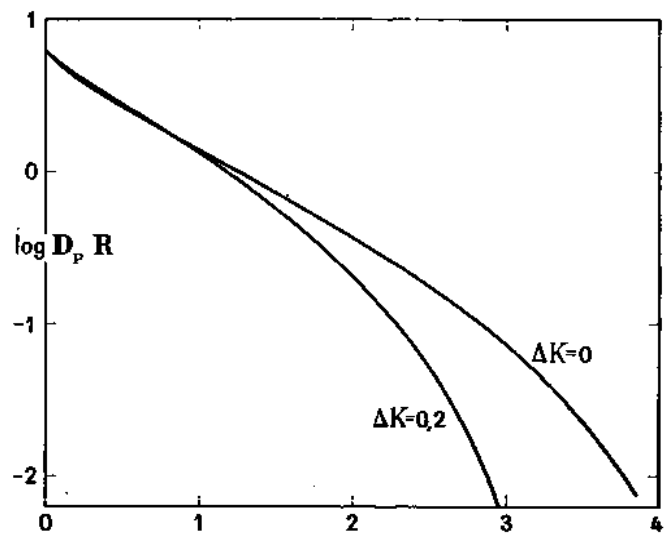}
\caption{Usual homologous model ($\Delta K=0$) and homologous model modified by the first proper difference ($\Delta K=0.2$).}
\label{fig:13}
\end{figure}

This is naturally explained by the fact that the perturbations between the stars are more efficient in the center of the cluster, where the density is higher; as a consequence this region becomes close to the final state, represented by the homologous model, earlier.

\subsubsection{Results: second proper difference}

The second proper value is:
\begin{equation}
s = 5.12.
\end{equation}

We only give the differences of the parameters:
\begin{eqnarray}
\frac{\Delta b}{\Delta K} = +1.71 \qquad \frac{\Delta c}{\Delta K} = +6.90,\qquad\\
\frac{\Delta \fb{R}_e}{\Delta K} = -3.48 \quad \frac{\Delta \fb{M}_e}{\Delta K} = -0.760 \quad \frac{\Delta \fb{L}_e}{\Delta K} = +0.182.\nonumber
\end{eqnarray}

The profile of the projected density (not shown here) shows that the second proper difference, like the first one, mostly consists in a variation of the external radius.

\subsection{Stability of the homologous model}

Up to now, we have implicitly assumed that the clusters naturally tend toward the homologous model. For this to be true, the model has to be stable, i.e. any difference with respect to this model is decreasing. On the other hand, the stability of the homologous model, if demonstrated, will be, if not a rigorous proof, at least a very strong clue that the model is indeed the final state toward which all clusters tend.

For the homologous model to be stable, it is necessary and sufficient that all the proper values $s$ yield a positive real part. Thus, we are going to study the complete family of the proper values $s$, thanks to a very approximate calculation, yet sufficient for the goal we seek.

We first assume that the central expansion (4.34) of $\fb{F}$ is valid up to $\fb{E}=-1$. (This approximation, as all which will follow, has been suggested and checked by the exact computation of the first two proper functions.) Hence, for $\fb{E}=-1$, we have:
\begin{eqnarray}
\frac{\Delta\fb{F}}{\Delta K}(-1) &=& \e{1/2},\\
\frac{\Delta\fb{F}'}{\Delta K}(-1) &=& -\frac{1}{2}\e{1/2}.\nonumber
\end{eqnarray}

Between $\fb{E}=-1$ and $\fb{E}=0$, we consider the exact equation (7.25). By calculating the derivative of this equation, neglecting the terms in $\fb{F}^{(-1)}$ and $\fb{F}^2$ (because $\fb{F}$ vanishes at the boundary), replacing (7.19) etc ..., and subtracting the equation of the homologous model, we obtain
\begin{eqnarray}
0 &=& \Bigg\{ \int_{-\infty}^{\fb{E}}\fb{F}_1\fb{Q}_1\, \dd\fb{E}_1 \cdot \Delta\fb{F}'' \\
&& + \fb{F}''\Delta\left(\int_{-\infty}^{\fb{E}}\fb{F}_1\fb{Q}_1\, \dd\fb{E}_1\right) + \int_{-\infty}^{\fb{E}}\fb{F}_1\fb{Q}_1'\, \dd\fb{E}_1 \cdot \Delta\fb{F}'\nonumber\\
&& + \fb{F}' \Delta\left(\int_{-\infty}^{\fb{E}}\fb{F}_1\fb{Q}_1'\, \dd\fb{E}_1\right)\nonumber\\
&& + \fb{F}'\fb{Q}\Delta\left(\frac{3}{4}b-\frac{3}{2}c\right) + \left(\frac{3}{4}b-\frac{3}{2}c\right) \fb{Q}\Delta\fb{F}' \nonumber \\
&& + \left(\frac{3}{4}b-\frac{3}{2}c\right) \fb{F}'\Delta\fb{Q} - \fb{F}\fb{Q}'\Delta c - c\fb{Q}'\Delta\fb{F} \nonumber\\
&&  - c\fb{F}\Delta\fb{Q}' + s\fb{Q}'\Delta\fb{F} - s\fb{F}'\Delta\fb{Q}\Bigg\}\nonumber
\end{eqnarray}
\tn{The upper limit of the first integral is missing in the original version}

Furthermore (7.26) becomes, when neglecting the term in $\Delta\fb{M}_e$:

\begin{equation}
0 = (3\lambda + 1)\Delta b + (2-2\lambda)\Delta c + 4s \frac{\Delta \fb{R}_e}{\fb{R}_e}.
\end{equation}

The expansions (4.35) show that we have, near the center,
\begin{eqnarray}
\frac{\Delta\fb{R}}{\fb{R}} &=& -\frac{1}{\sqrt{2}}\e{\fb{U}/2}\ \Delta K,\\
\frac{\Delta\fb{Q}}{\fb{Q}} &=& -\frac{9\sqrt{3}}{8\sqrt{2}}\e{\fb{E}/2}\ \Delta K,\nonumber\\
\frac{\Delta\fb{Q}'}{\fb{Q}'} &=& -\frac{3\sqrt{3}}{2\sqrt{2}}\e{\fb{E}/2}\ \Delta K.\nonumber
\end{eqnarray}

We shall suppose that these expressions are valid up to the boundary. In (7.32), we can neglect the second and the fourth terms, because the integrals do not change much. We also neglect the terms in $\Delta b$, as they are small with respect to $\Delta c$. By taking $\Delta c$ from (7.33) and substituting it in (7.32), we obtain
\begin{eqnarray}
0 &=& \Bigg\{ \int_{-\infty}^{\fb{E}}\fb{F}_1\fb{Q}_1\, \dd\fb{E}_1 \cdot \Delta\fb{F}'' \\
&& + \left( \int_{-\infty}^{\fb{E}}\fb{F}_1\fb{Q}_1'\, \dd\fb{E}_1 -\frac{3\lambda+3}{3\lambda+1} c\ \fb{Q}\right) \Delta\fb{F}' \nonumber\\
&& + (s-c)\fb{Q} \Delta\fb{F} + (s-c) \bigg[\frac{9\sqrt{3}}{8\sqrt{2}}\ \e{\fb{E}/2}\ \fb{F}'\fb{Q} \nonumber\\
&& - \frac{\sqrt{2}}{1-\lambda}\left(\fb{F}\fb{Q}'+\frac{3}{2}\fb{F}'\fb{Q}\right)\bigg]\Delta K\nonumber\\
&& + c \bigg[\frac{6\lambda+4}{3\lambda+1}\frac{9\sqrt{3}}{8\sqrt{2}}\ \e{\fb{E}/2}\ \fb{F}'\fb{Q} + \frac{3\sqrt{3}}{2\sqrt{2}}\ \e{\fb{E}/2}\ \fb{F}\fb{Q}'\nonumber\\
&& - \frac{\sqrt{2}}{1-\lambda}\left(\fb{F}\fb{Q}'+\frac{3}{2}\fb{F}'\fb{Q}\right)\bigg]\Delta K\Bigg\}\nonumber
\end{eqnarray}

Finally, we neglect the term in $\Delta\fb{F}'$, whose factor is relatively small, and the term in $c$. The equation shrinks to:
\begin{equation}
0 = \Delta\fb{F}'' + (s-c) B_1\ \Delta F + (s-c) B_2\ \Delta K,
\end{equation}
where $B_1$ and $B_2$ are two functions of $\fb{E}$. This is a second order differential equation for $\Delta\fb{F}$; it must fulfill the boundary conditions (7.31), as well as the condition
\begin{equation}
\Delta\fb{F}(0) = 0.
\end{equation}

These three conditions would be simultaneously fulfilled only for certain values of $s$, which are the proper values we seek.

The numerical computation of the functions $B_1$ and $B_2$ shows that they can be quite well fitted, in the range (-1,0), with the expressions
\begin{eqnarray}
B_1 & \simeq & \frac{2.10}{(0.332-\fb{E})^2},\\
B_2 & \simeq & (0.265 + 0.643\ \fb{E})\ B_1.\nonumber
\end{eqnarray}

By replacing this in (7.36), we find that the general solution of this equation is
\begin{eqnarray}
\frac{\Delta\fb{F}}{\Delta K} &=& -0.265 - 0.643\ \fb{E}\\
&& + C_1 (0.332 - \fb{E})^{p_1} + C_2 (0.332 - \fb{E})^{p_2},\nonumber
\end{eqnarray}
where $C_1$ and $C_2$ are two arbitrary constants, and $p_1$, $p_2$ are the roots of the equation:
\begin{equation}
p^2 - p + 2.10 (s-c) =0,
\end{equation}
so that we have
\begin{eqnarray}
p_1+p_2 &=& 1,\\
p_1\ p_2 &=& 2.10 (s-c).\nonumber
\end{eqnarray}

By writing the boundary conditions (7.31) and (7.37) for the formula (7.39), and then eliminating $C_1$ and $C_2$, we obtain
\begin{equation}
\left|\begin{array}{rrr}
(1.332)^{p_1} & (1.332)^{p_2} & 1.271 \\
p_1 (1.332)^{p_1} & p_2 (1.332)^{p_2} & -0.181 \\
(0.332)^{p_1} & (0.332)^{p_2} & 0.265
\end{array}\right|=0.
\end{equation}

The equations (7.41a) and (7.42) make a system of equations for $p_1$ and $p_2$ that can be solved numerically. We do not give the details of this solution, which is long but without difficulty; we find that $p_1$ and $p_2$ are necessarily of the form
\begin{eqnarray}
p_1 = \frac{1}{2}+\xi i,\\
p_1 = \frac{1}{2}-\xi i,\nonumber
\end{eqnarray}
where $\xi$ is real, and $i = \sqrt{-1}$. The possible values of $\xi$ constitute an infinite series; the first ones are given in the table below, as well as the corresponding values of $s$, found from (7.41b).

\begin{center}
\begin{tabular}{r@{}lr@{}l}
\multicolumn{2}{c}{$\xi$} & \multicolumn{2}{c}{$s$} \\
\hline
\hline
1&.21 & 1&.22\\
3&.82 & 7&.47\\
5&.43 & 14&.6\\
8&.28 & 33&.2\\
9&.91 & 47&.3\\
12&.8 & 78&.3\\
14&.4 & 99&.6\\
17&.3 & 143&\\
18&.9 & 171&\\
21&.8 & 227&\\
\hline
\end{tabular}
\end{center}

That is, all the proper values $s$ are real and positive; therefore, \emph{the homologous model is stable}, and it is likely the final state toward which all the clusters tend.

The fact that $s$ never yields an imaginary part shows that the clusters tend toward the homologous model through a simple ``relaxation'', with no oscillations.

The first two values of $s$ are in rough agreement with the exact values (7.27) and (7.29); the differences are not surprising, given the great number of approximations we made.

Note that the values of $s$ increase very fast which shows indeed that only the first values are interesting, in practice; the next ones corresponds to differences that vanish very quickly.

Finally, we note that these results allow us to define rigorously the ``relaxation time'' for the entire cluster: this would be the time required for the amplitude of the first proper difference to be divided by $e$.

\subsection{General picture of the evolution}

We have identified three distinct evolutionary phenomena:
\begin{enumerate}
\item the normal homologous evolution;
\item the formation of the central singularity;
\item the decrease of the differences.
\end{enumerate}
We are now going to compare their timescales.

The normal homologous evolution can be illustrated through the variation of the total mass given by (5.15) and plotted as a straight line on \reff{14}. The cluster disappears after a time $T_1$ given in (5.13):
\begin{equation}
T_1 = \frac{1}{\gamma_0 c} = \frac{2.452}{\gamma_0}.
\end{equation}

The formation of the central singularity can be seen as the variation of $\Delta M$, difference in mass with respect to the homologous model. This variation is given in (7.7a) and (7.16), in homologous variables; we switch to non-homologous variables by means of (5.12) and (5.15). The resulting curve is plotted in \reff{14} (for all the curves of this figure, the mass or the mass difference is normalized with respect to its own value at $T=0$).

\begin{figure}
\includegraphics[width=\columnwidth]{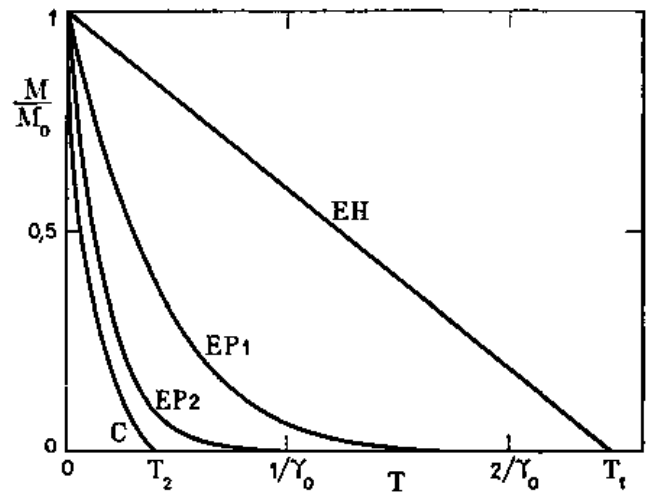}
\caption{Compared evolutions: EH = homologous evolution \tn{\emph{\'evolution homologique}}; EP1 = first proper difference; EP2 = second proper difference \tn{\emph{\'ecart propre}}; C = formation of the central singularity.}
\label{fig:14}
\end{figure}

We immediately see that the formation of the central singularity is relatively rapid: it ends at the time $T_2$, computed from (7.18) and (5.12):
\begin{equation}
T_2 = \frac{0.4051}{\gamma_0}
\end{equation}
and thus occurs for only the first sixth of the lifetime of the cluster.

Finally, the decrease of the proper differences will also be represented as variations of the mass differences. In homologous variables, we have, from (7.23),
\begin{equation}
\frac{\Delta \fb{M}_e}{\Delta \fb{M}_{e_0}} = \e{-s\fb{T}},
\end{equation}
and we switch back to non-homologous variables by means of (5.12) and (5.15), which gives
\begin{equation}
\frac{\Delta M_e}{\Delta M_{e_0}} = \left(1-\frac{T}{T_1}\right)^{1+s/c}.
\end{equation}
This variation is plotted in \reff{14}, for the first two proper differences. We see once more, that the decrease of the differences is rapid with respect to the homologous evolution.

In reality, the three evolutionary phenomena overlap. The initial amplitude of the differences with respect to the homologous model cannot be determined in the framework of this theory; it could only be found thanks to a detailed study of the formation and the initial phases of the evolution (see Chapter~IX). However, there is no reason for the initial state of the cluster to already be close to the homologous model, and thus the initial differences in the various quantities are likely of the same order of magnitude as the quantities themselves. Therefore, we can suppose that the curves of \reff{14}, normalized to unity for $T=0$, roughly represent the relative importance of the various effects.

The evolution of a cluster can be describe in a broad outline as follows: the central density first increases, faster and faster; simultaneously, the differences with respect to the homologous model decrease everywhere in the cluster. After a time $T_1/6$, the central density has became infinite; the remaining difference is practically limited to the first proper difference; the cluster has only lost 1/6 of its mass, through escapes. Then, the first proper difference continues to decrease; after a time $T_1/3$, it has almost vanished; the cluster takes the form of the homologous model and keeps it until it disappears, at the time $T_1$.

We will see in the next Chapter how these results can be used to estimate the age of globular clusters.

\section{Application to globular clusters}

We are going to compare the theoretical results from the previous Chapters to observational data of globular clusters. The first goal of this comparison is to check that the theory is compatible with the facts (but this would not be a proof of the correctness of the theory, for the reasons detailed in Chapter~I); the comparison will give us information that is out of reach of the direct observation, in particular about the mass-luminosity relation and the age of the clusters.

\subsection{Artificial cluster}

In order to get a concrete picture of the theoretical model, an ``artificial cluster'' has been built by calculating the spacial coordinates of the stars, from a table of random numbers, so that it reproduces the density law of the homologous model. More precisely, let $n_1$, $n_2$, $n_3$ be three random number, taken between 0 and 1, with an uniform probability density function; the spherical coordinates $\fb{R}$, $\theta$, $\varphi$ of the star are computed by means of
\begin{eqnarray}
\fb{M}_\fb{R} &=& n_1 \fb{M}_e,\\
\cos{\theta} &=& 2n_2 -1,\nonumber\\
\varphi &=& 2\pi n_3.\nonumber
\end{eqnarray}

\reff{15} plots the two-dimensional projection of the cluster obtained. The number of stars is 1320. The figure covers only a fraction of the surface of the cluster (see the scale at the bottom of the figure), but, however, contains all the stars; this is because the projected density is extremely small in the external regions.

\begin{figure2}
\includegraphics[width=16cm]{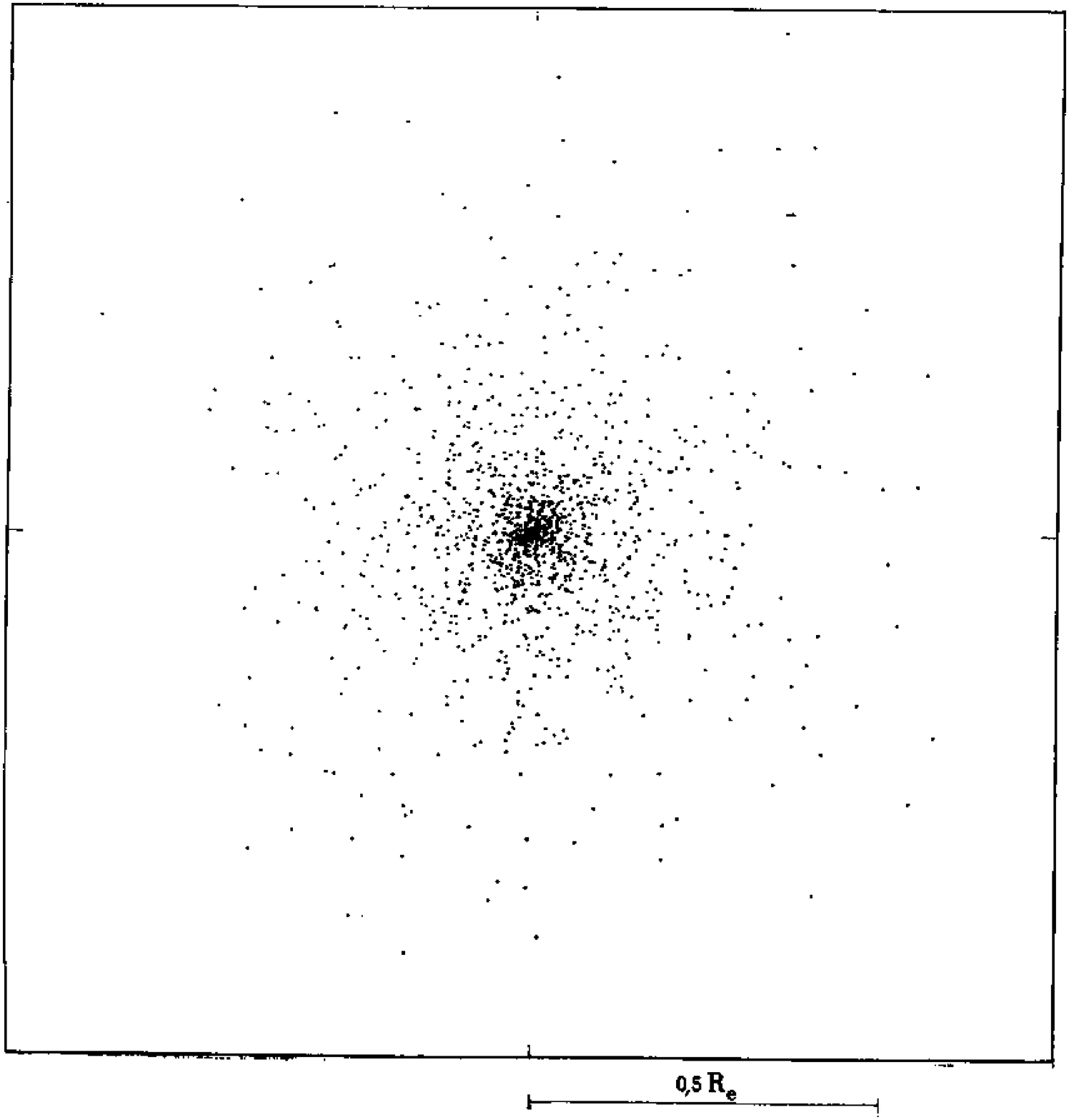}
\caption{Homologous model: artificial cluster.}
\label{fig:15}
\end{figure2}

\reff{16} is a zoom-in on the central region of the cluster.

\begin{figure2}
\includegraphics[width=16cm]{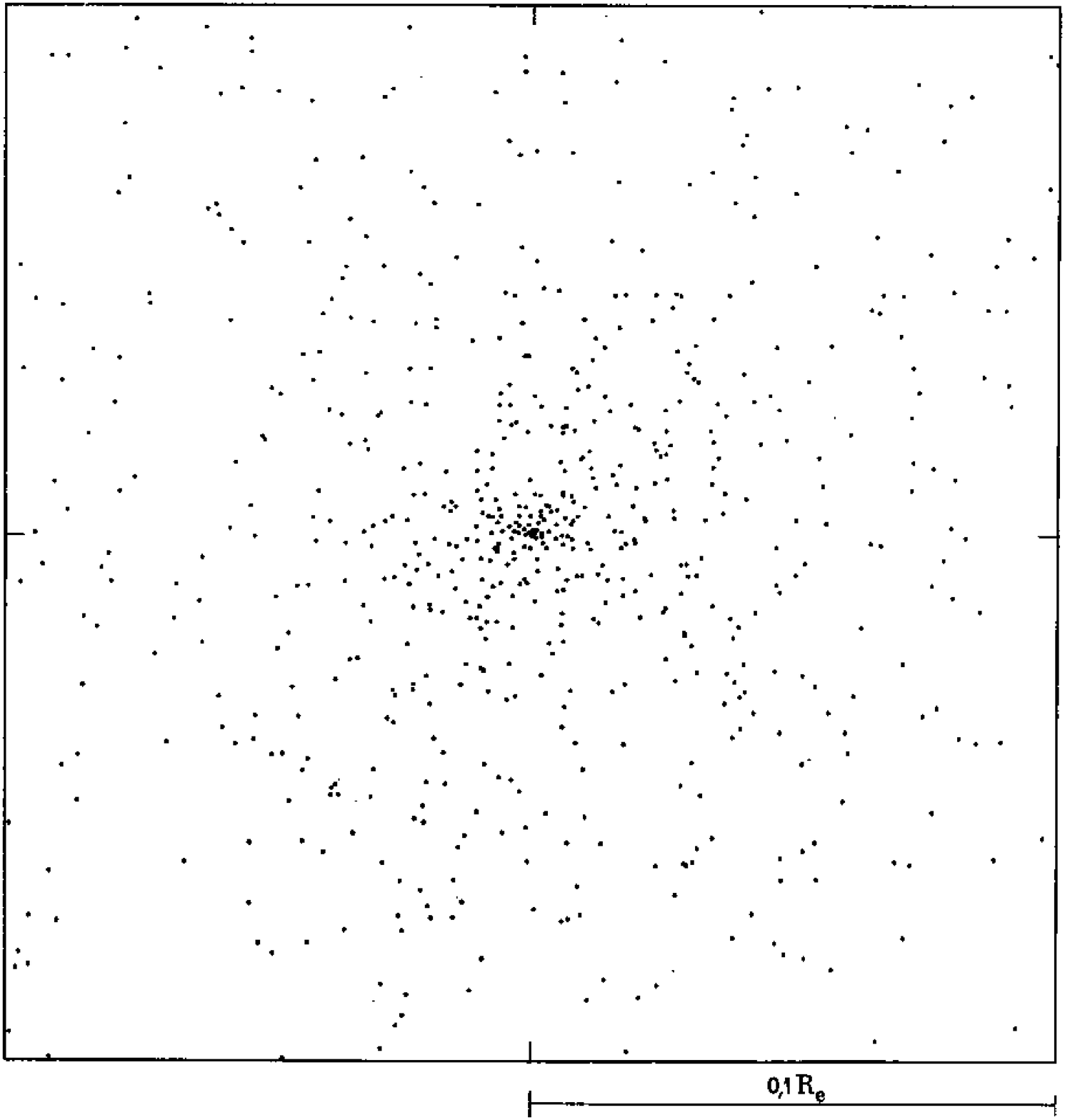}
\caption{Zoom-in on the center of \reff{15}.}
\label{fig:16}
\end{figure2}

Globally, this artificial cluster resembles well the real clusters; the similarity would not be perfect because the artificial cluster of \reff{15} corresponds to the case of equal masses, while the real clusters yields an extended mass spectrum, of which we only observed the upper end. The central condensation in \reff{16} is likely to be too high \citep{King1961a}.

\subsection{Comparisons of the projected densities}

A more precise comparison of the homologous model with real clusters can be obtained when plotting the density profiles. It seems better to do the comparison for the projected densities, and not the spacial densities. Indeed, the theoretical curves can be obtained with an arbitrary accuracy and going from the theoretical spacial density to the theoretical projected density is errorless; on the contrary, going from the observed projected density to the spacial density strongly amplifies the errors, that were already not negligible. This is shown in \reff{17} where the spacial density of the cluster M3, derived from the observations thanks to two different methods \citep{Kholopov1955, Oort1959} is plotted.

\begin{figure}
\includegraphics[width=\columnwidth]{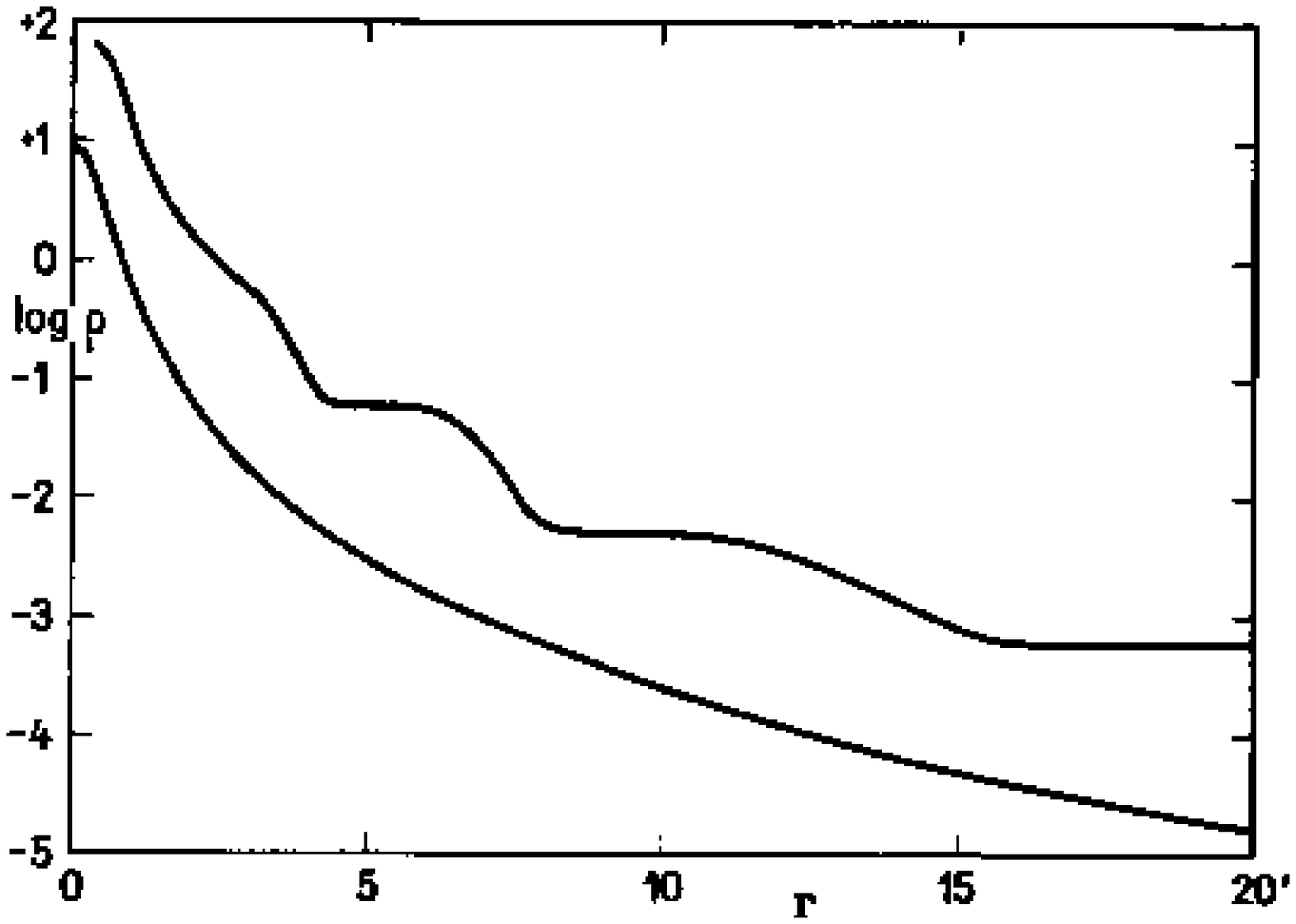}
\caption{Spacial density of the cluster M3, from Kholopov (top) and Oort \& van Herk (bottom). The curves are vertically shifted.}
\label{fig:17}
\end{figure}

Generally, it seems better to transform the observational data as little as possible, therefore to compare theory and observations not at an intermediate level, but rather at the very level of the observations.

We shall use, not the projected density itself, but its product with the distance to the center: $\rho_P r$. This has several advantages:
\begin{enumerate}
\item in the theoretical model (case of equal mass stars), this quantity tends to a finite value in the center, while the projected density becomes infinite.
\item $\rho_P r$ varies more slowly than the projected density (recall \reff{5} and \reff{9});
\item the observations are often counts of stars in concentric rings of constant width. The number of stars in a ring is
\begin{equation}
\Delta n = \frac{1}{m} \int_{r_1}^{r_2} 2\pi \rho_P\ r \, \dd r.
\end{equation}
Because this function is smooth and does not yield a strong curvature, we can write
\begin{equation}
\Delta n = \frac{2\pi}{m} \Delta r\ (\rho_P r)_{\overline{r}},
\end{equation}
with:
\begin{equation}
\Delta r = r_2 - r_1, \qquad \overline{r} = \frac{r_1 + r_2}{2}.
\end{equation}
That is, the observed numbers $\Delta n$ immediately give, up to a factor, the values of $\rho_P r$.
\end{enumerate}

\subsubsection{Projected density: star counts (M3)}

\citet{Sandage1954, Sandage1957a} has made detailed counts of stars in the cluster M3 = NGC~5272; the results have been published by \citet{Oort1959}. These counts corresponds to the stars brighter than a given magnitude, and situated in concentric rings. It would be necessary, in principle, to derive the numbers of stars in successive magnitude ranges by subtraction; however, attempts showed that this strongly increases the spreading of the points, because the errors accumulate, while the number decreases. Furthermore, because the luminosity function rapidly grows with the magnitude \citep[see][]{Sandage1957a}, the stars brighter than a given magnitude are almost all gathered in a small range of magnitude, and thus all have similar masses (we assume a well-defined mass-luminosity relation). It is therefore better and allowed to accept that the observed numbers correspond, for each limit in magnitude, to stars of a given mass.

\begin{figure2}
\includegraphics[width=16cm]{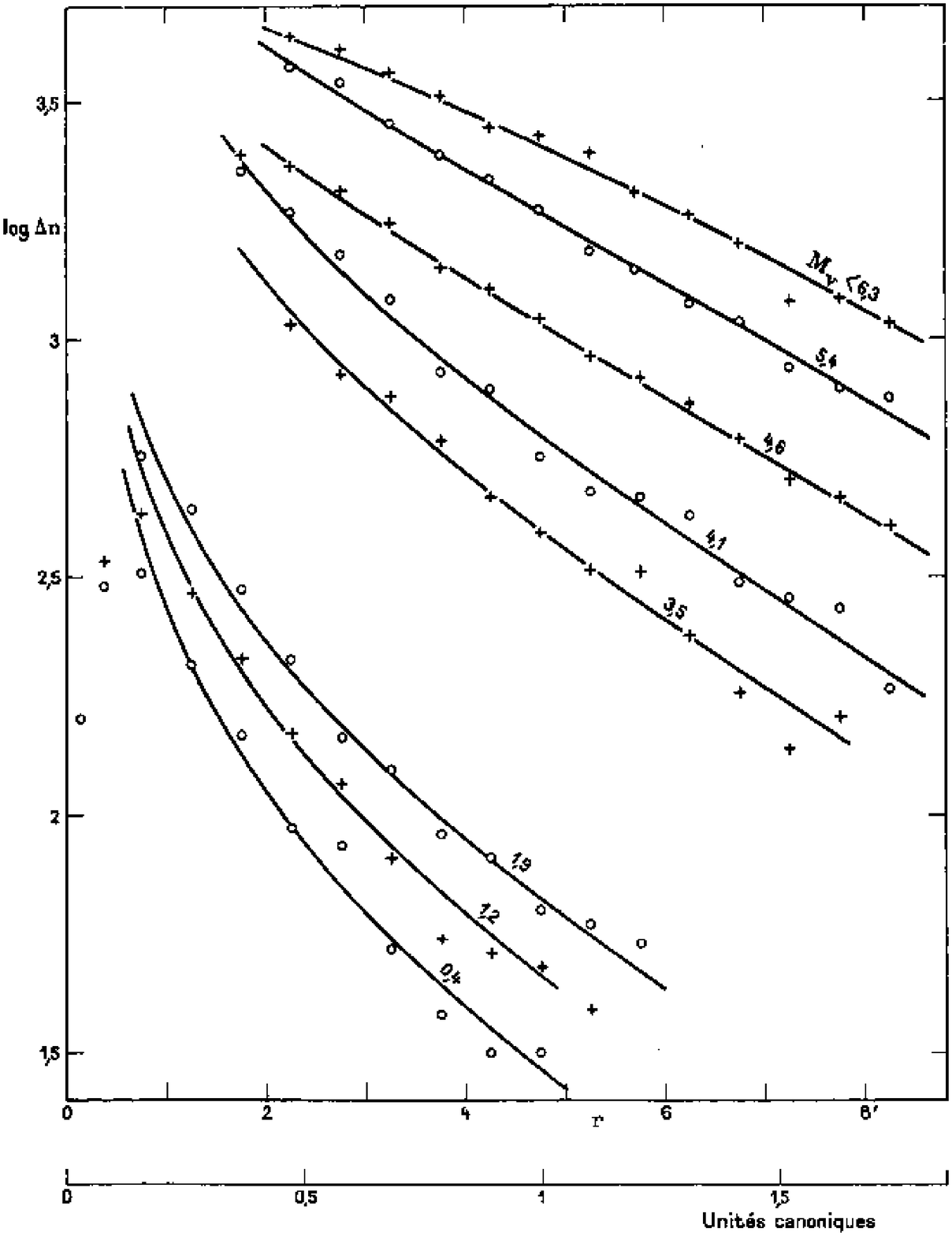}
\caption{M3: comparison between observations (points) and theory (lines). \tn{The scale reads \emph{Unit\'es canoniques}: canonical units.}}
\label{fig:18}
\end{figure2}

\begin{figure2}
\includegraphics[width=16cm]{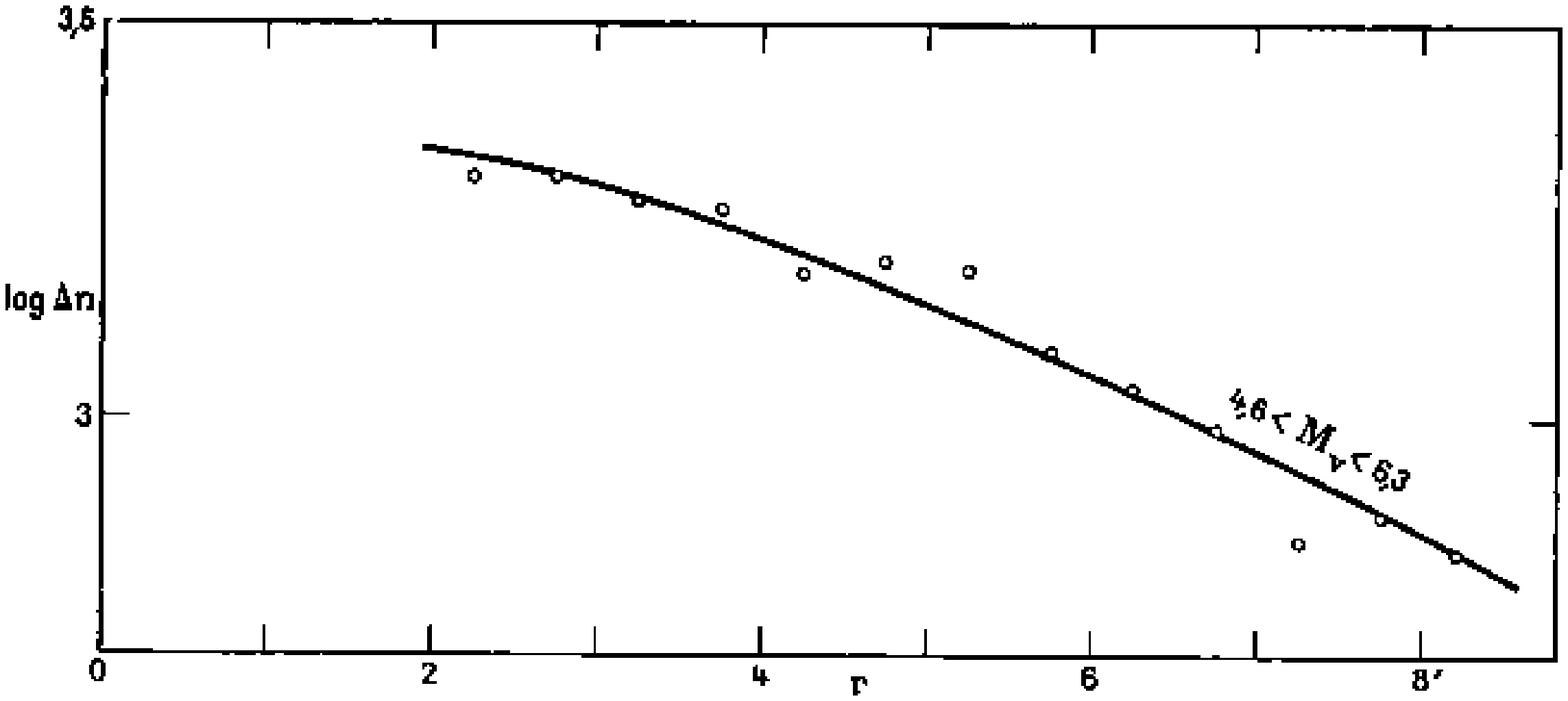}
\caption{Continuation of \reff{18}.}
\label{fig:19}
\end{figure2}

That is, the points on \reff{18} mark the observed numbers of stars brighter than a given magnitude $M_V$, in rings of width 30'' situated at the mean distance $r$ from the center. These numbers must be compared to the theoretical curves obtained in Chapter~VI (\reff{9}). By trial and error, we find that the best match is obtained for
\begin{equation}
1' \sim 0.21 \textrm{ canonical units}.
\end{equation}
We note that this gives out an external radius
\begin{equation}
r_e = 4.703 \textrm{ canonical units} \sim 22.4' \sim 79 \textrm{ pc}.
\end{equation}

For each group, the curve is fitted, by playing with two parameters: the relative mass $\mu$, and factor $C$, i.e. the number of stars in the group; a variation of $C$ corresponds simply to a vertical shift of the curve.

The values of $\mu$ and $C$ determined this way are given in \reft{5}, columns 4 and 5. The curves are plotted in \reff{18}; we see that the agreement is as good as possible, given the accuracy of the observations. Only the brightest stars are not well-retrieved near the center of the cluster, which can easily be explained by two reasons: (1) we have seen in Chapter~VI that the simplified theoretical model is likely to be wrong near the center, where it predicts a too high density of massive stars; (2) the observation underestimates the number of stars in the very populated central regions.

\begin{table} 
\caption{}
\label{tab:5}
\begin{tabular}{ccccccc}
$M_V$ & $M_V$ & $\overline{M_V}$ & $\mu$ & $\log{(C)}$ & $\log{\left(\frac{m}{\msun}\right)}$ & $M_\textrm{bol}$ \\
min & max & &&&&\\
\hline
\hline
$-\infty$ & 0.4 &  -0.5 &  1.55 &  1.328 &  0.090 &  -0.6\\
$-\infty$ & 1.2 &  -0.1 &  1.50 &  1.538 &  0.076 &  -0.2\\
$-\infty$ & 1.9 &  0.5 &  1.45 &  1.698 &  0.061 &  0.4\\
$-\infty$ & 3.5 &  2.2 &  1.42 &  2.472 &  0.052 &  2.1\\
$-\infty$ & 4.1 &  2.8 &  1.40 &  2.672 &  0.046 &  2.7\\
$-\infty$ & 4.6 &  3.3 &  1.10 &  2.894 &  1.941 &  3.2\\
$-\infty$ & 5.4 &  3.9 &  1.05 &  3.125 &  1.921 &  3.8\\
$-\infty$ & 6.3 &  4.5 &  0.80 &  3.246 &  1.803 &  4.4\\
4.6 & 6.3 &  5.5 &  0.60 &  2.986 &  1.678 &  5.4\\
\hline
\end{tabular}
\end{table}

\subsubsection{Mass-luminosity relation of the stars}

\reft{5}, column 3, gives the mean absolute magnitude $\overline{M_V}$ of each group, computed from the luminosity function \citep{Sandage1957a}. We notice that the difference between this mean magnitude and the maximum magnitude is never big, which confirms what has been said above.

The values of $\mu$ and $\overline{M_V}$ allow us to plot the mass-luminosity relation of the cluster (\reff{20}); $\mu$ is proportional to the mass $m$ of the stars, according to (6.32). The bolometric correction has been uniformly taken equal to -0.1. In order to extend the relation as much as possible for the faintest stars, we made an exception and considered the group of stars whose magnitude is between 4.6 and 6.3 (in the case of the narrower group $5.4<M_V<6.3$, the dispersion of the points is too large and we cannot plot a curve). This group is compared to the closest theoretical curve in \reff{19}; the parameters are given in the last row of \reft{5}.

\begin{figure}
\includegraphics[width=\columnwidth]{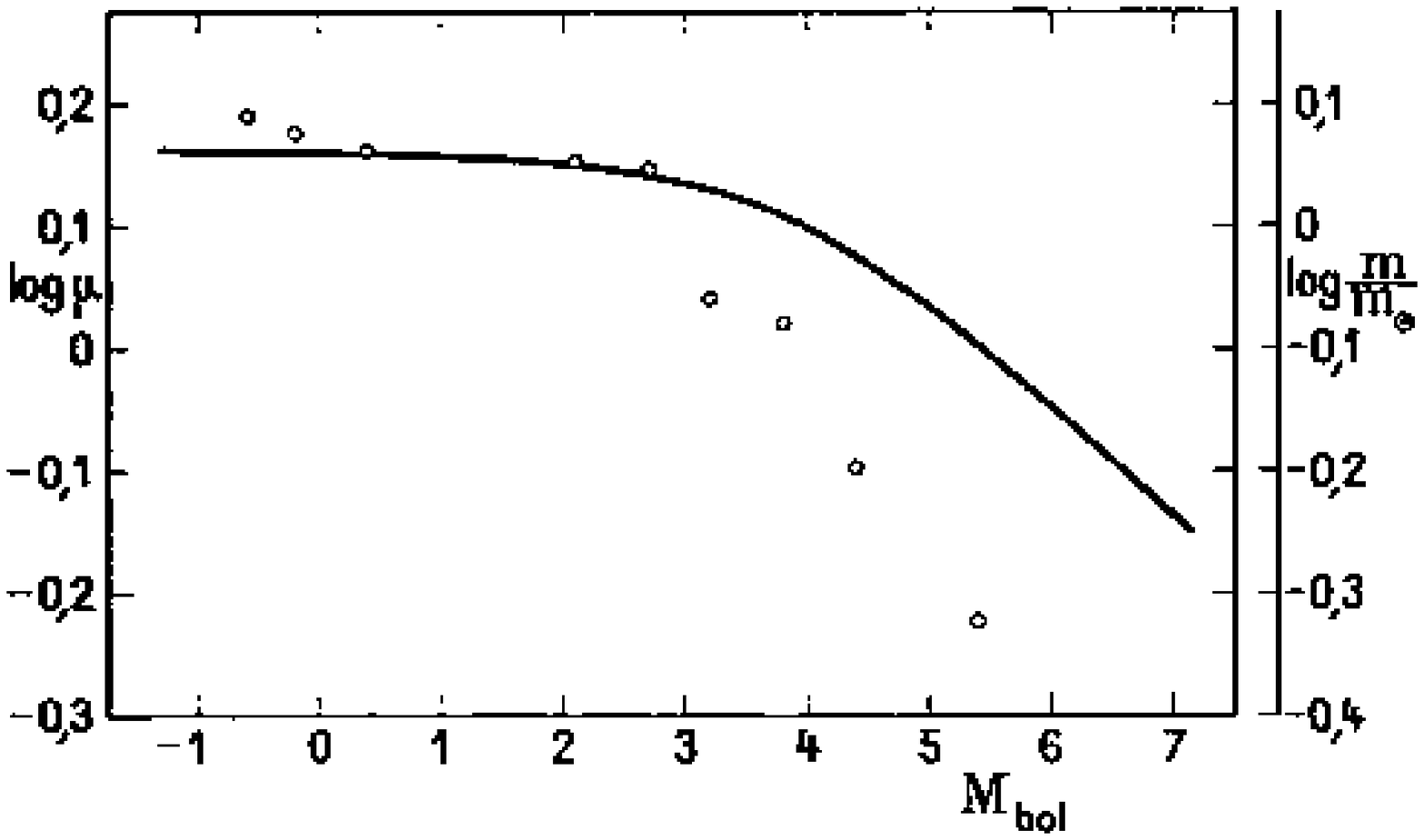}
\caption{Mass-luminosity relation in M3, from the spacial distribution of the stars (dots) and from the theory of stellar evolution (line).}
\label{fig:20}
\end{figure}

The points in \reff{20} show a well-defined relation\footnote{\citet[Figure~8]{vonHoerner1957} has obtained a similar curve, from the same observations, but with a different theoretical model. Our results indicate a larger variation of the mass as a function of the luminosity.}. We observe a strong bend at $M_V = 3$, which corresponds exactly to the bend in the Hertzsprung-Russell diagram, and which marks the beginning of the zone of rapid evolution of the stars; the stars with a magnitude below 3 almost have all the same mass. The slight increase of the points on the left-hand side is likely not real and comes from errors when computing $\mu$.

It is interesting to compare this mass-luminosity relation, deduced from the purely mechanical considerations, to those provided by the stellar evolution theory. \citet[Table~6]{Sandage1957b} has computed the relation between the present-day magnitude and the initial magnitude of the stars of M3, thanks to a semi-empirical method based on the observed HR diagram and on the theoretical evolutionary tracks. We get the mass from the initial magnitude, assuming that the initial mass-luminosity relation is that of the main sequence, i.e.
\begin{equation}
\frac{L}{\textrm{L}_\odot} = \left(\frac{m}{\msun}\right)^4.
\end{equation}

This way, we get the present-day mass-luminosity relation, the solid line on \reff{20}; the two relations have been vertically shifted so that their horizontal parts match.

The two curves are clearly in disagreement. The position of the bend is more or less the same; but on the low luminosities end, the ``dynamical'' masses decrease much faster than the mass found from the evolution.

It is hard to tell which curve is wrong. Neither is very reliable. The finding of the dynamical masses relies on an approximate calculation (Chapter~VI), which may lead to large errors when (as is the case for globular clusters) the mass spectrum is very broad. On the other hand, the hypothesis stating that the initial mass-luminosity relation of the cluster would match the main sequence, should be taken carefully, because of different chemical abundances.

We will explore the possibilities of progress in the computation of the dynamical masses in the last Chapter.

The adjustment of the vertical scales gives an interesting piece of information: when compared to (6.32), we find that
\begin{equation}
\frac{\overline{m^2}}{\overline{m}} = 0.80 \msun.
\end{equation}
This value can also be computed from the mass spectrum of the cluster. From the initial luminosity function $\psi(M_V)$ given by \citet[Table~2 and Figure~2]{Sandage1957a} and the mass-luminosity relation of \citet{Kuiper1942}, we obtain the initial mass spectrum. The present-day spectrum is derived by assuming that all the stars heavier than $1.44 \msun$ have been evolving down to this value through their transformation into white dwarves. The effect of escape can be neglected. We find
\begin{eqnarray}
\overline{m} &=& 0.353 \msun,\\
\overline{m^2} &=& 0.265 \msun,\nonumber\\
\overline{m^2}/\overline{m} &=& 0.75 \msun,\nonumber
\end{eqnarray}
in very good agreement with the value (8.8).

This previous finding is quite sensitive to the mass chosen for the white dwarves, $m_b$ \tn{The subscript $b$ stands for \emph{naine blanche}: white dwarf.}. For $m_b = 1\msun$, we find $\overline{m^2}/\overline{m} = 0.61 \msun$; for $m_b = 2\msun$, we find $\overline{m^2}/\overline{m} = 0.98 \msun$. The comparison of the two findings seems to confirm that the white dwarves have the limit mass of Chandrasekhar, $m_b = 1.44 \msun$. However, here again, the uncertainties of the theory and of the determination of the mass spectrum question the validity of this conclusion.

\subsubsection{Projected density: brightness measurements (47~Tuc)}

\citet{Gascoigne1956} have obtained the curves of projected density of the clusters 47~Tucanae and $\omega$~Centauri from a photometric method. The disadvantage of this procedure is to provide only the total brightness, without distinguishing the stars of different magnitudes; however, it allows for a precise measurement of the density from the center of the cluster up to quite remote distance.

The case of $\omega$~Centauri will be studied later. \reff{21} plots the observed values for 47~Tuc = NGC~104 and the closest theoretical curve, which corresponds to:
\begin{eqnarray}
1' &=& 0.088 \textrm{ canonical units},\\
\mu &=& 1.33,\nonumber\\
C &=& 0.555 \textrm{ stars of the tenth magnitude per minute}.\nonumber
\end{eqnarray}
The external radius is then:
\begin{equation}
r_e = 53.4'.
\end{equation}

\begin{figure}
\includegraphics[width=\columnwidth]{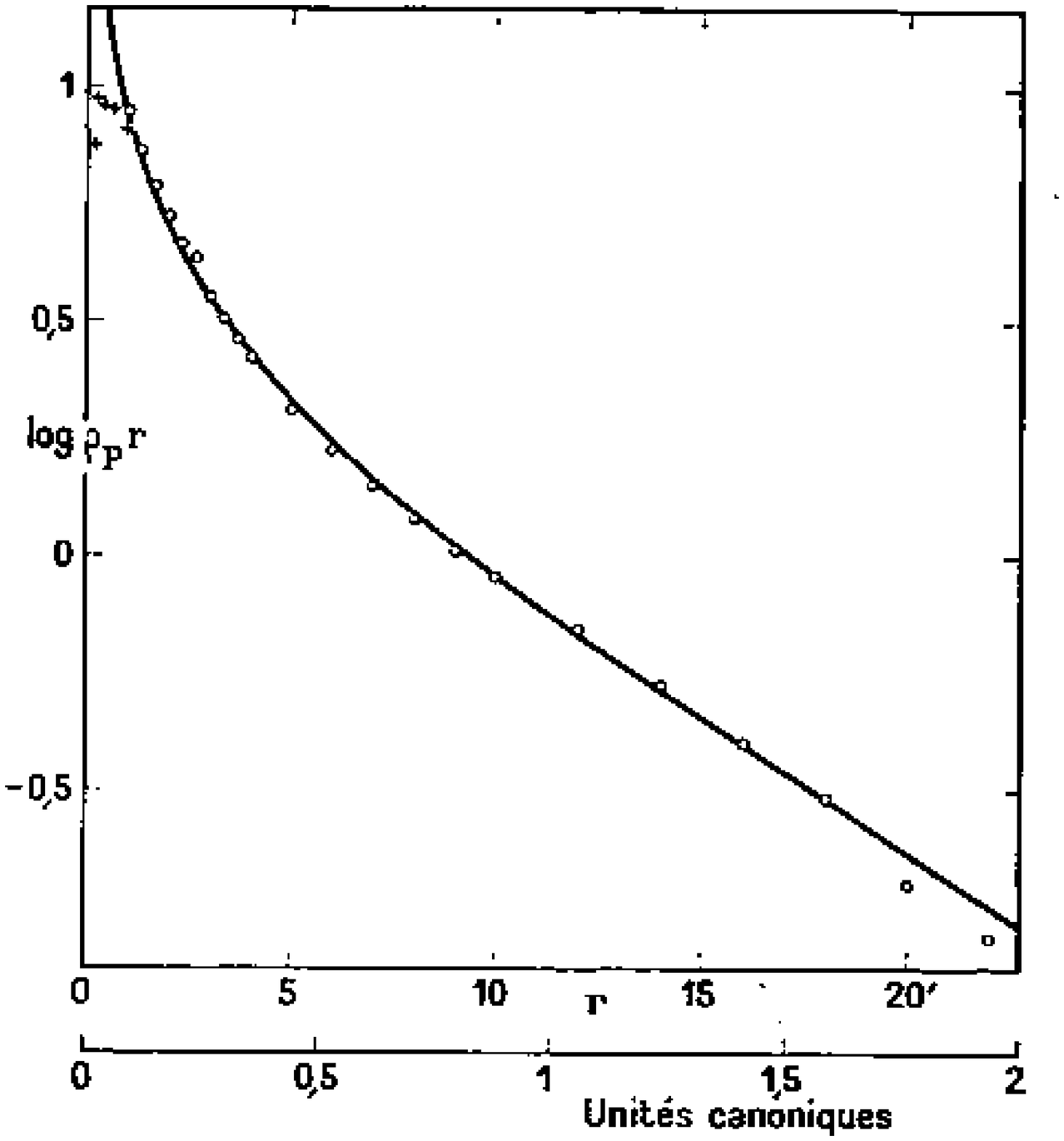}
\caption{47~Tuc: comparison between observations (dots) and theory (line).}
\label{fig:21}
\end{figure}

The value of $\mu$ found here can been seen as a particular ``mean mass'', obtained by weighting the stars proportionally to their luminosity. It is only slightly lighter than the mass of the heaviest stars ($\mu\simeq 1.45$ from the results obtained for M3) which contribute to the major part of the total brightness.

\reff{21} shows good agreement (the small systematic deviations of the points from the theoretical curve can be explained: the data of Gascoigne \& Burr does not directly come from observations, but has rather been read on a smoothed curve). Near the center, the theoretical curve is too high, for reasons already listed in relation to M3.

Gascoigne \& Burr have measured the total luminosity in a series of circles of small radius around the center of the cluster. For each circle, we can compute the mean value of $\rho_P r$, by using (8.3), with $r_1 = 0$; the crosses on \reff{21} mark the value obtained. We clearly see that $\rho_P r$ tends toward a well-defined central value. This seems to be a good confirmation of the shape predicted by the theory near the center, and in particular of the fact that the central density is infinite\footnote{However, \citet{King1961b} has observed the central region of several clusters in detail and concluded that a high but finite central density exists.}

\subsection{Mass-radius relation}

The relation (3.7) links the mass of a cluster with its radius and the mean value of the negative curvature of the galactic field along its orbit. Unfortunately, the exact motions of the globular clusters in the Galaxy are unknown; the radial velocities can be measured with a good precision \citep{Mayall1946, Kinman1959}, but the proper motions stand at the limit of the observational possibilities, and have been obtained for 9 clusters only \citep{Gamalej1948}, with a low precision. It seems even not possible to know wether the orbits of the clusters are in average rather radial or circular \citep{vonHoerner1955, Kurth1960}. That is why we will consider the curvature of the galactic field to be the same for all the clusters, for want of anything better.

The radius and the mass of the globular clusters should be linked, at least approximately, through a relation like (3.8). We are going to check this using observational data. To limit the effect of observational errors, we will only use a series of homogeneous observations, done by the same author, under the same circumstances.

The distances have been taken from \citet{Kinman1958}. In order of preference, we take the distances derived from the position of the main sequence, from the magnitude of the variable stars, or from the magnitude of the 25 brightest stars. If none of these three estimates exists, the cluster is not considered.

\citet{Christie1940} made very precise measurements of the apparent magnitudes of the clusters. From them, we derive the absolute photographic magnitudes $M_{pg}$; the possible effect of absorption is cancelled, because the distances themselves are derived from the apparent magnitudes of the stars.

Then, we assume that the ratio of the total mass to the total luminosity is the same for all the globular clusters; by taking the mass estimate of M3 by \citet{Sandage1957a} as reference, we find that the mass $\mc{M}_e$ of a cluster is given as a function of its absolute magnitude $M_{pg}$ by
\begin{equation}
\log{\mc{M}_e} = -0.4\ M_{pg} + 2.00.
\end{equation}

This is confirmed by the study of \citet{Zeliakh1957} who, after a detailed discussion on several clusters, arrived at a quite similar value for the constant: 1.94 instead of 2.00.

Finally, we take the apparent external radii measured by \citet{Shapley1935}; they appeared to be more precise that those published more recently by \citet{Mowbray1946}. To derive the real radii, we should, in principle, use the distances corrected for absorption. However, as indicated by \citet{Shapley1935, Shapley1949}, absorption induces two opposite effects: an overestimate of the distance and an underestimate of the apparent radius. A quantitative study shows that these two effects balance each other almost exactly \citep{Parenago1949, Lohmann1952}. Therefore, we can completely neglect the possible existence of absorption. For that matter, this is very fortunate because the absorption is not quite well known.

\reft{6} gives the list of the 35 clusters for which the three measurements exist, with their mass (in solar masses), and their radius (in parsecs). The mass-radius diagram is plotted in \reff{22}: we see that, despite the dispersion of the points, a defined relation arises. The main source of error comes from the estimation of the external radius, which is difficult to observe precisely; this error alone is enough to explain the dispersion of the points, and it is even remarkable that it does not induce a larger dispersion. (This shows the quality of the measurements of Shapley \& Sayer; if one uses the radii from \citealt{Mowbray1946}, the dispersion is much larger.)

\begin{table} 
\caption{}
\label{tab:6}
\begin{tabular}{ccc}
NGC & $\log{(\mc{M}_e)}$ & $\log{(r_e)}$ \\
\hline
\hline
288 & 4.64 & 1.40\\
1904 & 5.12 & 1.37\\
2419 & 5.08 & 1.71\\
4147 & 4.44 & 1.03\\
5024 & 5.12 & 1.66\\
5139 & 5.80 & 1.90\\
5272 & 5.40 & 1.64\\
5897 & 4.60 & 1.50\\
5904 & 5.28 & 1.60\\
6093 & 5.04 & 1.52\\
6121 & 4.48 & 1.30\\
6171 & 4.36 & 1.34\\
6205 & 5.20 & 1.38\\
6218 & 4.80 & 1.48\\
6229 & 4.88 & 1.38\\
6254 & 5.04 & 1.53\\
6266 & 5.28 & 1.60\\
6273 & 5.04 & 1.50\\
6284 & 4.56 & 1.49\\
6293 & 4.84 & 1.39\\
6333 & 5.00 & 1.53\\
6356 & 5.16 & 1.70\\
6402 & 5.00 & 1.69\\
6626 & 4.84 & 1.46\\
6638 & 4.88 & 1.32\\
6656 & 5.08 & 1.54\\
6715 & 5.36 & 1.71\\
6723 & 5.00 & 1.30\\
6779 & 4.68 & 1.38\\
6809 & 4.84 & 1.46\\
6864 & 5.48 & 1.74\\
6981 & 4.64 & 1.42\\
7078 & 5.44 & 1.60\\
7089 & 5.56 & 1.63\\
7099 & 4.76 & 1.29\\
\hline
\end{tabular}
\end{table}

\begin{figure}
\includegraphics[width=\columnwidth]{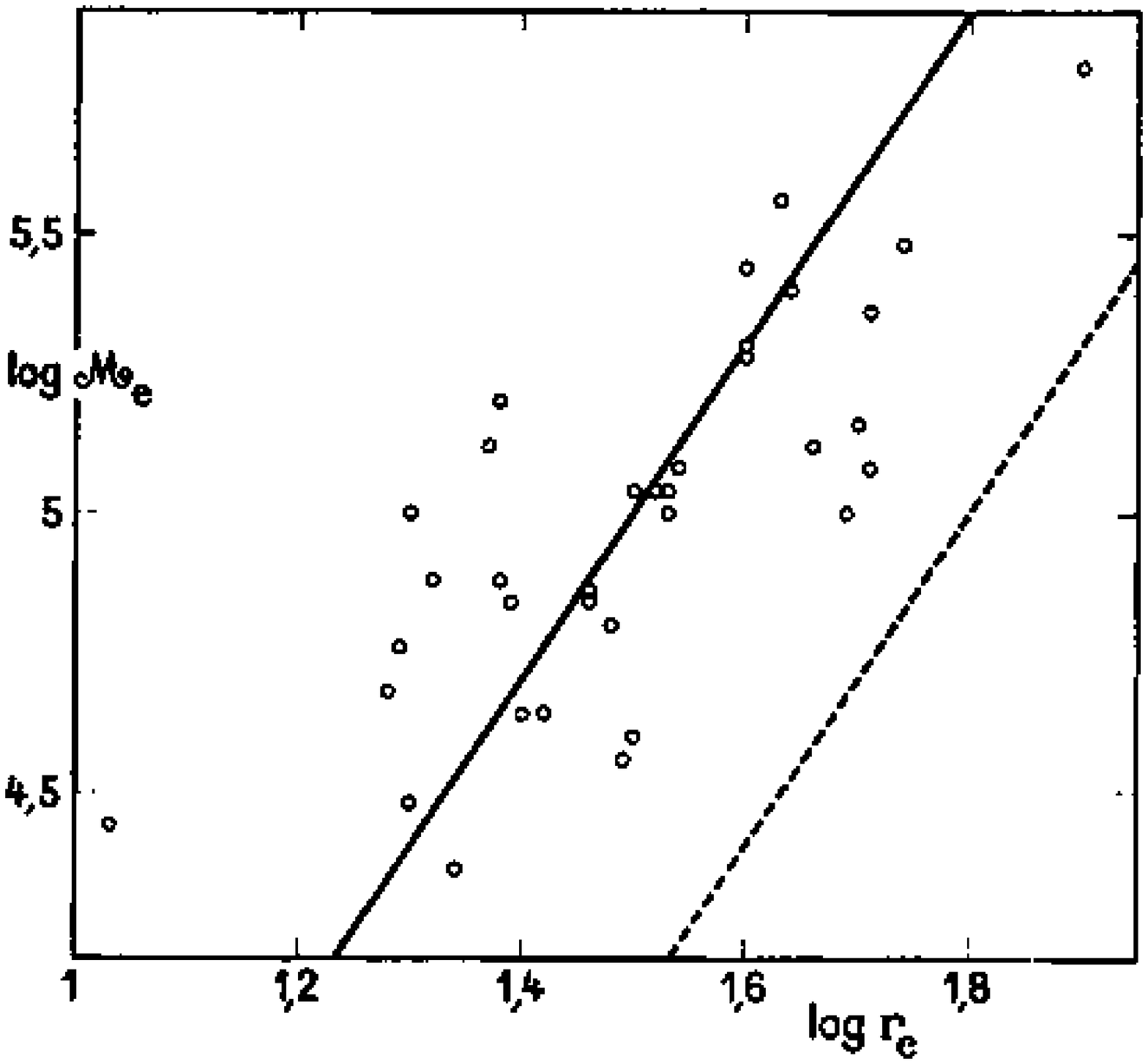}
\caption{Mass-radius relation of the globular clusters.}
\label{fig:22}
\end{figure}

In addition, the small dispersion justifies the hypothesis of constant curvature of the galactic field for all the clusters.

The largest error being on $r_e$, we shall compute the means in the horizontal direction. This way, we find that the data points are well fitted with the relation
\begin{equation}
\log{(r_e)} =\frac{1}{3} \log{(\mc{M}_e)} - 0.17\ (\pm 0.11).
\end{equation}
i.e., 
\begin{equation}
\frac{\mc{M}_e}{r_e^3} = 3.2 \U{M_\odot/pc^3},
\end{equation}
in agreement with the theoretical formula (3.8).

The isolated point in the bottom-left corner of the figure is NGC~4147. This cluster is easily distinguishable from the others; Shapley \& Sayer qualify it as ``abnormal''. The point on the top-right corner is $\omega$~Cen, also abnormal, as we will see later.

\subsubsection{Correction of the external radii}

The observed external radius is, in fact, systematically smaller than the real radius. Indeed, the projected density decreases very rapidly toward the outskirts and becomes very small long before reaching the boundary of the cluster. This clearly appears in the artificial cluster of \reff{15}: a direct estimate of the radius of this cluster, from its aspect, gives a value of the order of 0.6 times the real radius, and this without any background effect. \reft{1} indicates that only 3\% of the total mass of the cluster is situated (in projection) beyond half of the radius. That is, even without absorption, the radius may be underestimated by a ratio of about 1/2. A good confirmation of this value can be obtained in the case of the two clusters studied above: for M3 and 47~Tuc, the observed radii are 11' and 28.2', i.e. 0.49 and 0.53 times the real radii derived from the theory and the projected density profiles, and given in (8.6) and (8.11), respectively. Furthermore, there is no absorption for these clusters.

Thus, we will systematically make a correction, assuming that \emph{the real radii are twice as large as the observed radii}, given by Shapley \& Sayer. As a consequence, the relation (8.14) must be replaced with
\begin{equation}
\frac{\mc{M}_e}{r_e^3} = 0.40 \U{M_\odot/pc^3}.
\end{equation}
This relation between the mass and the real external radius is plotted in \reff{22} as a dashed line. All the observed radii stand on its left side, as expected.

\subsubsection{Curvature of the galactic field}

When comparing (3.7) and (8.15), we find the mean curvature of the galactic field for the clusters:
\begin{equation}
\overline{\frac{\partial^2 U_G}{\partial x^2}} = -5.4\times 10^{-16} \yr^{-2} = -520 \U{km^2 s^{-2} kpc^{-2}}.
\end{equation}

For comparison, we can compute the curvature in the solar neighborhood; it derives from the Oort's constants $A$ and $B$ (the numerical values are taken from \citealt{Allen1955}):
\begin{equation}
\frac{\partial^2 U_G}{\partial x^2} = (B-A) (3A+B) = -1370 \U{km^2 s^{-2} kpc^{-2}}.
\end{equation}

The mean curvature of the galactic field for the globular clusters is thus about 3 times smaller than in the solar neighborhood. This result is plausible, because the clusters travel in average at a quite long distance from the center of the Galaxy, and not in its plane. A more detailed comparison of the value (8.16) with the known structure of the Galaxy would be interesting, but is out of the scope of this work.

\subsection{Evolution}

Knowing the mass $\mc{M}_e$ and the radius $r_e$ of a real cluster, we can compute the parameters $\beta$ and $\gamma$ of the homology, using the transformation relations (6.31b), (6.34a), (2.32c), (2.43a). We get
\begin{eqnarray}
\beta &=& G \left(\frac{\mc{M}_e}{\fb{M}_e}\right)\left(\frac{r_e}{\fb{R}_e}\right)^{-1},\\
\gamma &=& (16\pi^2\ \overline{m})^{-1}\ G^{-3/2}\ \left(\frac{\mc{M}_e}{\fb{M}_e}\right)^{-1/2} \left(\frac{r_e}{\fb{R}_e}\right)^{-3/2}.\nonumber
\end{eqnarray}

This allows us to go from any quantity of the homologous model to the corresponding physical quantity, using the transformation formulae (2.31), (2.32), (2.43), (2.48), (6.31) and (6.34). In particular, we obtain:
\begin{equation}
\frac{\dd t}{\dd \fb{T}} = G^{-1/2} \frac{\overline{m}}{\overline{m^2}}\ \frac{1}{\ln{(n)}}  \left(\frac{\mc{M}_e}{\fb{M}_e}\right)^{1/2} \left(\frac{r_e}{\fb{R}_e}\right)^{3/2}.
\end{equation}

This relation gives the time scale of the evolution of the cluster, because $\fb{T}$ is, as we have seen (Chapter~V), a parameter that measures the level of evolution of the cluster, while $t$ is the real time. $\fb{M}_e$ and $\fb{R}_e$ are given in (5.5). We use the value $\overline{m^2}/\overline{m}$ found from the observation of M3 (Equation 8.8). The factor $\ln{(n)}$ does not vary much for the different globular clusters; we use the typical value $n=10^5$. Finally, the radius is linked to the mass through (8.15), found from the observations. Thus, we find the very simple relation:
\begin{equation}
\frac{\dd t}{\dd\fb{T}} = 1.76 \times 10^5\ \mc{M}_e,
\end{equation}
where $t$ is in years and $\mc{M}_e$ in solar masses.

(6.22) provides the escape rate of the stars of given mass, which also reads
\begin{equation}
\frac{1}{n_m}\frac{\dd n_m}{\dd\fb{T}} = -\theta.
\end{equation}
Recall that $n_m$ is the number of stars whose mass is between $m$ and $m+\dd m$. $\theta$ is given in \reft{2} and can be fitted with the relation:
\begin{equation}
\theta \simeq 0.4078\ (3-2\mu).
\end{equation}

Then, we can compute the variation of the total mass, by using (6.25b), (6.32) and (6.26). We find
\begin{equation}
\frac{\dd\mc{M}_e}{\dd\fb{T}}=-0.4078\ \mc{M}_e.
\end{equation}
We note that this result does not depend on the mass function.

When comparing with (8.20), the mass $\mc{M}_e$ cancels, and we get
\begin{equation}
\frac{\dd\mc{M}_e}{\dd t}=-2.3 \times 10^{-6} \U{M_\odot/yr}.
\end{equation}

Thus: \emph{the mass of a globular cluster decreases by $2300 \msun$ per billion years}; this rate is almost the same for all the globular clusters, and it also remains constant with time. The simplicity of this result is to be noted.

In fact, the rate is not exactly constant. The term $\ln{(n)}$ changes with time, which would lead to slight slowdown of the evolution near the end. On the other hand, the term $m_0 = \overline{m^2}/\overline{m}$ increases because the lightest stars escape faster. This effect is opposite to the previous one and likely stronger. For example, in the cluster M3 which is still at the beginning of its evolution, the value observed is $m_0 = 0.80 \msun$; near the end of its evolution (in about $100 \Gyr$), only the heaviest stars will remain and we will get $m_0 = 1.44 \msun$, thus a multiplication of the evolution rate by a factor 1.8.

\subsection{Mass function of the globular clusters}

The evolution law of the masses of the globular clusters partially sets the present-day mass distribution of these masses. We are going to see whether a confirmation of (8.24) can be obtained this way.

From the apparent magnitudes and the apparent distance moduli provided by \citet{Lohmann1952} for 94 clusters, we compute the absolute magnitudes and then the masses using (8.12).

The mass function we obtain is plotted in \reff{23}, using the dotted line on the left-hand side. The two most massive clusters are off the figure: M2 and $\omega$~Cen of masses $3.4\times 10^{5} \msun$ and $5.2\times 10^{5} \msun$. We must also investigate the effect of observational selection. By looking at the apparent magnitudes $m_{pg}$ of the clusters, we see a clear upper limit at $m_{pg} = 12.5$; we assume that this value marks the highest observable magnitude. As a consequence, the clusters of given absolute magnitude $M_{pg}$ are observed only if their apparent distance modulus is smaller than $12.5-M_{pg}$. That is, we can compute, for each value of $M_{pg}$, the observable fraction of the clusters, as soon as we know the distribution of clusters as a function of the apparent modulus. This distribution (that we assume to be independent of the absolute magnitude) is easily found thanks to the brightest clusters, for which observational selection does not play any role.

\begin{figure}
\includegraphics[width=\columnwidth]{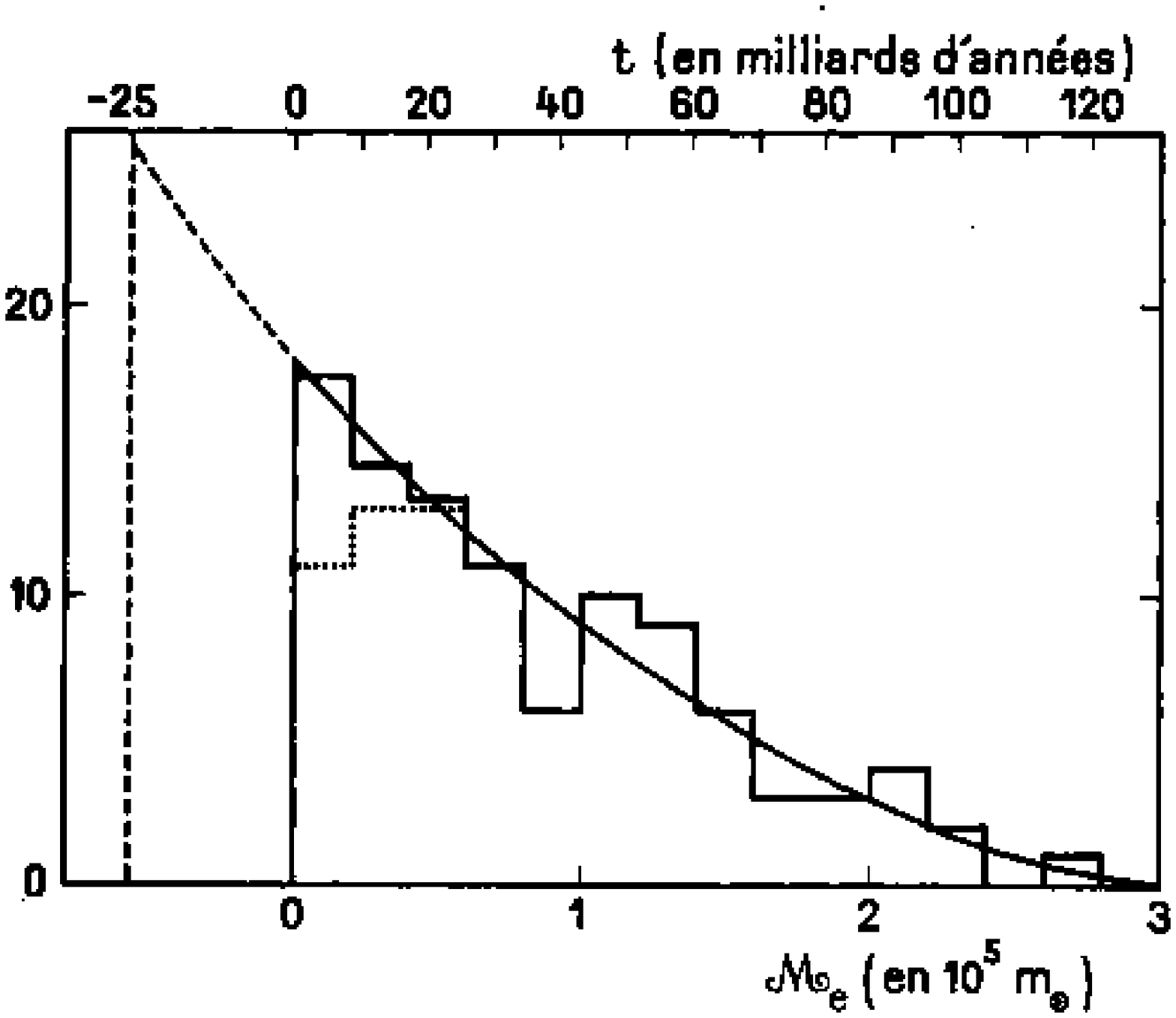}
\caption{Mass function of the globular clusters, observed (dotted line), corrected for observational selection (solid line), initial (dashed line). \tn{The upper axis reads \emph{$t$ en milliards d'ann\'ees}: $t$ in billion years.}}
\label{fig:23}
\end{figure}

The solid line in \reff{23} is the corrected mass function. It can be fitted with a smooth curve. In particular, we note that this curve takes a finite value for $\mc{M}_e = 0$. But if we assume that all the globular clusters formed in the past and that no more are born now; and if the evolution of the masses is as
\begin{equation}
\frac{\dd\mc{M}_e}{\dd t} \propto \mc{M}_e^\nu,
\end{equation}
we can easily find that the present-day mass function must be proportional to $\mc{M}_e^{-\nu}$ near the origin \tn{i.e. for $\mc{M}_e \to 0$}. Therefore, this mass function must be zero for $\mc{M}_e = 0$ if $\nu$ is negative, and become infinite if $\nu$ is positive; it can take a finite value only if $\nu=0$. Therefore, the observed mass function seems to confirm the evolution law (8.24).

The validity of this point is slightly weakened because it relies on the number of the least massive clusters, that are the least well observed; but we can argue that the correction for observational selection (\reff{23}) is not very important; the conclusion would have been the same without the correction.

Note also that (8.25) is not fanciful but in fact covers all the evolutionary models already studied by several authors. The formulae of \citet{Chandrasekhar1943b, Chandrasekhar1943c} assume $\nu = +1$; the model from \citet{King1958b} corresponds to $\nu = -2.5$ and the similar model of \citet{vonHoerner1958} to $\nu = -3.55$.

The evolution formula (8.24) translates into a simple shift of the mass function toward the left; or alternatively, into a shift of the zero-mass toward the right. The upper axis of \reff{23} indicates the position of this zero-mass at different times, past and future. In particular, the \tn{vertical} dashed line on the left-hand side marks the position of the zero-mass point at the time of birth of the clusters, assuming it occurred $25 \Gyr$ ago (see the next Section). By extrapolating the observed curve a little, we obtain the initial mass function. We find that about 62 globular clusters have disappeared since the origin of the galaxy. Nowadays, one cluster disappears every $500 \Myr$, in average. Every cluster that has survived has lost $57000 \msun$ as escaping stars since its birth. For all the clusters, the mass of the escapers since their birth is: $8 \times 10^6 \msun$.

In the future side, we find that half of today's clusters will be gone in $30 \Gyr$; but the most massive ones will survive much longer: $\omega$~Centauri will reach $230 \Gyr$.

Up to now, we have neglected the phenomenon of gas ejection by the massive stars. \citet{vonHoerner1958} showed that this phenomenon is only relevant at the beginning of the life of the cluster: 30\% of the initial mass is lost in the gaseous form within the first billion years, then only 6\% during the next $5 \Gyr$. The results obtained above are therefore valid, as soon as we put aside the initial evolutionary phase, lasting about one billion years.

\subsection{Age of the globular clusters}

When a cluster reaches the homologous state, any sign of its past (in particular its initial structure and its age) is lost. That is, one should specially care about the clusters that still show significant differences with the homologous model. According to the theory (see Equation 8.20), the most massive clusters are those which evolve the slowest, therefore, those where the largest differences should remain. That is, we are interested in $\omega$~Centauri, the most massive of all globular clusters.

But precisely, $\omega$~Centauri yields a peculiar structure, very different from those of the other clusters. The difference already arises as its concentration class in Shapley's classification is VII, while all the other high luminosity clusters lie in classes I to V. The difference is even more visible in the work of \citet{Gascoigne1956} who observed $\omega$~Cen and 47~Tuc under the same circumstances. The Table~II and the Figure~2 from these authors perfectly show that the light from a small circle of radius $r$ surrounding the center of the cluster is proportional to $r^2$ in the case of $\omega$~Cen, and to $r$ for 47~Tuc. In other words: in 47~Tuc, the projected density goes as $1/r$ near the center; but in $\omega$~Cen, it is almost constant. \emph{The central density of $\omega$~Cen is finite}.

The theoretical interpretation comes immediately: 47~Tuc has already reached the stage in its evolution where the central density becomes infinite; $\omega$~Cen, being more massive, has not reached it yet. (It seems to be the only globular cluster in this case.)

This confirms the theory, and we now have a simple way of estimating the age of the clusters. When comparing (2.32e), (7.44), (7.45) and (8.20), we find that the time required for the infinite central density to appear is
\begin{equation}
t_2 = 0.713\times 10^5\ \mc{M}_{e_0} = 0.854\times 10^5\ \mc{M}_{e_2},
\end{equation}
where $t_2$ is in years, $\mc{M}_{e_0}$ is the initial mass of the cluster and $\mc{M}_{e_2}$ its mass at the time when the infinite central density appears. The present-day masses of 47~Tuc and $\omega$~Cen, obtained as in the previous Section, are: $2.8 \times 10^5 \msun$ and $5.2\times 10^5 \msun$. We conclude that the age of these clusters (assuming it is the same) must be between
\begin{equation}
24\times 10^9 \yr \qquad \textrm{and}\qquad 44\times 10^9 \yr.
\end{equation}

Furthermore, the most recent estimates of the age of the globular clusters from the stellar evolution theory \citep{Sandage1961} are
\begin{eqnarray}
22\times 10^9 \yr && \textrm{for M13},\\
26\times 10^9 \yr && \textrm{for M3 and M5}.\nonumber
\end{eqnarray}

That is, two completely independent methods lead to values in agreement for the age of the globular clusters. This result is very encouraging, both for the theory of the dynamical evolution of the clusters and for the theory of stellar evolution. It also emphasizes the discrepancy underlined by \citet{Sandage1961} between the ages of the globular clusters and the recent models of an expanding Universe.

According to theory, $\omega$~Cen should have more differences with respect to the homologous model; the first proper difference should still exist with a non negligible amplitude. We have seen (see \reff{13}) that this difference affects mostly the external radius. And we precisely note that that the radius of $\omega$~Cen is far too large, given its mass (\reff{22}).

Finally, the strong ellipticity of $\omega$~Cen is likely the remnant of an initial difference from spherical symmetry, that has not enough time to vanish yet.

\section{Conclusions}

We are going to summarize the major results obtained in this work, and then list the directions in which it would be good to develop the research, in order to improve the theory and extend the range of its applications.

\subsection{Results obtained}

By making two hypotheses: (1) isotropy of the velocities and (2) equal masses, we have obtained the system of equations (2.25), which, supplemented by the right boundary conditions, allows us to compute the evolution of a cluster from a given initial state. We have looked for an homologous solution, i.e. a model that remains self-similar, the evolution being limited to the scaling of the various physical quantities; we have found that this model exists and is unique. Its mass and its radius are finite. The projected density matches the observations as much as their precision allows. Near the center, the homologous model yields unexpected properties: the central density and the central potential are infinite; a continuous flux of negative energy goes toward the center where it is absorbed by the formation of multiple stars. Furthermore, the escape of stars through the boundary of the cluster leads to a linear decrease of the total mass with time.

Then, we have supposed that the main population of the cluster is mixed with a second population, much less numerous, of stars with a different mass, and we have computed the distribution of this secondary population and its escape rate. These results allow for an approximate solution in the case of any mass function. By comparing with the observations, we obtain in particular the approximate mass-luminosity relation of the stars in clusters.

Finally we have studied the evolution of clusters close to the homologous model. We found that if the central density is finite, it rapidly increases and becomes infinite after a certain time. Generally, any difference with the homologous model decreases with time; a cluster becomes almost identical to the homologous model after the first third of its lifetime. In particular, these results lead to an estimate of the age of the globular clusters.

\subsection{Desirable improvements and extensions}

\begin{enumerate}
\item The most debatable hypotheses of the present model is indubitably the isotropy of the velocities. We could, as a first step, make the distribution function more general by adding an anisotropic term, small compared to the main term; this would allow us to quantify the error made when using the isotropy hypothesis. In a more ambitious second step, we would eliminate any restriction for the distribution function $f(E,A)$; but in this case, the equation of local evolution would become extremely complex (see \citealt{Rosenbluth1957}, Equation~31).
\item In the study of the approach to the homologous model made in Chapter~VII, we assume that the cluster is already similar enough to the homologous model that we can linearize the equations of the differences. It would be very interesting to also study the evolution of clusters far from the homologous model; this would allow us to describe the beginning of the evolution of a cluster, from a given initial state. To do so, we would have to come back to the real \tn{physical} variables and directly solve (2.25), by computing step by step the successive forms taken by the cluster in time\footnote{A similar calculation has been done, in the much simpler case of a homogeneous plasma, by \citet{MacDonald1957}.}. The results could be applied to a detailed explanation of the structure of $\omega$~Cen and to a precise estimate of its age. They would also allow for a more rigorous demonstration of the statement that clusters tend toward the homologous model, whatever their initial state is.
\item The previous paragraph implies the knowledge of the initial state of the clusters, therefore the building of a more or less approximate theory for the formation of the clusters and the initial phase during which the steady state is established. We could assume, for example, that the protocluster is an homogeneous sphere with a density slightly higher than the critical density (therefore, favoring a contraction), with negligible internal velocities. The evolution equations for this initial phase would be completely different from those we have considered here; on the one hand, the steady state is not reached and thus the distribution function does not only depend on $E$ and $A$, and on the other hand, because this phase is very short, the effect of perturbations can be neglected. Fortunately this implies that there is no need for the knowledge of the time of formation of the stars.
\item The hypothesis of equal mass stars is likely to be far from reality. In Chapter~VI, we have obtained a more general solution but at the cost of a quite arbitrary approximation. Hence, we should extend the investigations to the case of any mass function. But then a big difficulty appears: the escape rates of the stars of different masses not being the same, their relative importance always changes and the structure of the cluster varies too; as a consequence \emph{a homologous solution cannot exist}. In other words, if one introduces the mass as an additional variable, one must also consider the time variable, and the complexity of the calculations is suddenly hugely increased.

There are two possibilities to overcome this. First, we could seek a better model than the homologous one, that would allow us to separate the time variable; this model would necessary count, in addition to the two dimensional parameters, one or more other parameters depending on time. Second, the mass function could be initially described as a sum of \tn{Dirac's} $\delta$ distributions. The simplest case of the sum of two $\delta$ distributions (i.e. a cluster made of the mix of stars of two different masses, in any proportion), would already be a great improvement and would allow us to estimate the actual effect of a spread of masses.

\item We have only considered the value $\lambda=1/3$ of the power of the radius-mass relation (3.11), with the perspective of globular clusters. The calculations could be re-done with different values of $\lambda$, and in particular, with $\lambda = \infty$, which correspond to the isolated cluster, and seems to better fit the cases of galaxies and galaxy clusters. The other values of $\lambda$ do not seem to correspond to existing objects, but could be useful to describe some intermediate states.

\item The hypothesis of spherical symmetry is valid for most of the clusters; but some are clearly elliptical. Furthermore, we could foresee the application of the models to other objects (see point 8 below). Thus, we should study the non-spherical models. The ellipticity can originate from a global rotation; a first approximation would consider that the rotation is small and only implies a correction term. We would likely find a model similar to the homologous one, but slightly flattened. It would be particularly interesting to find whether the escaping stars take away a little or a lot of angular momentum, and thus, if the flattening increases or decreases.

We should also study the case of an angular momentum distribution without spherical symmetry, but still with no global rotation. Such an asymmetry would probably decrease, with a relaxation time of the same order as those found in Chapter~VII for the radial differences.

Finally, it would be good to consider the effect of the asymmetry in the galactic field (see Chapter~III). But this effect is likely not so important for the applications, because it only affects the external regions, which are invisible because of their very low density \citep{King1961a}.

\item From the observational side, it seems that much remains to be done. Accurate counts per luminosity class, as those made by Sandage for the cluster M3, would be very useful for other clusters. Furthermore, errors could probably be reduced by repeating the counts for several photographs of the same cluster. (\citealt{Tayler1954} noted big differences between the plates obtained with two different telescopes.)

The comparisons of accurate observation with a sufficiently elaborated theory would particularly allow for:
\begin{itemize}
\item a precise physical determination of the mass-luminosity relation; to push this relation as far as possible toward the low luminosities, one should particularly study the closest clusters. In the future, space observatories would allow us to reach much higher magnitudes;
\item an estimate of the age of the clusters, and perhaps information about their formations, when analyzing the differences between the present-day structure of the clusters and their theoretical final structure. However, these differences are small for most of the clusters, so that this analysis would require a high precision for the observations. One should start with the most massive clusters, where the largest differences remain;
\item a study of the galactic field and of the orbits of the globular clusters.
\end{itemize}

\item Here, we have only focussed on globular clusters. The models obtained could be applied to other objects: galactic clusters \tn{star clusters in the disk of the galaxy, as opposed to the globulars that are further away in the halo}, elliptical galaxies and cores \tn{central, spherical regions} of spiral galaxies, galaxy clusters. But in every case, difficulties appear, making the application of the theory more doubtful or more complex. That is, we think that the theory should first focus on describing properly the relatively simple case of globular clusters (which it is still far from doing) before extending its ambitions to more involved problems.

The galactic clusters would be extremely interesting, for many reasons: being closer, less massive stars are visible there; their ages are not all the same and thus we can observe all the phases of the dynamical evolution; and it is possible to measure individual velocities. Unfortunately, there are two major difficulties: the number of stars is small, so that statistical fluctuations forbid an accurate estimate of the projected density and of the other quantities; and above all, as \citet{Spitzer1958a} showed, the evolution of the cluster is largely influenced by passages near interstellar clouds.

\item Finally, we note some minor issues: the enhancement of the computation of the perturbations between the stars; the study of the accumulation of central energy as multiple stars and the final evolution of the cluster (see the end of Chapter~V); the influence of the motion of the cluster within the galaxy on the escape of stars (this could be simplified as a three-body problem: galactic center, cluster, star).
\end{enumerate}

One sees that there is no shortage of work. Stellar dynamics is a rising science, where almost everything remains to be done. It looks today much more like a collection of isolated attempts than like a homogeneous doctrine. These attempts have only led to isolated results, with limited range; the major problems, far from being solved, have remained almost untouched.

As we have seen, the issues do not arise from a poor knowledge of the physical mechanisms involved, but only from the complexity of the calculations. But today, a new, extremely powerful, weapon exists to overcome this kind of issues: the electronic computers. That is, we think that stellar dynamics should now leave the relative state of neglect where it has been left and spark off the interest and the long-term efforts of a growing number of researchers. It seems worthless to underline the importance of the results that could be obtained: the understanding of the collective dynamical phenomena is directly linked to the solving of the major cosmogonical and cosmological problems, and thus to our understanding of the Universe. We hope that the present essay, although it has a very limited scope, will contribute to the illumination of the new perspectives and give to others the wish to explore them.

\section*{Acknowledgments}

My acknowledgements go to Prof.~A.~Danjon, director of the Institut d'Astrophysique; to Prof.~E.~Schatzman, my supervisor from whose teaching I got the initial idea of this work, and who always encouraged and helped me; to Messieurs L.~Malavard and J.F.~Denisse, who accepted to be my research sponsors; to J.~Arsac, who created and organized the Service de Calcul Num\'erique de l'Observatoire de Meudon and taught my colleagues and myself the computational techniques on an electronic device; to the staff of this department, as well as the Bureau de Calcul de l'Institut d'Astrophysique, to Mrs Hernandez, Mrs Lagorce, Messieurs Charbey and de Postel, who assisted me for the technical realization of this work.

\onecolumn
\begin{center}
List of the main notations\\
The number is those of the equation where the notation is used first.\\
\begin{tabular}{ll|ll|ll|ll|ll|ll}
\hline
\hline
$a$ & 2.9 & $F$ & 2.23 & $L_e$ & 5.5 & $q$ & 2.19 & $t$ & 2.1 & $\fb{Z}$ & 5.1 \\
$A$ & 2.2 & $\fb{F}$ & 2.31 & $\mc{L}$ & 2.39 & $Q$ & 2.24 & $T$ & 2.24 &  & \\
$b$ & 2.36 & $F_2$ & 6.1 & $m$ & 2.1 & $\fb{Q}$ & 2.32 & $\fb{T}$ & 2.32 & $\alpha_1$ & 4.3 \\
$c$ & 2.36 & $\fb{F}_2$ & 6.5 & $M$ & 2.41 & $r$ & 2.7 & $U$ & 2.3 & $\alpha_2$ & 4.3 \\
$c_2$ & 6.6 & $g$ & 2.12 & $\fb{M}$ & 2.43 & $r_e$ & 2.8 & $\fb{U}$ & 2.32 & $\beta$ & 2.31 \\
$C$ & 6.10 & $H$ & 4.26 & $\fb{M}_e$ & 3.12 & $r_m$ & 2.13 & $\fb{U}_e$ & 3.5 & $\gamma$ & 2.31 \\
$D$ & 2.24 & $\fb{H}$ & 4.31 & $\fb{M}_P$ & 2.51 & $R$ & 2.24 & $U_G$ & 3.1 & $\gamma_2$ & 6.5 \\
$\fb{D}$ & 2.32 & $K$ & 4.33 & $\fb{M}_\fb{R}$ & 2.45 & $\fb{R}$ & 2.32 & $\fb{U}_0$ & 7.1 & $\theta$ & 6.22 \\
$D_P$ & 2.47 & $K_2$ & 7.10 & $\mc{M}$ & 2.38 & $\fb{R}_e$ & 3.12 & $\fb{U}_{\infty}$ & 5.7 & $\lambda$ & 3.11 \\
$\fb{D}_P$ & 2.48 & $K_D$ & 4.7 & $\mc{M}_e$ & 3.3 & $s$ & 7.23 & $v$ & 2.3 & $\mu$ & 6.4 \\
$E$ & 2.2 & $L$ & 2.41 & $n$ & I & $S$ & 2.27 & $x$ & 2.1 & $\rho$ & 2.5 \\
$\fb{E}$ & 2.31 & $\fb{L}$ & 2.43 & $n_m$ & 6.24 & $\fb{S}$ & 2.32 & $y$ & 2.1 & $\rho_p$ & 2.46 \\
$f$ & 2.2 & & & $p$ & 6.21 & &  & $z$ & 2.1 & &  \\
\hline
\end{tabular}\\
The symbol $'$ indicates the derivation with respect to $E$ or $\fb{E}$.\\
The subscript $e$ corresponds to the values taken by the quantities at the boundary of the cluster.\\
The symbols in capital letters ($F$, $D$, $R$, ...) generally represent the ``normalized variables''; the bold font ($\fb{F}$, $\fb{D}$, $\fb{R}$, ...) represent the ``canonical variables'' (see Chapter~II).
\end{center}
\begin{multicols}{2}


\end{multicols}
\end{document}